\RequirePackage{fix-cm}  
\documentclass[pre,amsmath,amssymb,amsfonts,superscriptaddress,lengthcheck,twocolumn,floatfix,longbibliography]{revtex4-2}
\usepackage{graphicx}
\usepackage{subfigure}
\usepackage{verbatim}
\usepackage{dcolumn}
\usepackage{bm}
\usepackage{epsf}
\usepackage[T1]{fontenc}
\usepackage[utf8]{inputenc}
\usepackage{multirow}
\usepackage{array, boldline, rotating}
\usepackage{xcolor}
\usepackage[colorlinks=true,citecolor=blue,linkcolor=blue,urlcolor=blue]{hyperref}%

\usepackage{xr}
\usepackage{upgreek}

\newcommand{\diag}{\mathop{\mbox{diag}}}
\newcommand{\s}[1]{{\color{black}#1}}

\renewcommand{\eqref}[1]{(\ref{#1})}

\begin{document}

\title{Quantum annealing in the NISQ era: railway conflict management}

\author{Krzysztof Domino}
\email{kdomino@iitis.pl}
\affiliation{Institute of Theoretical and Applied Informatics, Polish
  Academy of Sciences, Ba{\l}tycka 5, 44-100 Gliwice, Poland}
\author{M\'aty\'as Koniorczyk}
\email{koniorczyk.matyas@wigner.hu}
\affiliation{Wigner Research Centre, H-1525 Budapest, Konkoly-Thege M. \'ut 29-33, Budapest, Hungary}
\author{Krzysztof Krawiec}
\email{krzysztof.krawiec@polsl.pl}
\affiliation{Silesian University of Technology, Faculty of Transport and Aviation Engineering}
\author{Konrad Ja\l{}owiecki}
\email{dexter2206@gmail.com}
\affiliation{Institute of Physics, University of Silesia, Katowice, Poland}
	\author{Sebastian Deffner}
	\email{deffner@umbc.edu}
	\affiliation{Department of Physics, University of Maryland, Baltimore County, Baltimore, Maryland 21250, USA}
	\affiliation{Instituto de F\'isica `Gleb Wataghin', Universidade Estadual de Campinas, 13083-859, Campinas, S\~{a}o Paulo, Brazil}
\author{Bart\l{}omiej Gardas}
\email{bartek.gardas@gmail.com}
\affiliation{Institute of Theoretical and Applied Informatics, Polish Academy of Sciences, Ba{\l}tycka 5, 44-100 Gliwice, Poland}
\date{\today}

\begin{abstract}
We are in the Noisy Intermediate-Scale Quantum (NISQ) devices' era, in which quantum hardware has become available for application in real-world problems. However, demonstrating the usefulness of such NISQ devices are still rare. In this work, we consider a practical railway dispatching problem:  delay and conflict management on single-track railway lines. We
examine the issue of train dispatching consequences caused by the arrival of an already delayed train to the network segment being considered.
This problem is computationally hard and needs to be solved almost in real-time. We introduce a quadratic unconstrained binary optimization (QUBO) model of this problem, compatible with the emerging quantum annealing technology. The model's instances can be executed on present-day quantum annealers. As a proof-of-concept, we solve selected real-life problems from the Polish railway network using D-Wave quantum annealers. As a reference, we also provide solutions calculated with classical methods, including those relevant to the community (linear integer programming) and a sophisticated algorithm based on tensor networks for solving Ising instances. Our preliminary results illustrate the degree of difficulty of real-life railway instances for the current quantum annealing technology. Moreover, our analysis shows that the new generation of quantum annealers (the advantage system) perform much worse on those instances than its predecessor.
\end{abstract}

\maketitle

\section{Introduction}

Concentrated efforts all around the globe \cite{Raymer2019QST,Riedel2019QST,Yamamoto2019QST,Sussman2019QST,Roberson2019QST} are pursuing the development of viable quantum technologies. However, the technological challenges are immense, and it may still take some time before the first fault-tolerant quantum computers may become available for practical applications \cite{Sanders2017}. Thus, it is of instrumental importance to not only build a quantum literate workforce \cite{Aiello2021QST}, but to also ensure investments are made in realistic and societally beneficial avenues for development \cite{Roberson2021QST}.

Despite the fact that the first demonstrations of quantum advantage have been published \cite{Arute2019Nature}, currently available hardware is still prone to noise. Thus, it has been argued that we are in the era of noisy-intermediate scale quantum (NISQ) technologies \cite{Preskill2018quantum}. For instance, the D-Wave quantum annealer promises to deliver scalability beyond current classical hardware limitations. However, exploiting NISQ technologies often requires a different mathematical modeling framework. For instance, the D-Wave quantum annealer accepts an Ising spin-glass instance as its input and outputs solutions encoded in spin configurations. High-quality solutions are expected to be computed by these devices in a reasonable time, even for problems of the size which already bear practical relevance (currently, up to $5000$ variables on a sparse graph~\cite{dattani.szalay.19}). More importantly, a NISQ computer may not (yet) be able to outperform classical computers, however seeking and demonstrating amendable applications provides the instrumental guiding principle for the development of purpose-specific devices with genuine quantum advantage.

To date, at least in public domain research, most of the studied `` quantum'' problems are not directly relevant to a particular industrial application, but rather concern the solution of ``classical'' generic, computationally hard problems, such as, e.g. the traveling salesman problem or the graph coloring problem \cite{Gardas2019PRA}, see e.g. Ref.~\cite{sax2020approximate} for a comprehensive review. The present work belongs to a more practically motivated line of research: it is dedicated to making quantum computing more broadly accessible by demonstrating its applicability to a problem that can be easily understood especially by the non-physics audience -- conflict management in railway operation. Railway operations involve a broad range of scheduling activities, ranging from provisional timetable planning over rolling stock circulation planning, crew scheduling and rostering, etc. to operational train dispatching in case of disturbances, such as, e.g. severe weather, or unplanned events, technological breakdown. Many of these tasks require solving computationally expensive and overall challenging combinatorial problems. Various consequences of improper planing, e.g. incorrect dispatching decisions can be severe in terms of resources (e.g., time costs, passengers' satisfaction, financial loss).

In the domain of transportation research the applicability of quantum annealing has only been demonstrated for very few problems. For instance, Stollenwerk et al.~\cite{8643733} recently addressed a class of simplified air traffic management problems of strategic conflict resolution. Their preliminary results show that some of the challenging problems can be solved efficiently with the D-Wave $2000$Q machine. As to the railway industry, to the best of our knowledge, a preliminary version of the present work~\cite{2010.08227} was the first to apply a quantum computing approach to a problem in railway optimization.

The main purpose of the present paper is to elucidate how railway management problems can be solved with the currently available hardware. Naturally, we do not expect the current generation of the D-Wave annealer to outperform the best available classical algorithms. Rather, the present work is of pedagogical and instructive value as it provides an entry point for transportation research into quantum computing, and demonstrates the applicability of a NISQ computer. To this end, we solve the delay and conflict management on an existing Polish railway, whose real-time solution is of paramount importance for the local community.

Realizing that especially in the NISQ era there is still a significant language barrier between foundational quantum physics and real-life applications, the present paper strives to be as introductory and self-contained as possible. In particular, Sec.~\ref{sec::lit_dispatching} provides a brief review of railway conflict management as well as quantum annealing. Our model of the ``real'' problem is then outlined in Sec.~\ref{sec::ourmodel}, before we discuss our findings in Sec.~\ref{sec::numerical_studies}. The discussion is concluded with a few remarks on future research directions in Sec.~\ref{sec::conclusions}.
In the supplemental material we give a fully detailed description of our model. We include there also the description of a possible linear integer programming formulation that we use for comparison. The Supplemental Material contains a number of additional partcular solutions.

\section{Railway dispatching problem on single-track
lines}\label{sec::lit_dispatching}

Railway dispatching problem management is a complex area of transportation research.  Here we focus on the delay management on single-track railways. This problem concerns the operative modifications of train paths upon disturbances in the railway traffic. Incorrect decisions may cause the dispatching situation to deteriorate further by propagating the delay, resulting in unforeseeable consequences. Henceforth, we discuss this problem's details and survey the relevant part of the literature. Although we focus on single-track railway lines, some considerations may also be applicable to multi-track railways~\cite{cacchiani2014overview}.

\subsection{Problem description}
\label{problemdescr}

Consider a part of a railway network in which the traffic is affected by a disturbance. As a result, some trains cannot run according to the original timetable. Hence, a new, feasible timetable should be created promptly, minimizing unwanted consequences of the delay.
To be more specific, we are given a part of a railway network (referred to as the \emph{network}), such as e.g. those depicted in Figs.~\ref{fig::small_line} and \ref{fig::line}. The network is divided into \emph{block sections} or \emph{blocks}~\footnote{This
term originated in the railway signalling terminology. In general, it refers to a section of the railway line between two signal boxes.}:  sections which can be occupied by at most one train at a time. The block sections are labeled with numbers in the figures.
We focus on single-track railway lines.
These include \emph{passing sidings} (referred to as \emph{sidings}): parallel
tracks, typically at stations; the blocks labeled with upper indices in parantheses in the figures. Via the sidings, trains heading in opposite directions can
meet and pass (M-P), while trains heading in the same directions can meet
and overtake (M-O).

All trains run according to a \emph{timetable}. Examples of timetables are illustrated in Figs~\ref{fig::diagram} and \ref{fig::small_diagram} in the form of train diagrams, and will be explained later in Section~\ref{sec::line}. We assume that the initial timetable is \emph{conflict free} and that it meets all feasibility criteria. The criteria may vary \cite{TORNQUIST2007342, Lamorgese2018} depending on the railway network in question. The possible variants include technical requirements such as speed limits, dwell times, and other signaling-imposed requirements, as well as rolling stock circulation criteria, and passenger demands for trains to meet.
The railway delay management problem can be viewed from various perspectives, including that of a passenger, the infrastructure manager, or a transport operation company~\cite{TORNQUIST2007342, Lamorgese2018, Jensen2016}. Here, we look at this problem from the perspective of the infrastructure manager, who is to make the ultimate decision about the modifications and is in the position to prioritize the requirements.

In what follows, we assume that -- for whatever reason -- a delay occurs.
Hence, some trains' locations differ from the scheduled ones. A \emph{conflict} is \s{an} inadmissible situation in which at least two trains are supposed to occupy the same block section. For instance, if an already delayed train would continue its trip according to the original plan shifted in time with the delay while  the other trains would run according to the original timetable, multiple trains could meet in the \s{same} block, as illustrated in Fig.~\ref{fig::s_conflict}.

The objective of conflict management is thus to redesign the timetable to avoid conflicts (like in Fig.~\ref{fig::s_solution} in our example), and minimize delays. The
overall delay of a train is the sum of two types of delays. A \emph{primary
delay} is caused by an initial disturbance directly. Obviously  there is a  minimal amount of time needed for a train to reach further destinations, e.g., due to speed limits, hence, the primary delay of a train propagates through the network; it would be unavoidable even in the  absence of any other trains.
The primary delay is thus a lower bound on the overall delay.
The delay of a train beyond the primary delay is termed as the secondary delay.
These are induced by conflicts that have to be resolved by appropriate dispatching decisions.
The objective of the optimization of these decisions is the minimization of a suitable function of secondary delays, e.g., their maximum or a weighted sum.  Note that there are many other
practically relevant options for the objective function~\cite{8795577}, e.g.
the total passenger delay or the cost of operations, and some of these are also
inline with our approach.

The mathematical treatment of railway delay and conflict management
leads to NP-hard problems \footnote{\s{In computational complexity theory, NP-hardness (non-deterministic polynomial-time hardness) is the defining property of a class of problems that are informally "at least as hard as the hardest problems in NP, that is the class of problems that can be solved in polynomial time on a non-deterministic Touring machine \cite{van1991handbook}."}}; certain simple variants are
NP-complete~\cite{cai_fast_1994}. It is broadly accepted that these
problems are equivalent to job-shop models with blocking
constraints~\cite{Szpigel1973}, given the release and due dates of the
jobs and depending on the requirements of the model and additional
constraints such as recirculation or no-wait. The correspondence
between the metaphors is the following. Trains are the jobs and block
sections are the machines. Concerning the objective functions, the (weighted)
total tardiness or make-span is typically addressed, which is the
(weighted) sum of secondary delays or the minimum of the largest
secondary delay in the railway context. So with the standard notation
of scheduling theory~\cite{PinedoBook}, our problem falls into the
class $J_m|r_i,d_i,block|\sum_jw_jT_j$.

\subsection{Existing algorithms}

The following summary of railway dispatching and conflict
management is focused on the works that are closely related to the problem
addressed by us. A more comprehensive review of the huge
literature on optimization methods applicable to railway
management problems can be found in many related publications,
notably in Refs.~\cite{Lamorgese2018, 8795577, Cordeau1998, tornquist2006,
  Dollevoet2018, Corman20151274, CACCHIANI2012727}.

On a single-track line, the possible actions that can be taken to
reschedule trains are the following: allocating new arrival and departure times,
changing tracks and platforms, and reordering the trains by adjusting
the meet-and-pass plans~\cite{Hansen2010, Lamorgese2018, 8795577}.  An
important issue in modeling single-track lines is the handling of
sidings (stations). As pointed out in~\cite{lange_approaches_2018},
there are three approaches:
In the \emph{parallel machine approach}, in which its is assumed that each track within the
 siding corresponds to a separate machine in the job shop, thereby
 losing the possibility of flexible routing i.e., changing track
 orders at a station afterward.   In the \emph{Machine unit approach}
   parallel tracks  are treated as additional units of the same machine.
 Finally, in the \emph{buffer approach} the sidings at the same location are handled as
   buffers without internal structure, therefore not warranting the
   feasibility of track occupation planning at a station.
We adopt the buffer approach in our model.

As to the nature of the decision variables, two major classes of models
can be identified:
\begin{itemize}
\item \emph{Order and precedence variables} prescribe the order in which
  a machine processes jobs, i.e., the order of trains passing a
  given block section in the railway dispatching problem on single-track
  lines.
\item \emph{Discrete time units}, in which the decision variables belong to
discretized time instants; the binary variables describe whether or not the event
happens at a given time.
\end{itemize}

These two approaches lead to different model structures, which
are hard to compare. The \emph{discrete time units} approach
appears to be more suitable for a possible QUBO formulation, but it
leads to a large number of decision variables and thus worse
scaling. On the other hand, the \emph{order and precedence variables approach} can
lead to a representation of the problem with alternative
graphs~\cite{mascis2002job, dariano2007branch}, which is an intuitive
picture. The solution of this problem representation leads to mixed-integer programs that can
possibly be solved with iterative methods (such as branch-and-bound),
which makes them unsuitable for a reformulation to QUBO.  Time-indexed
variables, on the other hand, can result in pure binary problems that
can be transformed into QUBOs~\cite{venturelli2016job}, so we follow
the latter approach.

Returning to \s{Ref.}~\cite{lange_approaches_2018}, the authors considered the problem adopting the \emph{parallel machine approach} and the \emph{machine unit approach},
with \emph{order and precedence variables}; addressing the problem $J_m|r_i,d_i,block,rcrc|\sum_jT_j$ in  the
standard notation of scheduling theory. In comparison, in the present work will adopt
slightly different constraints and objectives, namely,
$J_m|r_i,d_i,block|\sum_jw_jT_j$.
 As to decision
variables, we opt for discrete time units and time-indexed variables \s{\footnote{For
the sake of completeness, in the Supplemental material we demonstrate that the problem can also be encoded with
precedence variables and handled by a linear solver}}.

In \s{Ref.}~\cite{zhou_single-track_2007}, Zhou and Zhong considered the
problem of timetabling on a single-track line. The starting times
of trains and their stops are given, and a feasible schedule is to be designed to minimize the total running time of (typically passenger) trains. Although their problem, notably its objective function and the input, is different, the constraints are similar to those of our problem. The authors also deal with
conflicts, dwell times, and minimum headway times for entering a segment of the railway line. They handle the problem with reference to
resource-constrained project scheduling. Their decision variables are
the discretized entry and leave times of the trains at the track
segments, binary precedence variables describing the order of the
trains passing a track segment, and time-indexed binary variables
describing the occupancy of a segment by a given train at a given
time. They introduce a branch-and-bound procedure with an efficiently
calculable conflict-based bound in the bounding step to supplement the
commonly used Lagrangian approach. They demonstrate its
applicability to scheduling of up to $30$ passenger trains for a $24$-hour periodic planning horizon on a line with $18$ stations in China.

Harrod~\cite{harrod_modeling_2011} proposed a discrete-time railway
dispatching model, with a focus on conflict management.  In this
work, the train traffic flow is modeled as a directed hypergraph, with
hyperarcs representing train moves with various speeds. This may be
confined to an integer programming model with time-, train-, and
hypergraph-related variables and a complex objective function
covering multiple aspects. The model is demonstrated on an imaginary
single-track line with long passing sidings at even-numbered block
sections of up to 19 blocks in length. An intensive flow of trains at
moderate speeds is examined. The model instances are solved by CPLEX
in the order of 1000 seconds of computation time. As a practical
application, a segment of a busy North American mainline is used, on
which the model produced practically useful results. Bigger examples were
also experimented with, leading to the conclusion that the approach is
promising but that it needs more specialized technology than a standard mixed-integer programming (MIP)
solver to be efficient.

Meng and Zhou~\cite{meng_simultaneous_2014} describe a simultaneous
train rerouting and rescheduling model based on network cumulative
flow variables. Their model also employs discrete-time-indexed
variables. They implement a Lagrangian relaxation solution algorithm
and make detailed experiments showing that their approach performs
promisingly on a general n-track railway network. In the introduction of their article they tabulate numerous timetabling and dispatching
algorithms.

This brief survey of the extensive literature confirms that the
problem of railway dispatching and conflict management is indeed a
good candidate for demonstrating new computational technology
capable of solving hard problems. Only a very few models have
been put into practice. The size and complexity of realistic
dispatching problems make it challenging for the models to solve them with current
computational technology.

\subsection{Quantum annealing and related
methods}\label{sec::lit_quantum_annealing}

Let us now turn our attention to the main tools used in the present
study: quantum annealing techniques. These have their roots
in adiabatic quantum computing, a new computational
paradigm~\cite{kadowaki.nishimori.98}, which, under additional
assumptions, is equivalent~\cite{1366223} to the gate model of quantum
computation~\cite{nielsen.chuang.10} (provided that the specific
interactions between quantum bits can be realized
experimentally~\cite{biamonte.love.08}). Thus this emerging
technology promises to tackle complicated (NP-hard in
fact~\cite{npising}) discrete optimization problems by encoding them
in the ground state of a physical system: the Ising spin glass
model~\cite{Lidar18}. Such a system is then allowed to reach its
ground state ``naturally'' via an adiabatic-like
process~\cite{Lanting14}. An ideal adiabatic quantum computer would in this way
provide the exact optimum, whereas a \s{\emph{real}} quantum annealer is a
physical device that has noise and other inaccuracies. \s{Hence, currently existing quantum annealers are paradigmatic examples of the NISQ era \cite{Preskill2018quantum}. Their output is only} a
sample of candidates that is likely to contain the optimum. Hence,
quantum annealing can be regarded as a heuristic approach, which will
become increasingly accurate and efficient as the technology
improves.

\subsubsection{Ising-based solvers}
\label{sec::Ising_based_solvers}

  \begin{figure*}[t]
	\includegraphics[width=2\columnwidth]{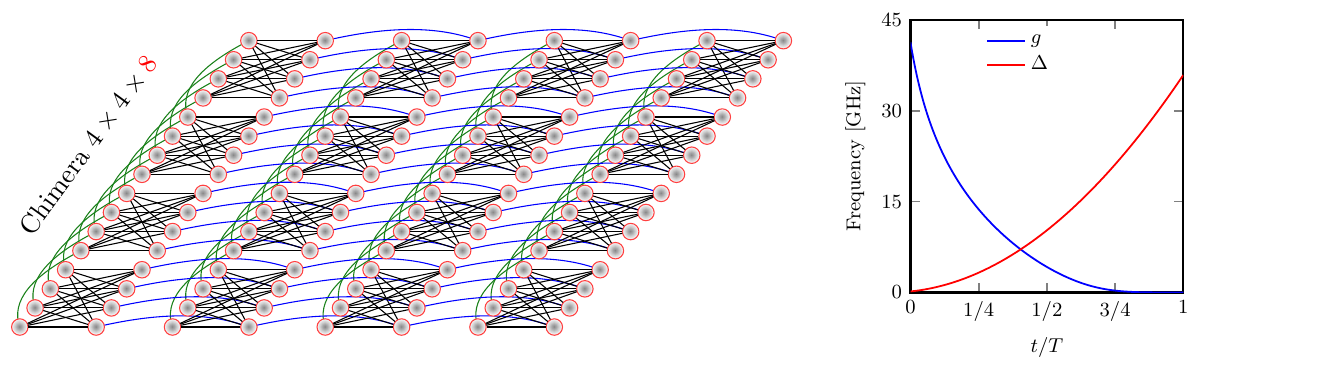}
	\caption{{\bf D-Wave processor specification}.
		Left: An example of the Chimera topology, here composed of $4 \times 4$ ($C_4$) grid consisting of
		clusters (units cells) of $8$ qubits each. The total number of variables (vertices) for this graph is
		$N=4\cdot 4\cdot 8 = 128$. A graph's edges indicate possible interactions between qubits. The maximum number
		of qubits is $N_{\text{max}}=2048$ for the Chimera $C_{16}$ topology, whereas the total number of connections between them
		is limited to $6000 \ll N_{\text{max}}^2$.
		Right: A typical annealing schedule controlling the evolution of a quantum processor, where $T$ denotes the time to complete
		one annealing cycle (the annealing time). It ranges from microseconds
		($\sim{}2\mu$s) to milliseconds ($\sim{}2000\mu$s). The parameters $g$ and
		$\Delta$ are used in~\eqref{eq:dwave}.}
	\label{fig:DW}
  \end{figure*}

The Ising model, introduced originally for the microscopic explanation
of magnetism, has been in the center of the research interests of physicists ever
since. It deals with a set of variables $s_i \in \{+1, -1\}$
(originally corresponding to the direction of microscopic magnetic momenta associated
with spins). The configuration of $N$ such variables is thus
described by a vector $\mathbf{s}\in \{+1, -1\}^N$. The model then
assigns an energy to a particular configuration:
\begin{equation}
  \label{eq:IsingClEnerg}
  E(\mathbf{s}) =\sum_{(i,j)\in E} J_{i,j}s_is_j + \sum_{i\in V} h_is_i,
\end{equation}
where $(V,E)$ is a graph whose vertices $V$ represent the spins, the
edges $E$ define which spins interact, $J_{i,j}$ is the strength of
this interaction, and $h_i$ is an external magnetic field at spin
$i$. Although the early studies of the model dealt with
configurations in which the spins were arranged in a one-dimensional
chain so that the coupling $J$ was non-zero for nearest neighbors
only, the model has been generalized in many ways, including the most general
setting of an arbitrary $(V,E)$ graph, i.e., incorporating the possibly of non-zero
couplings for any $i,j$ pair. Such a system is referred to as a spin
glass in physics. \s{For comprehensive reviews of generalizations of the Ising model we refer to the literature \cite{Wu1982RMP,Castellani2005JSM}.

From an operations research point of view, the physical model is interesting , since} it
describes is a computational resource for optimization.
The idea originated from the fact that in
physics, the minimum energy configuration determines many properties of
a material.

In mathematical programming, it is often more convenient to deal with
$0$-$1$ variables. By introducing new decision variables
${\mathbf{x}}\in \{0,1\}$ so that
\begin{equation}
  \label{eq:spin2bit}
  x_i = \frac{s_i+1}{2},
\end{equation}
and the matrix
\begin{equation}\label{eq::x_2_s}
      Q_{i,i} = 2\left( h_i - \sum_{j=1}^n J_{i,j}\right), \quad
      Q_{i,j} = 4J_{i,j},
\end{equation}
the minimization of the energy function is equivalent to solving a QUBO problem:
\begin{equation}
  \label{eq:qubogen}
    \text{min}\, \mathbf{x}^T Q \mathbf{x}  ,
    \quad \text{s.t.} \quad  \mathbf{x}\in \{0,1\}^N.
\end{equation}
Therefore, minimizing the Ising objective in \s{Eq.}~\eqref{eq:IsingClEnerg} is
equivalent to solving a QUBO.  Moreover, the matrix $Q$ can always be
chosen to be symmetric, as $Q=(Q'+Q'^T)/2$ defines the same objective.
Solvers based on Ising spin glasses are actual devices (or specialized
algorithms simulating them or calculating their relevant properties)
that can handle models of this form only. The technology offers the
possibility of efficiently tackling computationally hard problems
(when formulated as a QUBO, which is possible for all linear or
quadratic $0$-$1$ programs~\cite{glover_quantum_2019}). It
does have limitations with respect to size and accuracy, as will be
illustrated in the present case study, but it is likely that the technology will
continue to improve.

Simultaneously, with the rapid development of quantum annealing technology,
probabilistic \emph{classical} accelerators have been under massive development.
In recent years, we have witnessed significant progress in the field of programmable
gate array optimization solvers (digital annealers~\cite{FDA19}), optical Ising
simulators~\cite{OIS20}, coherent Ising machines~\cite{CIM17},  stochastic cellular
automata~\cite{STATICA}, and, in general, those based on memristor electronics~\cite{MHN20}.

It is therefore vital to develop modeling strategies to make
operational problems suitable for such models and to create novel techniques
for analyzing the obtained results. This should progress similarly to
how the powerful solvers for linear programs first
started appearing: modeling strategies for linear programs as well as sensitivity analysis had been developed ahead of the creation of the hardware.

\subsubsection{Quantum annealing}
\label{subsec:qannealing}

An essential step in finding the minimum of an optimization problem
[encoded in \s{Eq.}~\eqref{eq:IsingClEnerg}] efficiently is to map it to
its quantum version.  The mapping assigns a two-dimensional complex
vector space to each spin, and a complete spin configuration becomes
an element of the direct (tensor) products of these spaces. An
orthonormal basis (ONB) is assigned to the $-1$ and $+1$ values of
the variables; thus the product of these vectors will be an ONB
(called the ``computational basis'') in the
whole $\mathbb{C}^{2^N}$. The vectors with unit Euclidean norms are
referred to as ``states'' of the system; they encode the physical
configurations. The fact that the state can be an arbitrary vector,
 and not only an element of the computational basis means that the quantum
annealer can simultaneously process multiple configurations, i.e., inherent
parallelism.

As to the objective function, the spin variables are replaced by their
quantum counterpart: Hermitian matrices acting on the given spin's
$\mathbb{C}^2$ tensor subspace.
The product of spins is meant to be the direct
(tensor) product of the respective operators. Thus, the objective
function~\eqref{eq:IsingClEnerg} turns into a Hermitian operator,
referred to as the problem's Hamiltonian:
\begin{equation}
  \label{eq:Hproblem}
  \mathcal{H}_{\text{p}} : =
  E(\hat {\bf \sigma} ^z) =
  \sum_{\langle i, j\rangle \in \mathcal{E}} J_{ij} \hat\sigma_i^z \hat\sigma_j^z + \sum_{i\in\mathcal{V}}h_i \hat\sigma_i^z,
\end{equation}
whose lowest-energy eigenstate is commonly called the ``ground
state.'' Above, $\sigma_{i}^z$ denotes the Pauli $z$-matrix associated with \s{$i$th} qubit.
In the present case, it is an element of the computational
basis, so it represents also the optimal configuration of the
classical problem. Note that the energy of a physical system is
related (via eigenvalues) to a Hermitian operator, called its Hamiltonian.
Although it seems to be a significant complication
to deal with $\mathbb{C}^{2^N}$ instead of having $2^N$ binary
vectors, it has important benefits, the most remarkable of which is that they model realistic physical systems.

The main idea behind quantum annealing is based on the celebrated
adiabatic theorem~\cite{avron.elgart.99}. \s{Assume} that a quantum system can be prepared in the
ground state of an initial (``simple'') Hamiltonian
$\mathcal{H}_0$. Then it will slowly evolve to the ground state of the
final (``complex'') Hamiltonian $\mathcal{H}_{\text{p}}$ in~\eqref{eq:Hproblem}
that can be harnessed to encode the solution of
an optimization problem~\cite{Lidar18}. In particular, the dynamics of
the current D-Wave $2000$Q quantum annealer is governed by the
following time-dependent Hamiltonian~\cite{Lanting14,Ozfidan19}:
\begin{equation}
\label{eq:dwave}
\mathcal{H}(t)/(2\pi\hbar)= -g(t) \mathcal{H}_0  -\Delta(t) \mathcal{H}_{\text{p}'},
\quad
t \in [0, T].
\end{equation}
Here the original problem's Hamiltonian in~\eqref{eq:Hproblem} must be
converted into a bigger one $\mathcal{H}_{\text{p}'}$ whose graph is
compliant with the existing hardware can realize: the ``Chimera
graph'' in case of DWave 2000Q; see Fig.~\ref{fig:DW}. The original problem's graph will be
the minor of this graph.  This procedure, called ``minor embedding'',
is standard in quantum annealing procedures (see also
Section~\ref{sec::simple_ex} for a simple graphical representation
of this \emph{Chimera embedding}).

In fact, many relevant optimization problems are defined on
dense graphs. Fortunately, even complete graphs can be embedded into
a Chimera graph~\cite{choi.08}. There is, however, substantial
overhead, which effectively limits the size of the computational graph
that can be treated with current quantum
annealers~\cite{CIvDW,Lanting18}. This is, nevertheless, believed to
be an engineering issue that will most likely be overcome in the near
future~\cite{dattani.szalay.19,onodera.ng.19}.  After the Chimera
embedding, the Hamiltonian describing the system reads as follow:
\begin{equation}
\label{eq:Hp}
\mathcal{H}_{\text{p}'} = \sum_{\langle i, j\rangle \in \mathcal{E}} J'_{ij} \hat\sigma_i^z \hat\sigma_j^z + \sum_{i\in\mathcal{V}}h'_i \hat\sigma_i^z,
\quad
\mathcal{H}_0 = \sum_{i} \hat \sigma_{i}^x,
\end{equation}
where $\sigma_{i}^x$ is the $x$ Pauli matrix associated with \s{$i$th} qubit.
The annealing time $T$ varies from
microseconds ($\sim{}2\mu$s) to milliseconds ($\sim{}2000\mu$s)
depending on the specific programmable schedule~\cite{Lanting14}. As shown in
Fig.~\ref{fig:DW}, during the evolution, $g(t)$ varies from
$g(0) \gg 0$ (i.e., all spins point in the $x$-direction) to
$g(T)\approx 0$, whereas $\Delta(t)$ is changed from
$\Delta(0)\approx 0$ to $\Delta(T) \gg 0$ (i.e.,
$\mathcal{H}(T) \sim \mathcal{H}_{\text{p}'}$). The Pauli operators
$\hat\sigma_i^z$, $\hat\sigma_i^x$ describe the spin's degrees of
freedom in the $z$- and $x$-direction, respectively.  Note that the
Hamiltonian $\mathcal{H}_{\text{p}}$ is classical in the sense that all its
terms commute (which is the result of their multiplication being
independent of the order). Thus, as mentioned previously, its
eigenstates translate directly to classical variables, $q_i=\pm 1$,
which are introduced to encode discrete optimization problems.

The annealing time, $T$ in \s{Eq.}~\eqref{eq:dwave}, is an important
parameter of the quantum annealing process: it must be chosen so
that the system reaches its ground state while the adiabaticity is at
least approximately maintained.  The adiabatic theorem gives us a
guidance in this respect. In the spectrum of the Hamiltonian
in \s{Eq.}~\eqref{eq:dwave}, there is a difference between the energy of the
ground state(s) and the energy of the state(s) just above it in energy scale. This
difference is known as the (spectral) ``gap'', and its
minimum value in the course of the evolution determines the required
computation time if certain additional conditions hold. Roughly
speaking, the bigger the gap, the faster the quantum system reaches
its ground state (the dependence is actually quadratic; see Ref.~\cite{PhysRevA.65.012322} for a detailed
discussion). Thus, if the run time is not optimal, there is a finite
probability of reading out an excited state instead of the true
ground state.

Therefore, the ideal approach would be to calculate the optimal time,
and only then let the system evolve for as long as it is
necessary. However, the spectrum is actually unknown, and its
determination is at least as hard as solving the optimization problem
itself.  Hence, in practice, a reasonable annealing time is \s{determined from an educated guess}, and the evolution is repeated reasonably many times,
resulting in a \emph{sample} of possible solutions (over different
annealing times as well as other relevant parameters).  The one with
the lowest energy is considered to be the desired solution, albeit
there is a finite probability that it is not the ground state. A
quantum annealer is thus a probabilistic and heuristic solver.
Concerning the benchmarking of quantum annealers, consult also~\cite{PhysRevX.5.031026}.

As a side note, it should be stressed that it is not always
possible to maintain the adiabatic evolution. As an example, consider
the second--order phase transition
phenomenon~\cite{QFT,Dziarmaga2005,Dziarmaga10}, in which even a
short--lasting lack of adiabaticity will result in the creation of
topological defects preventing the system from remaining in its
instantaneous ground state. This effect, on the other hand, is a clear
manifestation of the \s{quantum} Kibble--Żurek
mechanism \s{Ref.}~\cite{Kibble76,Kibble80,Zurek85,Francuz2016PRB,Deffner2017PRE,Gardas2017PRB,Gardas2018SR} and can be
used to detect departures from adiabaticity.

\subsubsection{Classical algorithms for solving Ising problems}

An additional benefit of formulating problems in terms of Ising-type
models is that the existing methods developed in statistical and
solid-state physics for finding ground states of physical systems can
also be used to solve a QUBO on classical hardware. Notably,
variational methods based on finitely correlated states (such as
matrix product states for 1D systems or projected entangled pair
states suitable for 2D graphs) have had a very extensive development in the
past few decades. A quantum information theoretic insight into
density matrix renormalization group methods
(DMRG~\cite{RevModPhys.77.259}) -- being the most powerful numerical
techniques in solid-state physics at that time -- helped in proving
the correctness of DMRG. These methods also led to a more general view of the
problem~\cite{PhysRevB.73.094423}, resulting in many algorithms that
have potential applications in various problems, in particular solving
QUBOs by finding the ground state of a quantum spin glass. We have
used the algorithms presented in \s{Ref.}~\cite{rams2018heuristic} to solve the models derived in the present
manuscript.

Neither the quantum devices nor the mentioned classical algorithms do always
provide the energy minimum and the corresponding ground state (as it
is not trivial to reach it~\cite{PhysRevA.100.042326}) but possibly another
eigenstate of the problem with an eigenvalue (i.e., a value of the
objective function) close to the minimum. The corresponding states are
referred to as ``excited states.'' In contrast to many problems of physics where it is only the true ground state is relevant, in optimization problems the excited stats are often also useful. Another important point in
interpreting the results of such a solver is the degeneracy of the
solution: the possibility of having multiple equivalent optima.

In analyzing these optima, it is helpful that for up to 50 variables, one
can calculate the exact ground states and the excited states closest
to them using a brute-force search on the spin configurations with
GPU-based high-performance computers. In the present work, we also use
such algorithms, in particular those introduced in
\s{Ref.}~\cite{jaowiecki2019bruteforcing} for benchmarking and evaluating our
results for smaller examples. This way we can compare the exact
spectrum with the results obtained from the D-Wave quantum hardware
and the variational algorithms.

\section{Our model}
\label{sec::ourmodel}

Here we describe our model in brief. The supplemental material
provides a description of all its details. First we formulate it as a
constrained quadratic $0-1$ program, then we turn it into a QUBO using
penalties.

\subsection{Integer formulation of the constaints}

Let us now again envise our single-track railway network.
First observe that it is only the leave times of trains from station blocks that the dispatchers decide upon. Let us denote the station blocks by $s \in \mathcal{S}$, and the set of trains by $j\in \mathcal{J}$. We will formulate the problem entirely in terms of the secondary delays: $d_s(j,s)$ stands for the value of the secondary delay of the train $j$ at the station block $s$. The detailed description in the Supplemental Material makes it clear that these values, along with the original timetable and the technical data (i.e. the railway network topology and time required for each train to pass a block) determine a modified timetable. In what follows, this description will be referred to as the \emph{delay representation}.

In order to approach a 0-1 model, we discretize the secondary delays requiring that
\begin{equation}\label{eq::dlim1}
d_s(j,s) \in \mathbb{N}, \quad 0 \leq d_s(j,s) \leq d_{\text{max}}(j),
\end{equation}
where a reasonable upper bound $d_{\text{max}}(j)$ can be obtained from some fast heuristics, and the time is measured in integer minutes from now on; a suitable scale for railway problems.
When formulating constraints, it is better to work with the actual (discretized) delay $d(j,s) = d_S(j,s) + d_U(j,s)$ of train $j$ at station block $s$.
At this stage we have a set of potential decision variables with finite ranges that already facilitate the formulation of a linear model for the problem, as done in the Supplemental Material.

As of constraints, we have considered the following ones which cover
the requirements of the particular railway operator. In the
Supplemental Material we describe them in very detail, here we just
give a brief summary:
\begin{description}
\item[The minimum passing time condition] ensures that no block sections are passed by any train faster than allowed:
\begin{equation}\label{eq::toutinequal_p}
d(j,\rho_j(s)) \geq d(j, s) - \alpha(j, s, \rho_j(s)).
\end{equation}
where $\rho_j(s)$ stands for the station block section succeeding $s$ in train $j$-s sequence, while $\alpha(j, s, \rho_j(s))$ is the largest reserve the train can achieve by passing the blocks following $s$ up to and together with the next station block $\rho_j(s)$ possibly faster than originally planned. This can be calculated in advance.
\item[The single-block occupation] ensures that at most one train can be present in a block section at a time.
\begin{equation}\label{eq::tau1_p}
\begin{split}
d(j',s) & \geq
d(j,s) + \Delta(j, s, j', s) \\
&+\tau_{(1)}(j,s, \rho_j(s)).
\end{split}
\end{equation}
where $\Delta$ is the difference of the leave times of trains $j$ and
$j'$ from the block $s$, whereas $\tau_{(1)}$ is minimum time for
train $j$ to give way to another train going in the same direction
in the route $s \rightarrow \rho_j(s)$.  This condition is to be tested in this form for the
pair of trains $(j,j')$ if $j$ leaves $s$ before $j'$, i.e. $d(j',s) \geq d(j,s) + \Delta(j, s, j', s)$;
otherwise it has to be applied so that the order of the trains is
reversed.
\item[The deadlock condition] ensures that no pairs of trains heading
  in the opposite direction will be waiting for each other to pass the
  same blocks:
\begin{equation}\label{eq::tau2_p}
d(j', \rho_j(s)) \geq  d(j,s) + \Delta(j, s, j', \rho_j(s))
+ \tau_{(2)}(j, s, \rho_j(s)),
\end{equation}
where $\tau_{(2)}(j, s, \rho_j(s))$ is the minimum time required for
train $j$ to get from station block $s$ to $\rho_j(s)$. Similarly to
the previous condition, Eq.~\eqref{eq::tau2_p} is to be applied for the
pairs of trains $(j,j')$ so that $j'$ is supposed to leave the block
$\rho_j(s)$ after the train $j$ leaves $s$; otherwise the order of the
trains is reversed.
\item[The rolling stock circulation condition] ensures the minimal
  technological time $R(j,j')$ for a given train set arriving as train
  $j$ at its terminating station $s_{j, \text{end}}$ before operaing
  again as train $j'$:
  	\begin{equation}\label{eq::circdelays_p}
		d(j',1)  > d(j,s_{j, \text{end}-1}) - R(j,j').
      \end{equation}

      Certainly this condition has to hold for pairs of trains
      $(j,j')$ which are operated with the same train set according to
      the rolling stock circulation plan.
\end{description}
Having described the constraints, now we formulate the model as a 0-1
program and define the objective function.

\subsection{0-1 formulation}

To turn our model into a 0-1 problem, we introduce our final decision
variables
\begin{equation}
x_{s,j,d} = \begin{cases}
1,& d(j,s)=d\\
0,&\text{otherwise}
\end{cases},\label{eq:qubovarsdef1}
\end{equation}
which take the value of $1$ if train $j$ leaves station block $s$ at
delay $d$, and zero otherwise. In this way we have 0-1 variables with
indices from a finite set.

As of the objective function we opt for a weighted sum of delays:
\begin{equation}
\label{eq::weighted_p}
f(\mathbf{x}) = \sum_{j \in J} \sum_{s \in S_j^*} \sum_{d \in A_{j,s}}
f(d, j, s) \cdot x_{j,s,d},
\end{equation}
where $f(d, j, s)$ are the weights. Here $S_j^* = S_j \setminus \{ s_{j, \text{end}} \}$, where $S_j$ stands for a sequence
of station blocks the train runs through, $s_{j, \text{end}}$ stands for last station of $j$, and $A_{j,s}$ is the respective
range of delays. It is easy to see that the weights $f(d, j, s)$ can
be chosen so that they depend on the secondary delays only; conider
the Supplemental Material for details.

As of the constraints, let us first assume that each train leaves each
station block once and only once (recall that we do not allow for
recirculation):
\begin{equation}\label{eq::Q1}
\forall_{j} \forall_{s \in S_j} \sum_{d \in A_{j,s}}
x_{s,j,d} = 1.
\end{equation}
The minimum passing time condition in Eq.~\eqref{eq::toutinequal_p} reads
\begin{equation}\label{eq::Q2}
\forall_{j} \forall_{s \in S_j^{**}}
\sum_{d \in A_{j,s}}
\left(
\sum_{ d' \in D(d) \cap A_{j, \rho_j(s)}} x_{j, s, d}
x_{j, \rho_j(s), d'} \right) = 0,
\end{equation}
where $D(d) = \{0, 1, \ldots, d - \alpha(j, s, \rho_j(s)) -1\}$, and
$S_j^{**} = S_j \setminus \{s_{j, \text{end}}, s_{j, \text{end}-1}\}$.

As of the single block occupation condition, from Eq.~\eqref{eq::tau1_p} it follows that
\begin{widetext}
\begin{equation}\label{eq::Q3}
\forall_{(j,j') \in \mathcal{J}^0 (\mathcal{J}^1)}
\forall_{s \in S^*_{j,j'}}
\sum_{d \in A_{j,s}}\left(\sum_{d' \in B(d) \cap A_{j', s}}
x_{j, s, d} x_{j', s, d'}\right) = 0,
\end{equation}
where $B(d) = \{d + \Delta(j, s, j', s), d + \Delta(j, s, j', s)+ 1,\ldots, d +
\Delta(j,
s, j', s) + \tau_{(1)}(j,s, \rho_j(s))-1
\}$ is a set of delays that violates the block occupation condition.

Concerning the deadlock condition, for two trains heading in the
opposite direction, we have from Eq.~\eqref{eq::tau2_p} that
\begin{equation}\label{eq::Q31}
\forall_{j \in \mathcal{J}^0(\mathcal{J}^1),  j' \in
\mathcal{J}^1(\mathcal{J}^0) }
\forall_{s \in S^*_{j,j'}}
\sum_{d \in A_{j,s}}\left(\sum_{d' \in C(d) \cap A_{j', \rho_j(s)}}
x_{j, s, d}
x_{j', \rho_j(s), d'}\right) = 0
\end{equation}
where
$C(d) = \{d(j,s) + \Delta(j, s, j', \rho_j(s)), d(j,s) + \Delta(j, s,
j', \rho_j(s)) + 1, \ldots, d(j,s) + \Delta(j, s, j', \rho_j(s)) +
\tau_{(2)}(j,s, \rho_j(s))-1\}$, and $\mathcal{J}^0$, $\mathcal{J}^1$ are explained in
Eq.~\eqref{sup:test} of the Supplemental Material.

The rolling stock circulation condition of Eq.~\eqref{eq::circdelays_p} can
be formulated as
\begin{equation}\label{eq::circ_p}
  \forall_{j,j' \in \text{terminal pairs}}
  \sum_{d \in A_{j,s_{(j, end-1)}}}
  \sum_{ d' \in E(d) \cap A_{j', 1}} x_{j, s_{(j, end-1)}, d}
  \cdot x_{j', s_{(j,' 1)}, d'}  = 0,
\end{equation}
where $E(d) = \{0, 1, \ldots, d - R(j,j')\}$.
\end{widetext}
The objective funtion in Eq.~\eqref{eq::weighted_p} together with the
constraints in Eqs.~\eqref{eq::Q1} -- \eqref{eq::circ_p} comprise a quadratic constrained 0-1 formulation
of our model.

\subsection{QUBO formulation: penalties}
\label{sec::quboimplem}

Having our problem formulated as a constrained 0-1 program, we need to
make it unconstrained to achieve a QUBO form --
see Eq.~\eqref{eq:qubogen}. This is usually done with penalty
methods~\cite{luenberger2015linear}. It has been shown
in~\cite{glover_quantum_2019} that all binary linear and quadratic
programs translate to QUBO along some simple rules.
(An alternative, symmetry-based
approach~\cite{PhysRevApplied.5.034007} to constrained optimization
has also been proposed in which the adiabatic quantum computer device
is supposed to use a tailored $\mathcal{H}_0$ term in its dynamics of model in
Eq.~\eqref{eq:Hp}. As such a modification of the actual device is not
available to us, we remain using penalty methods.)

The problems one faces with a quadratic 0-1 program require certain
specific considerations when adopting the penalty method. Let us
outline this approach with a focus on our problem. As we have a linear
objective function Eq.~\eqref{eq::weighted_p}, it can be written as
a quadratic function because the decision variables are binary:
\begin{equation}
  \label{eq:linobj}
 \min_{\mathbf{x}} f(\mathbf{x})= \min_{\mathbf{x}} \mathbf{c}^T\mathbf{x} =
 \min_{\mathbf{x}} \mathbf{x}^T\diag(\mathbf{c})\mathbf{x}.
\end{equation}
(A general QUBO can contain linear terms as well; however, the solver
implementations accept a single matrix of quadratic
coefficients~\cite{DwaveDoc}, so transforming linear terms into
quadratic ones is more a technical than a fundamental step.)

We need to meet the constraints set out in Eqs.~\eqref{eq::Q2} -- \eqref{eq::circ_p}
to make the solution feasible. These constraints are
regarded as \emph{hard constraints}. To obtain an unconstrained
problem, we define a penalty function in the following way. We add the
magnitude of the constrains' violation, multiplied by some well-chosen coefficient,
to the objective function.

In particular, we shall have quadratic constraints in the form of
\begin{equation}
  \label{eq:pairconstr}
  \sum_{(i,j)\in \mathcal{V}_{\text{p}}} x_ix_j=0,
\end{equation}
excluding pairs of variables that are simultaneously $1$.
We can deal with such a constraint by adding to our objective the following
terms:

\begin{equation}
  \label{eq:pairpen}
  \mathcal{P}_{\text{pair}}(\mathbf{x})=p_{\text{pair}}  \sum_{(i,j)\in
  \mathcal{V}_\text{p}}(x_ix_j+x_jx_i),
\end{equation}
where $p_{\text{pair}}$ is a positive constant.
Additionally, from Eq.~\eqref{eq::Q1}, we have additional hard
constraints in the form of:
\begin{equation}
  \label{eq:sumconstr}
\forall \mathcal{V}_\text{s}\quad  \sum_{i\in \mathcal{V}_\text{s}} x_i=1.
\end{equation}
These constraints yield a linear expression that
can be transformed into the following quadratic penalty function:
\begin{equation}
  \label{eq:sumpenq}
  \mathcal{P'}_{\text{sum}}(\mathbf{x}) = \sum_{\mathcal{V_\text{s}}} p_{\text{sum}}\left(\sum_{i\in
\mathcal{V}_\text{s}}
x_i-1\right)^2.
\end{equation}
Next we replace the $x_i$s with $x_i^2$s in
the linear terms, and omit the constant terms as they provide only
an offset to the solution, yielding:
\begin{equation}
  \label{eq:sumpen}
  \mathcal{P}_{\text{sum}}(\mathbf{x})=\sum_{\mathcal{V_\text{s}}} p_{\text{sum}}  \left(
  \sum_{{i,j}\in \mathcal{V}^{\times 2}, i\neq j}x_ix_j  -  \sum_{i\in
  \mathcal{V}_\text{s}}x_i^2
   \right).
\end{equation}
So our effective QUBO representation is
\begin{equation}
  \label{eq::effqubo}
  \min_{\mathbf{x}} f'(\mathbf{x})= \min_{\mathbf{x}} \left( f(\mathbf{x}) +
  \mathcal{P}_{\text{pair}}(\mathbf{x})
  + \mathcal{P}_{\text{sum}}(\mathbf{x}) \right),
\end{equation}
which can be written in the form of Eq.~\eqref{eq:qubogen}. We shall
have many constraints similar in form to in Eq.~\eqref{eq:pairconstr}
and Eq.~\eqref{eq:sumconstr}, so we have one summand for each constraint in
the objective. (It would also be possible to assign a separate
coefficient to each of the constraints.)

Recall that in the theory of penalty
methods~\cite{luenberger2015linear} for continuous optimization, it is
known that the solution of the unconstrained objective will tend to a
feasible optimal solution of the original problem as the multipliers
of the penalties ($p_{\text{sum}}$ and $p_{\text{pair}}$ in our case)
tend to infinity, provided that the objective function and the
penalties obey certain continuity conditions.  As in our case both the
objective and the penalties are quadratic, this convergence would be
warranted for the continuous relaxation of the problem. And even
though we have a 0-1 problem, if we had an infinitely precise solution
of the QUBO, increasing the parameters would result in
convergence to an optimal feasible solution.

However, somewhat similarly to the continuous case (in which the
Hessian of the unconstrained problem diverges as the parameters grow,
making the unconstrained problem numerically ill-conditioned), the
properties of the actual computing approach or device makes it more
cumbersome to make a good choice of multipliers.

The parameters $p_{\text{sum}}$ and $p_{\text{pair}}$ have to be be
chosen so that the terms representing the constraints in this energy
do not dominate the original objective function. If the penalties are
too high, the objective is just a too small perturbation on it, which
will be lost in the noise of the physical quantum computer or in the
numerical errors of an algorithm modeling it. If, however, the penalty
coefficients are too low, we get infeasible solutions. In the ideal
case there is a ``feasibility gap'' in the spectrum of solutions.

The multipliers can be assigned in an \emph{ad hoc} manner by
experimenting with the solution; however, a systematic, possibly
problem-dependent approach to their appropriate assignment (as in
case of classical penalty methods; see~\cite{luenberger2015linear})
would be highly desirable in order to make the QUBO more
reliable and prevalent.

Having a QUBO representation of the problem at hand, let us turn our
attention to the actual instances of our model and the results
obtained for them.

\section{Results}
\label{sec::numerical_studies}

In this section, we discuss certain possible situations in train dispatching on
the railway lines managed by the Polish state-owned
infrastructure manager \emph{PKP Polskie Linie Kolejowe S.A.} (\emph{PKP PLK}
in what follows). In particular, we consider two single-track railway lines:
\begin{itemize}
	\item Railway line No. 216 (Nidzica -- Olsztynek section)
	\item Railway line No. 191 (Golesz\'{o}w -- Wis\l{}a Uzdrowisko section).
\end{itemize}

Railway line No. 216 is of national importance. It is a single-track
section of the passenger corridor Warsaw - Olsztyn, which has recently been
modernized. There are both \emph{Inter-City} (\emph{IC}) and
regional trains operating on the Nidzica -- Olsztynek section of line
No. 216. In this paper, we consider an official train schedule (as for
April, 2020). The purpose of the analysis in this section is to
demonstrate the application of our methodology to a real-life
railway section.

Railway line No. 191 is of local importance. The main train service on
the No. 191 railway line is Katowice – Wis\l{}a G\l{}ebce, operated by a local
government-owned company called \emph{``Koleje \'Slaskie''} (in English Silesian
  Railways; abbreviated \emph{KS}).  There are a few \emph{Inter-City}
trains of higher priority there as well.  Since 2020, the traffic at
this section has been suspended due to comprehensive renovation works
(a temporary rail replacement bus service is in operation). Our
problem instances are based on the planned parameters of the line
after its commissioning, based on public procurement documents
\cite{PKPPLK}. On the basis of these parameters, a cyclic timetable
has been created. The aim of analyzing this case is to show the broader
application possibilities of the methodology.

\subsection{The studied network segment}\label{sec::line}
\label{subsec:lines}

\begin{figure*}[t]
    \subfigure[Train diagram for the timetable of the line in
	Fig.~\ref{fig::small_line}.\label{fig::small_diagram}]{
	\includegraphics[scale = 0.6, width=\columnwidth]{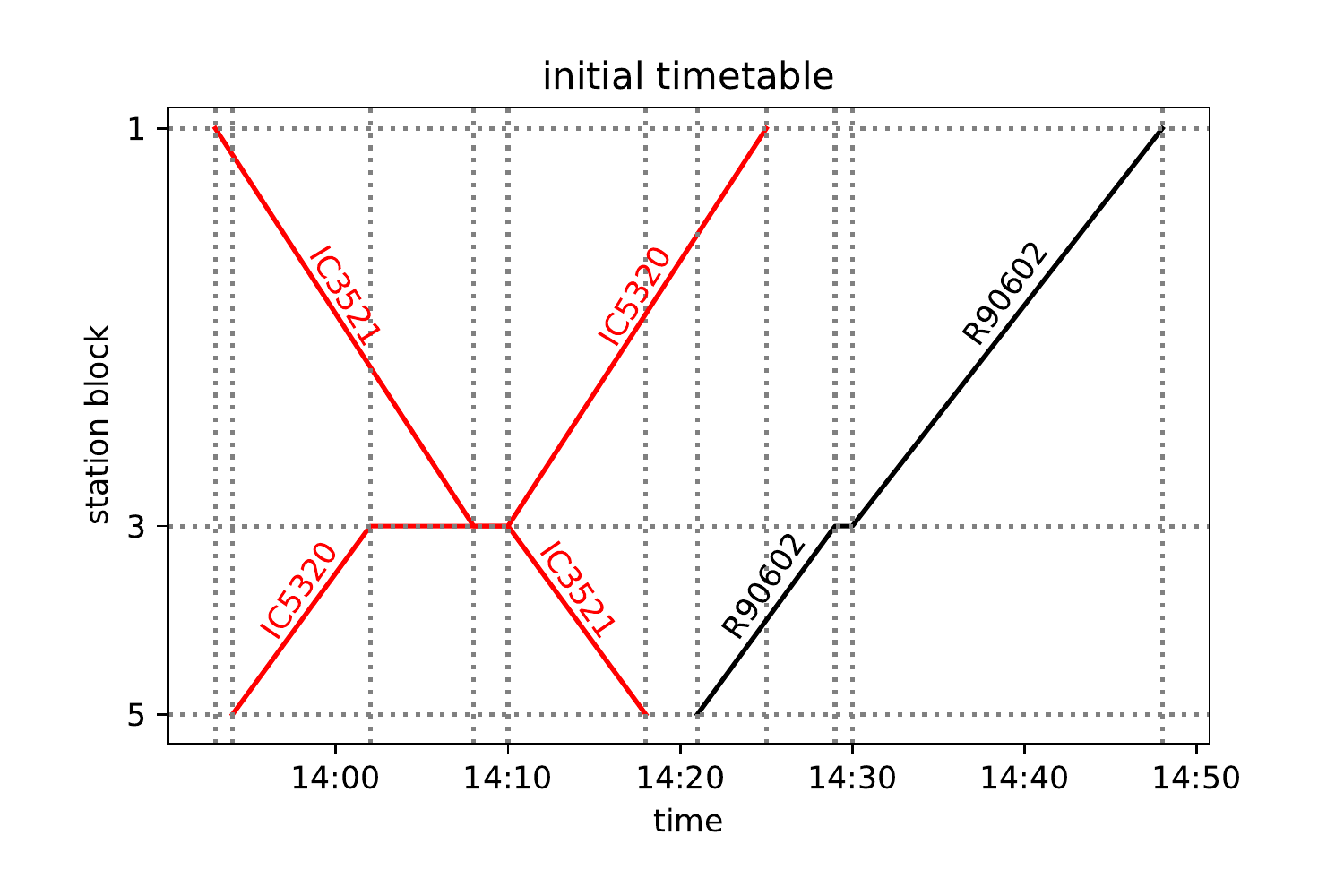}}
	\subfigure[Train diagram for the timetable of the line in
	Fig.~\ref{fig::line}.
	\label{fig::diagram}] {\includegraphics[scale =
      0.6, width=\columnwidth]{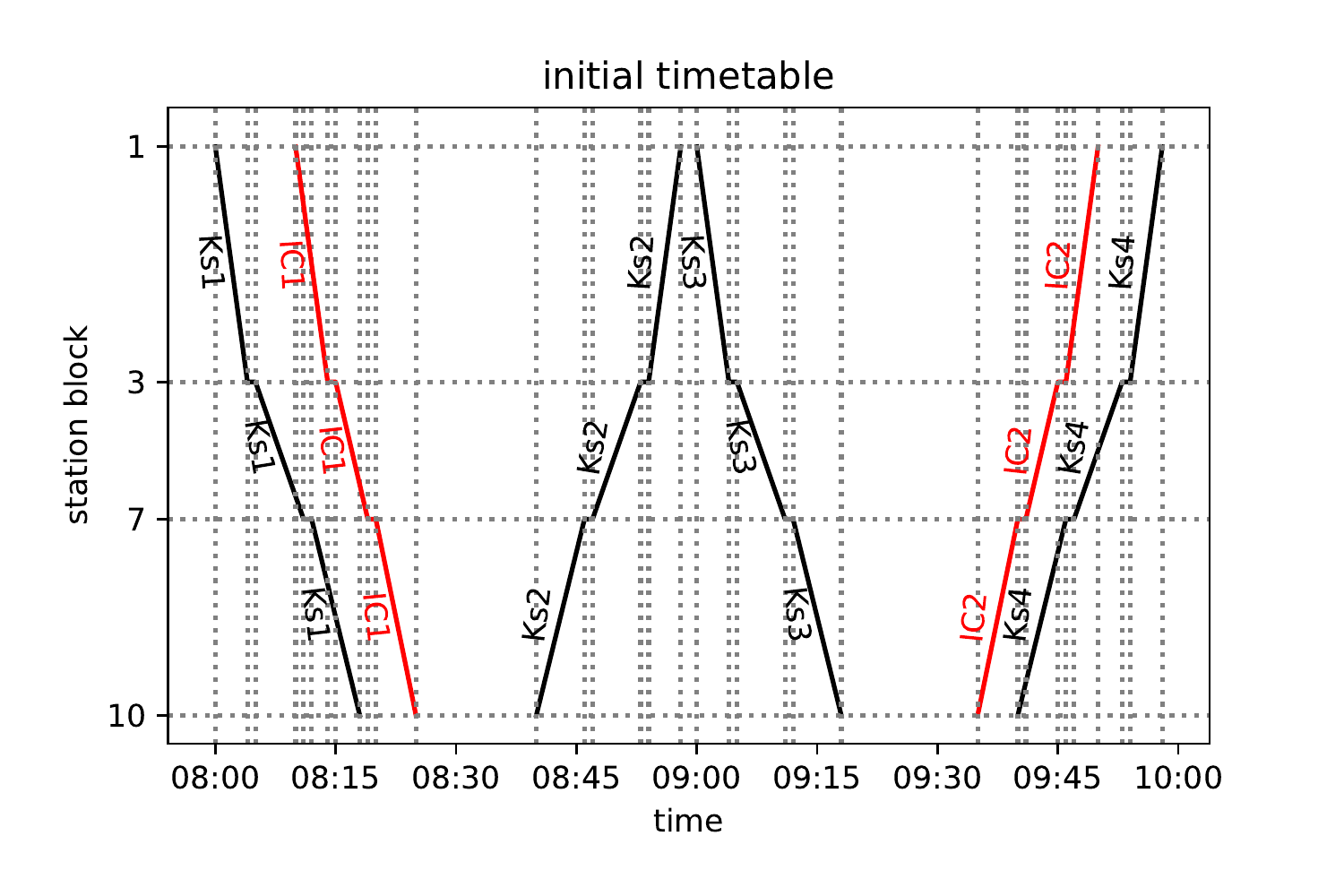}}
	\subfigure[Nidzica-Olsztynek section of railway line No. 216.\label{fig::small_line}]{\includegraphics[scale
	= 0.45, height=30mm, width=\columnwidth]{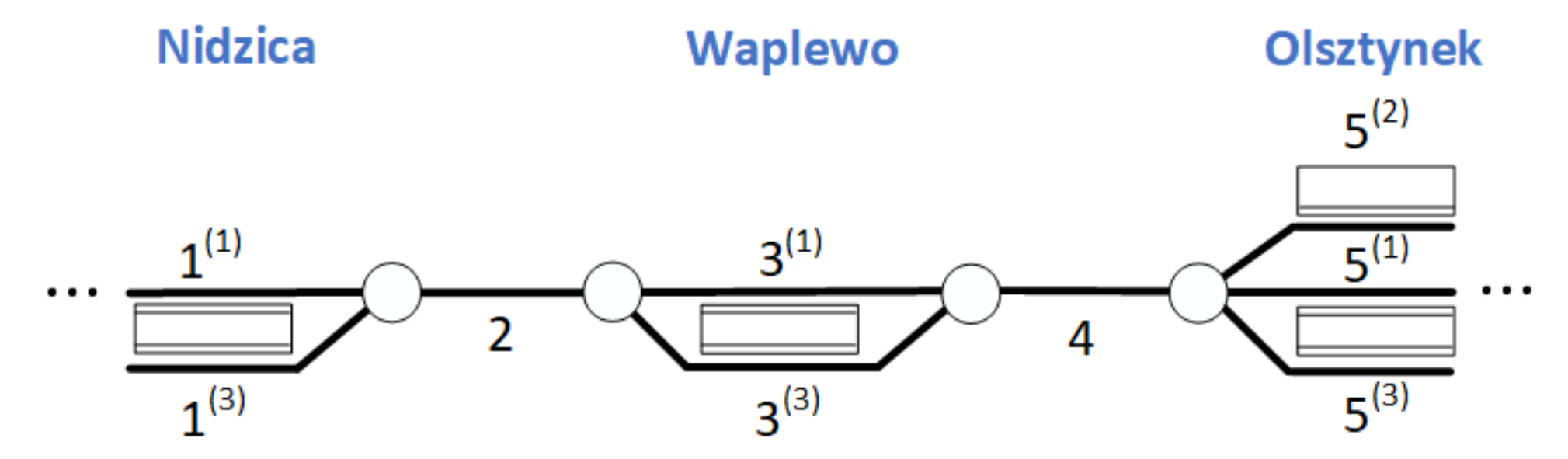}}
	\subfigure[Goleszów - Wisła Uzdrowisko section of railway line No. 191.\label{fig::line}]{\includegraphics[scale
	=
		0.30, height=30mm, width=\columnwidth]{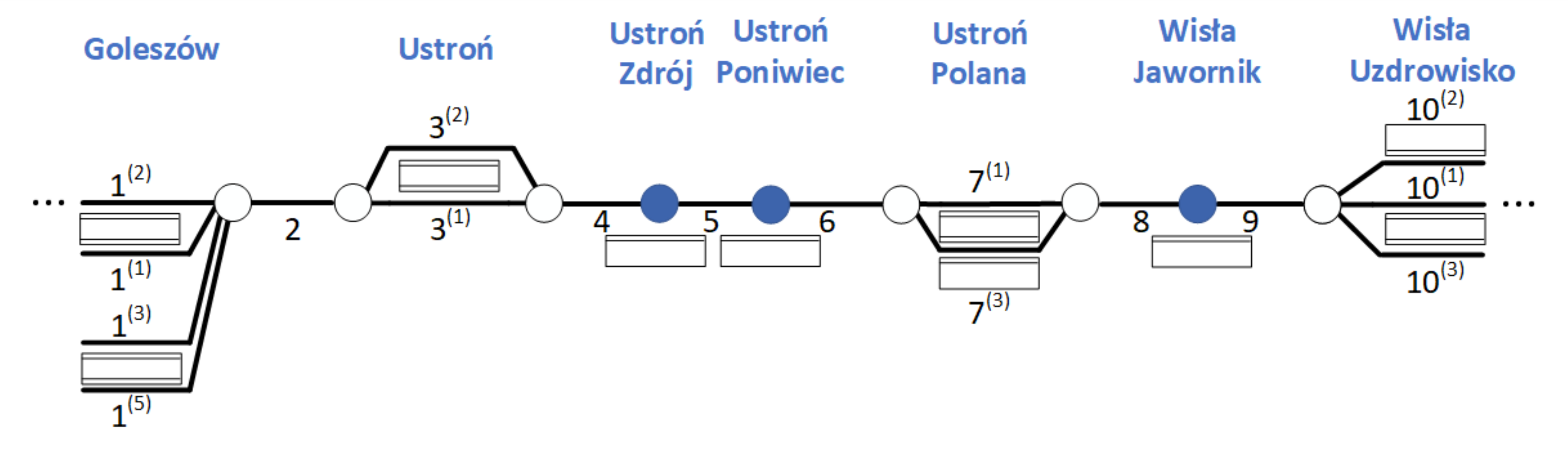}}
	\caption{The railway line segments and their initial (undisturbed) timetables addressed in our calculations. The train diagrams represent train paths by connecting characteristic points of the location of trains at certain times by straight lines. The numbers represent blocks; upper indices in parentheses refer to the tracks of sidings.
      }\label{fig::wline}
\end{figure*}

In Fig.~\ref{fig::small_line}, we present a segment of railway line No.
216 (Nidzica -- Olsztynek section), and in Fig.~\ref{fig::small_diagram} the
analyzed part of the real timetable is depicted in the form of a train diagram.

In Fig.~\ref{fig::small_line}, three stations are presented.  Block
$1$ represents Nidzica station, which has two platform edges numbered
according to the rules of \emph{PKP PLK}. Block $3$ represents Waplewo
station, with another two platform edges. Olsztynek station, with three
platform edges, is represented by block $5$.  The model involves two
line blocks with the labels $2$ and $4$. It is assumed that it takes
the same amount of time to get through a given station block
regardless of which track the train uses. To leave the
station, it is required that the subsequent block is free.

As to the trains, Fig.~\ref{fig::small_diagram} represents their
planned paths. Thee trains are modeled: two \emph{Inter-City}
trains in red and the regional train in black. The scheduled meet-and-pass situations take place in Waplewo and Olsztynek (which might change in case of a delay). IC$5320$ leaves  station
block $5$ (Olsztynek) at 13:54, has a scheduled stopover at
station block $3$ (Waplewo) from 14:02 to 14:10 to meet and pass
IC$3521$, and finally arrives at station block $1$ (Nidzica) at
14:25. As to the opposite direction, IC$3521$ leaves station block
$1$ at 13:53, stops at station block $3$ from 14:08 to 14:10, and
arrives at station block $5$ (Olsztynek) at 14:18. These two trains
depart at the same time from station block $3$ in opposite
directions. The third train considered is R$90602$. It is scheduled to
leave block $5$ at 14:20 and stops at station block $3$ (Waplewo) from
14:29 to 14:30, so it is scheduled to start occupying this
track $19$ minutes after both ICs left. It is behind IC$5320$
during the whole section, and does not meet the IC train at all, so
the original schedule is feasible and conflict free.

Now let us add a $15$-minute delay to the departure time of
IC$5320$ from station block $5$ and $5$-minute delay to that of
IC$3521$ from station block $1$. The passing times were originally
scheduled according to the maximum permissible speeds.  The minimum
waiting times at all the considered stations are $1$ minute regardless
of the train type.  This introduces the following situation: the two
\emph{Inter-City} trains and the regional train have a conflict at
line block $4$. This schedule will be referred to as the ``conflicted
diagram'' -- see Fig~\ref{fig::s_conflict}.  The resolution of this
conflict requires taking a decision at station blocks $3$ and $5$.

Let us now turn our attention to the other example. The line segment
(a part of railway line No. 191) is presented
in~Fig.~\ref{fig::line}, while the considered train paths of the
real timetable are shown in Fig.~\ref{fig::diagram}. There are four
stations and another three stops for the passengers modeled. Block $1$
represents Goleszów station, which has four platform edges. Block $2$
represents a line block between Goleszów station and Ustroń station
(which has two platform edges and is represented by block $3$).
Subsequently, we have three line blocks numbered $4$, $5$, and $6$,
with two stops for passengers: Ustroń Zdrój and Ustroń Poniwiec (with
one platform edge each). Next, we have station block $7$ -- Ustroń
Polana, which has two platform edges. Between this station and Wisła
Uzdrowisko station (numbered $10$ with three platform edges), there are
two more line blocks ($8$ and $9$) with one stop for passengers (Wisła
Jawornik).  We assume that it takes exactly the same time to get
through a block regardless of the track used.

There are six trains, two \emph{Inter-City} trains in red and four
regional (\emph{KS}) trains in black, presented in
Fig.~\ref{fig::diagram}. The regional trains serve all the stops and
stations, while the \emph{Inter-City} service stops only at stations.  We
consider Wisła Uzdrowisko (station block $10$) to be a terminus for
the \emph{Inter-City} trains (however, it does not apply to the
regional trains, which go farther). In this situation, there are no
meet-and-pass situations at intermediate stations (Ustroń and Ustroń
Polana) in the original timetable.  Both \emph{Inter-City} trains are
served by the same train set, and the minimum service time is
$R(j,j') = 20$ minutes at the terminus for ICs (block $10$); see Condition~\ref{cond::rollingstock}.

We analyze the following dispatching cases, selected to demonstrate the algorithm
behavior in various situations:

\begin{enumerate}
	\item A moderate delay of the \emph{Inter-City} train setting off from
	station
	block $1$; see
	Fig.~\ref{fig::large-conf}(a).
	\item A moderate delay of all trains setting off from station block $1$;
	see
	Fig.~\ref{fig::large-conf}(b).
	\item A significant delay of some trains setting off from
	station block $1$; see Fig.~\ref{fig::large-conf}(c).
	\item A large delay of the
	\emph{Inter-City} train setting off from
	station
	block $1$; see Fig.~\ref{fig::large-conf}(d).
\end{enumerate}
 The conflicted timetables of cases $1-4$ are presented in Fig.~\ref{fig::large-conf}.

\subsection{Simple heuristics}\label{sec::conventional_heu}

As a simple way of resolving conflicts, there are very simple heuristics prevalently known and used in the railway practice, the First Come First Served (FCFS) and the First Leave First Served principles. A slightly more complex one is AMCC (avoid maximum current
$C_{\text{max}}$)~\cite{mascis2002job}.
All of these heuristics are used to determine the order of trains when
passing the blocks (for implementation reasons, the trains are
analyzed in pairs). In FCFS and FLFS  the way is given to the train that first arrives --
or first leaves -- the analyzed block section. In practice, the
decisions based on both these heuristics are taken starting from the
most urgent conflict. Next, since passing and overtaking is
possible only at stations, so-called \emph{implied
  selections}~\cite{dariano2007branch} are determined. The procedure
is repeated as long as all the conflicts are solved.

The AMCC is a more complex approach, whose objective is to minimize the
maximum secondary delay of the trains; this objective will be referred
to as the ``AMCC objective'' in what follows. This is quite an
intuitive procedure, yet more sophisticated than FCFS and FLFS. To facilitate the comparison, stations are assigned an infinite
capacity. Of course, solutions requiring a capacity higher than that
of the given station must be rejected.

\begin{figure*}[t]
	\subfigure[The conflicted diagram. All the three trains would meet in block 4 as it can be seen from the intersecting train paths.\label{fig::s_conflict}]{\includegraphics[scale =
		0.6, width=\columnwidth]{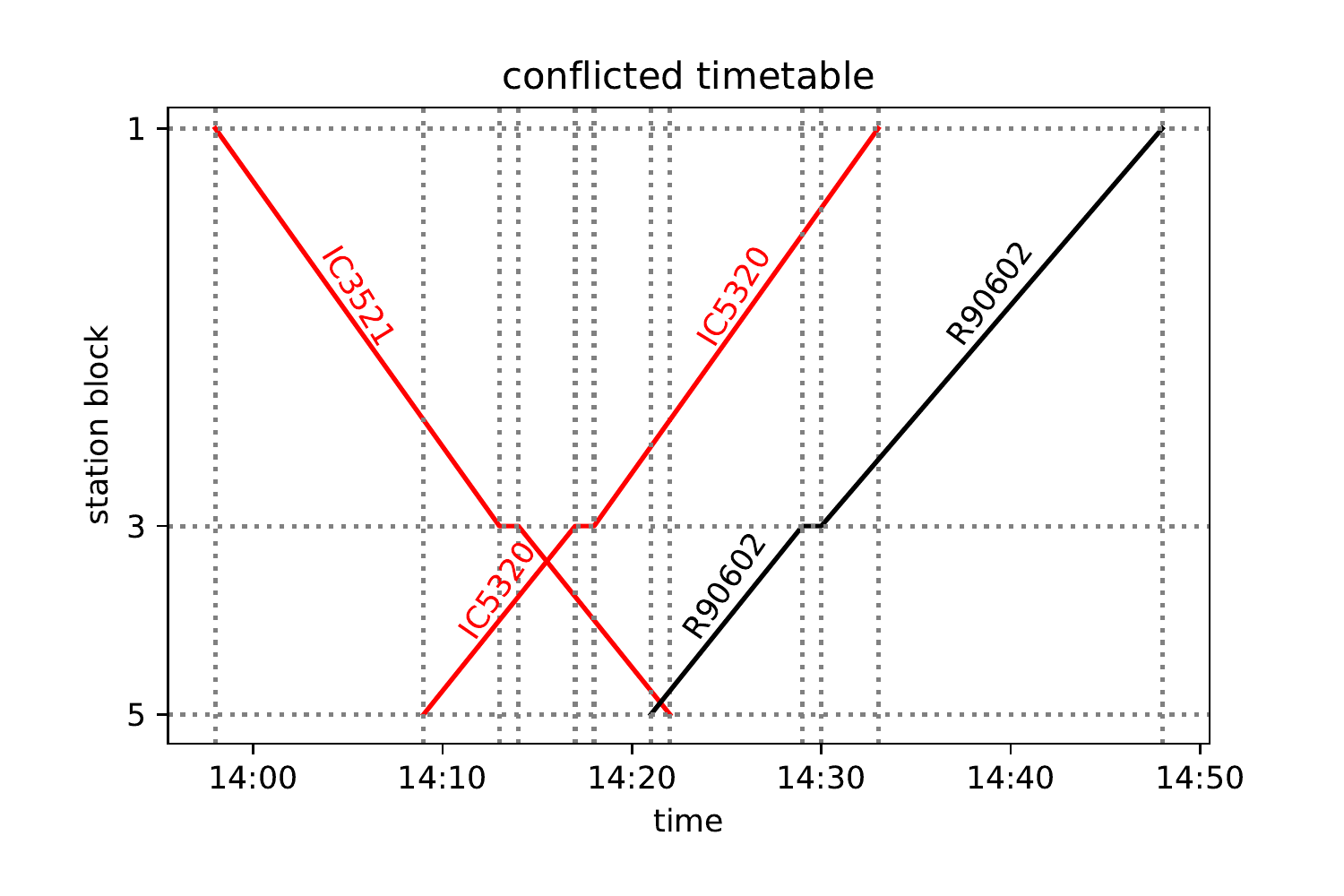}}
	\subfigure[The solution; FCFS, FLFS, and AMCC give the same outcome with a maximum
	seconday delay of $4$ minutes.\label{fig::s_solution}]
	{\includegraphics[scale = 0.6, width=\columnwidth]{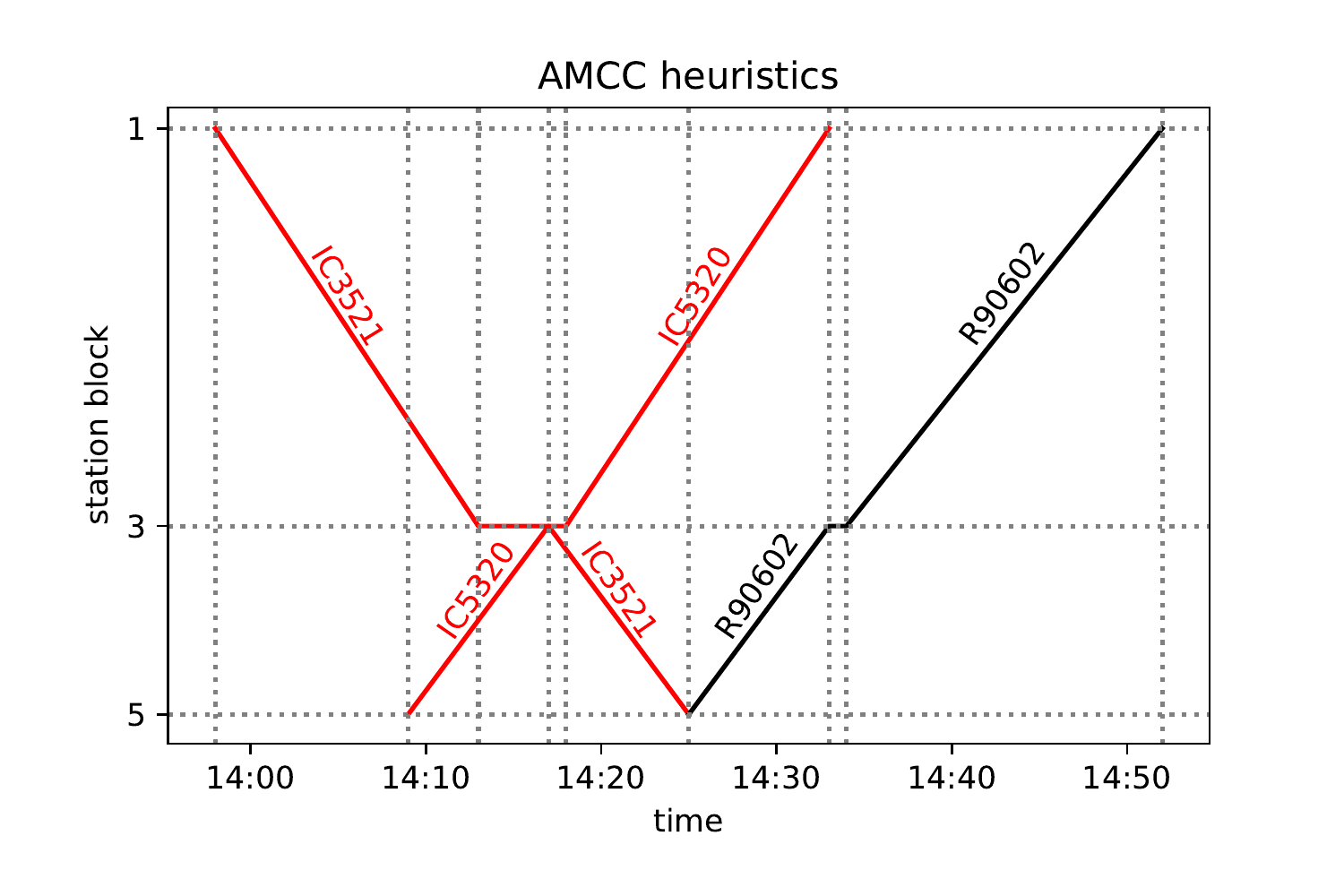}}\caption{A possible
	solution of the conflict on line No. 216.}\label{fig::small_case}
\end{figure*}

In the example presented in
Fig.~\ref{fig::small_line}, for the conflicted timetable in
Fig.~\ref{fig::s_conflict}, each of the heuristics returns the same
solution; this is presented in Fig.~\ref{fig::s_solution}. When
comparing the FCFS with the FLFS, observe that in the conflicted
timetable, three trains (IC$5320$, IC$3521$, R$90602$) are scheduled to
occupy block $4$ simultaneously, which is forbidden.

To avoid the conflict, IC$3521$ is allowed to enter this block with a $3$-minute delay at 14:17 (as soon as IC$5320$ leaves the block), thus leaving the
block at 14:25 instead of 14:22, which results in $3$ minutes of secondary
delay. Consequently, R$9062$ is allowed to enter the block not earlier than 14:25,
an additional $4$-minute delay as compared with the
conflicted timetable. Thus the maximum secondary delay is $4$ minutes, and
the sum of the delays on entering the last block is $7$ minutes. The maximum secondary
delay is $4$ minutes; it is the lowest possible one, so the solution
is optimal with respect to the AMCC objective.

The other example -- presented in Fig.~\ref{fig::line} -- is more
complex, yet it is still solvable by a state-of-the-art quantum
annealer. We do not discuss this example in detail; we describe only
the maximum secondary delays as the objective function. This is
presented in Table~\ref{tab::tarfet_f_class} for the discussed
heuristics. The upper limit used in the quantum computing is set to
$d_{max} = 10$ on this basis. (Observe that most of the secondary
delay values are within this limit.)  The respective train diagrams are
presented in Appendix A,
Figs.~\ref{fig::large-FCFS} - \ref{fig::large-AMCC}
The values of the AMCC objective function are presented in
Table~\ref{tab::tarfet_f_class}; AMCC appears to find the
optimum in these cases, thus providing a good enough reference for
comparisons, albeit with an objective function different from that of
ours. Our choice of the objective will be more flexible, thus leaving room for further non-trivial optimization.

\begin{table}
	\centering
\begin{tabular}{ccccc}
	 Heuristics & case $1$ & case $2$ & case $3$ & case $4$  \\
	\hline
	FLFS & 6 & 13 & 4 & 2 \\
	\hline
	FCFS & 5 & 5 & 5 & 2 \\
	\hline
	AMCC & 5 & 5 & 4 & 2 \\
	\hline
\end{tabular}
\caption{The maximum secondary delays, in minutes, resulting from simple heuristics.
  Observe that for each case, there are solutions far below $d_{\text{max}}
= 10$.}
\label{tab::tarfet_f_class}
\end{table}

\subsection{Quantum and calculated QUBO solutions}\label{sec::emulated_qubo}

Our approach based on QUBO concerns the objective function set out in~\eqref{eq::effqubo}.
This contains the feasibility conditions (\emph{hard constraints}) and
the objective function $f(\mathbf{x})$
of~\eqref{eq::penalty_prticular}. For the feasibility part, we need to
determine $\tau_{(1)}\left(j, s \right)$, the minimum time for
train $j$ to give way to another train going in the same direction,
and $\tau_{(2)}\left(j, s \right)$ - the minimum time for train $j$ to
give way for the another train going in the opposite direction (see
Condition~\ref{cond::2trains1way} and
Condition~\ref{cond::2trains2way}).

As noted before, the QUBO objective function introduces flexibility in
choosing the dispatching policy by setting the values of the penalty
weights for the delays of the trains. In this way, almost any train
prioritization is possible. To demonstrate this flexibility, we make
the penalty values proportional to the secondary delays of the trains
that enter the last station block. This is equivalent to the
secondary delay on leaving the penultimate station block. Each train
is assigned a weight $w_j$, yielding the form of~\eqref{eq::penalty}
\begin{equation}\label{eq::penalty_prticular}
f(\mathbf{x}) = \sum
w_j  \cdot \frac{d(j,s^*)-
	d_U(j,s^*)}{d_{\text{max}}(j)} \cdot x_{j,s^*,d},
\end{equation}
where the sum is taken over $j \in J$ and $d \in A_{j,s^*}$ with
$s^* = s_{(j, \, \text{end}-1)}$.  Note that this objective
coincides with that of the linear integer programming
approach~\eqref{eq::penalty_linear}, which will be used for comparisons.

The following train prioritization is adopted.  In the case of railway
line No. 216, the \emph{Inter-City} trains are assumed to have a
higher value of the delay penalty weight $w = 1.5$, while the regional
train is weighted $w = 1.0$. We give the higher priority to the
\emph{Inter-City} train, which is a reasonable approach resulting from
the train prioritization in Poland (and in many other countries). In
the other case (line No. $191$), the priorities of trains heading
towards block $10$ (Wisła Uzdrowisko) are lower, weighted $0.9$ for
all the other trains in this direction. However, train priorities for
the trains heading in the opposite direction (toward block $1$ --
Goleszów -- and beyond the analyzed section) have higher values: $1.0$
for the regional trains and $1.5$ for the \emph{Inter-Cities}.  Such a
policy is motivated by the reluctance of letting the delays propagate
across the Polish railway network -- that regional trains proceed
toward the main railway junction in the region's major city (Katowice)
and that the \emph{Inter-City} train service is scheduled toward the
state's capital city (Warsaw). Observe that $w_j$ is the highest
possible penalty for the delay of train $j$;
see~\eqref{eq::penalty_prticular}. In both these cases, the maximum of
$w_j$ is $1.5$. Hence, the penalties for a infeasible solution should
be higher (as discussed in Section~\ref{sec::quboimplem}. We set
$p_{\text{pair}} = p_{\text{sum}} = 1.75 > 1.5$.

Referring to~\eqref{eq::dlim}, we have the maximum secondary delay
$d_{\text{max}}$ parameter (for simplicity, we assume that $d_{\text{max}}$ is the same for
all trains and all analyzed station blocks). It cannot be smaller
than the delay value returned by the AMCC heuristics. However, since the AMCC may not
be optimal
in terms of our objective function, we need to leave a margin for some
larger
values of the maximum secondary delay. On the other hand, since the system size
grows with $d_{\text{max}}$, it must be limited enough to make the problem
applicable to state-of-the-art quantum devices and classical algorithms motivated by them. Specifically,
since
we do not analyze the delays at the last station of the analyzed segment of the
line, we
require as many as
$(\text{number of station blocks} - 1) \cdot (\text{number of trains}) \cdot
(d_{\text{max}}+1)$ qubits.

In the case of railway line No. $216$, we set $d_{\text{max}} = 7$, which is
considerably
larger than the AMCC solution. There are $48$ logical quantum
bits needed to handle this problem instance, making it suitable for both quantum annealing at the
current state of the
art, and the GPU-based implementation of the brute-force search for the low-energy spectrum (ground state and
subsequent excited state)~\cite{jaowiecki2019bruteforcing}, which is possible with up to $50$ quantum bits. The benefit of this possibility is
that it provides an exact picture of the spectrum, which can be used as a reference when evaluating the heuristic results of approximate methods (tensor networks) or quantum annealing. This may guide for the understanding of the results of the bigger instances, in which the brute-force exact search is not available.

There are many possible distinct solutions in the case of line No. $191$,
making the analysis more interesting from the dispatching point of view. We
set
$d_{\text{max}} = 10$: for a justification see Table~\ref{tab::tarfet_f_class}, and
observe that $d_{\text{max}}$ is considerably larger than the AMCC output. The
$d_{\text{max}} = 10$ yields $198$ logical quantum bits, which we were able to embed
into a present-day quantum annealer, the D-Wave device DW-$2000\text{Q}_5$, in most
cases.

Recall that current quantum annealing devices are imperfect and often output
excited states.
The clue of our approach is that the excited states (e.g.,
returned by the quantum annealer)
still represent the optimal dispatching solutions, provided that their
corresponding energies are relatively small. The reason for this is that what really needs to be determined is the order of trains leaving
from each station block (i.e., this is the decision to be
made). What is crucial here is to determine all the meet-and-pass and the
meet-and-overtake situations (in analogy with the determination of all the precedence
variables in the linear integer programming approach). The exact time of leaving
block sections
is of a secondary importance. Therefore, we consider those
excited states that describe the same order of trains as the ground
state, to be equivalent to the actual ground state encoding the
global optimum.  As discussed in Section~\ref{sec::quboimplem}, our QUBO
formulation problem ensures that those equivalent solutions are present in the low-energy spectrum.

\subsubsection{Exact calculation of the low-energy spectrum}

\begin{figure*}[t]
	\subfigure[$p_{\text{pair}} = 2.7,
	p_{\text{sum}} = 2.2$
	\label{fig::bf_spec_22_27}]{\includegraphics[scale
		=
		0.58,
		width=\columnwidth]{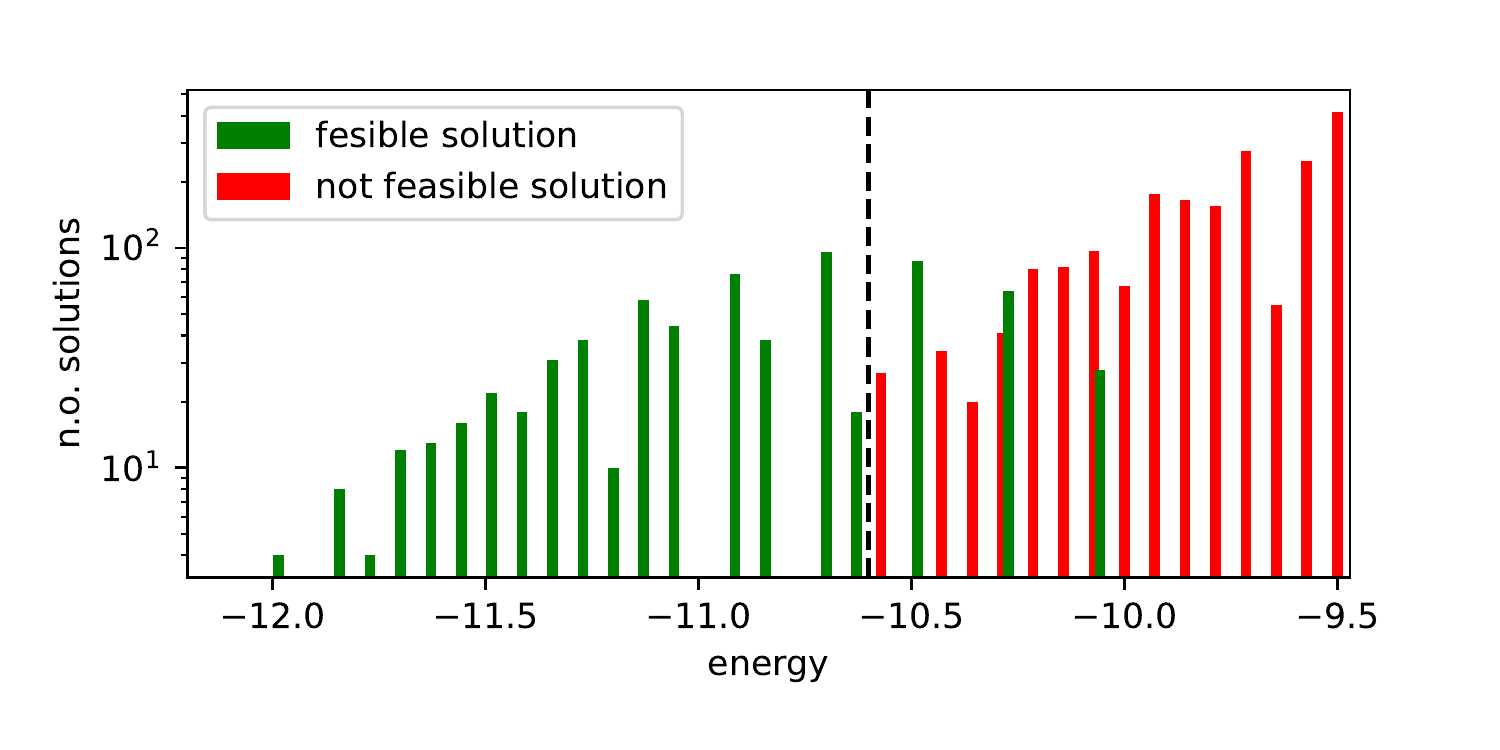}}
	\subfigure[
	$p_{\text{pair}} = p_{\text{sum}} =
	1.75$\label{fig::bf_spec_175_175}]{\includegraphics[scale
		=
		0.58,
		width=\columnwidth]{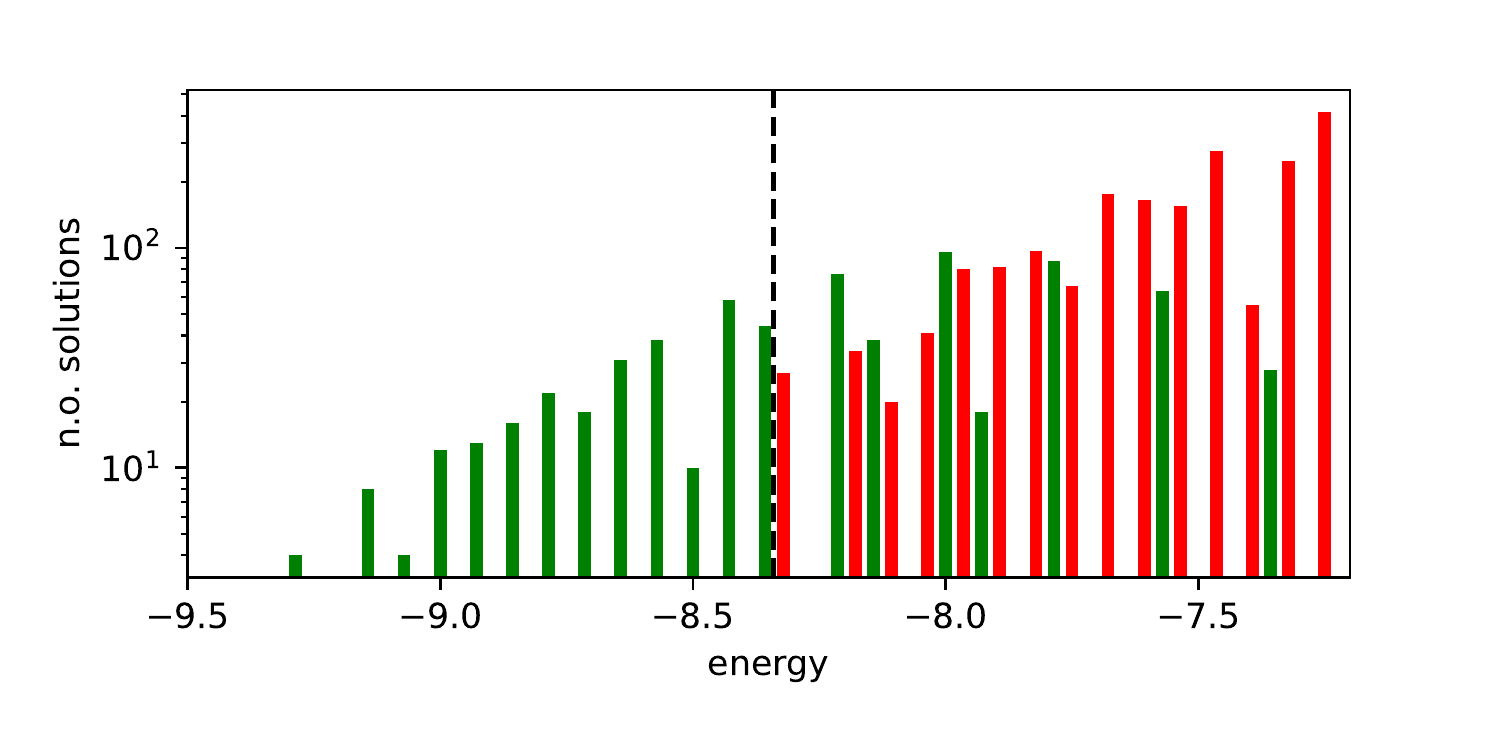}}
	\caption{Spectra of the
		low-energy solutions for two penalty
		strategies of the brute-force (exact) solution. The
		black line separates the phase in which only feasible solutions appear.
		Observe the mixing phase, in which both feasible and unfeasible solutions
		occur. Here $p_{\text{pair}}$ and $p_{\text{sum}}$ are penalties of the unconstrained problem expressed in the ``logical'' variables. The term $p_{\text{sum}} = \left(\sum_{i\in
\mathcal{V}_\text{s}}
x_i - 1\right)^2$, cf.~\eqref{eq:sumpenq}, ensures that each train leaves a station only once, whereas $p_{\text{pair}} = \sum_{(i,j)\in
  \mathcal{V}_\text{p}}(x_ix_j+x_jx_i)$, cf.~\eqref{eq:sumpen}, imposes the following: minimal passing time constrain, single block occupation constrain, deadlock constrain, and rolling stock circulation constrain.
		}\label{fig::brute_force}
\end{figure*}

To demonstrate the aforementioned idea, we first present the results of the
brute-force numerical calculations performed on a GPU
architecture~\cite{jaowiecki2019bruteforcing}. With this approach, the
low-energy spectra of the smallest instances have
been calculated exactly, providing some guidance for the understanding
of the model behavior and parameter dependence. The method is suitable
for small (i.e., up to $N \le 50$ quantum bits) but otherwise
arbitrary systems. To study (hard) penalties resulting from
non-feasible solutions, apart from
$p_{\text{pair}} = p_{\text{sum}} = 1.75$ in ~\eqref{eq::effqubo},
we use other, higher penalties that are not equal to each other,
$p_{\text{pair}} = 2.7$ and $p_{\text{sum}} = 2.2$.

Let us assume that the solution in Fig.~\ref{fig::s_solution} is the optimal
one.
Here the train IC$3521$ ($w = 1.5$) waits $3$ minutes at block $3$, while
regional train R$90602$
($w = 1.0$) waits $4$ minutes at block $5$, causing $4$ minutes of
secondary delay upon leaving block $3$. This adds $1.214$ to the objective.
Referring to the feasibility terms in~\eqref{eq::effqubo}, for a
feasible solution $\mathcal{P}'_{\text{sum}} = 0$, while the linear constraint
gives the negative offset to the energy. Referring to~\eqref{eq:sumpen}, as
we have three trains for which we analyze two stations, this negative offset is
$\mathcal{P}_{\text{sum}} = - 3 \cdot 2 \cdot
p_{\text{pair}}$.
Based on the feasibility terms set out~\eqref{eq::effqubo}, this yields $-10.5$
for $p_{\text{sum}} = 1.75$, and $-13.2$ for $p_{\text{sum}} = 2.2$. This
results in a ground state energy of $f'(\mathbf{x}) = -9.286$ and $f'(\mathbf{x}) =
-11.986$, respectively. Finally, in the ground state solution shown in
Fig.~\ref{fig::s_solution}, the IC$3521$ train can leave the station block $1$
with a
secondary delay of $0$, $1$, $2$, or
$3$, not affecting any delays of the trains leaving block $3$. All these
situations correspond to the ground state energy. Hence, our approach produces a $4$-fold
degeneracy of the ground state.

Low-energy spectra of the solutions and their degeneracy are presented in
Fig.~\ref{fig::bf_spec_175_175} and Fig.~\ref{fig::bf_spec_22_27}. All the
solutions that are equivalent to the ground state from the dispatching point of
view are marked in green. Infeasible excited state
solutions (in which some of the feasibility conditions set out
in~\eqref{eq::effqubo} are
violated) are marked in red. In this example, we do not have feasible
solutions that are not optimal, i.e., in which the order of trains at a station
is different from the one in the ground state solution.

\begin{figure*}[t]
	 \subfigure[QUBO parametrization:
	$p_{\text{pair}} = 2.7,
	p_{\text{sum}} =
	2.2$.\label{fig::dw_hist_27_22}]{
	\includegraphics[scale=0.9,width=\columnwidth]{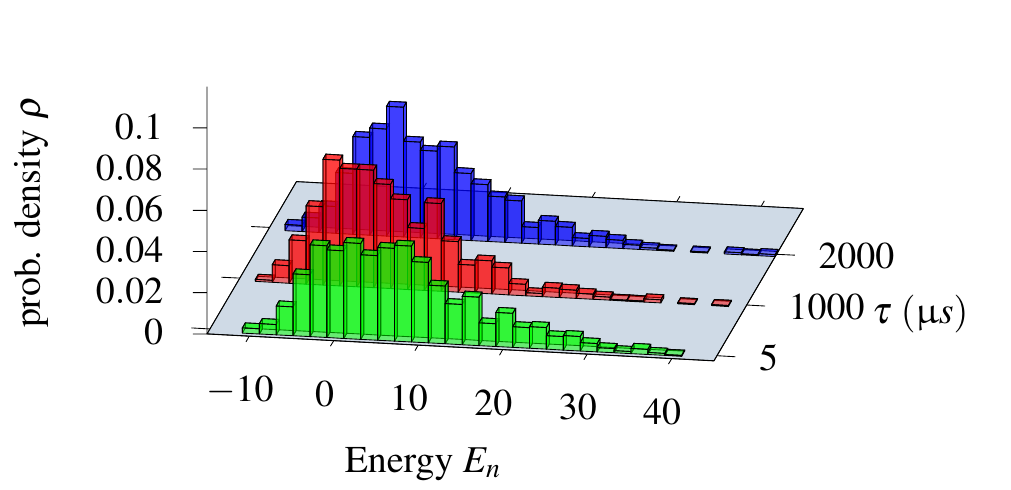}}
	\subfigure[
	$p_{\text{pair}} = p_{\text{sum}} =
	1.75$.\label{fig::dw_hist_175_175}]{
	\includegraphics[scale=0.9,width=\columnwidth]{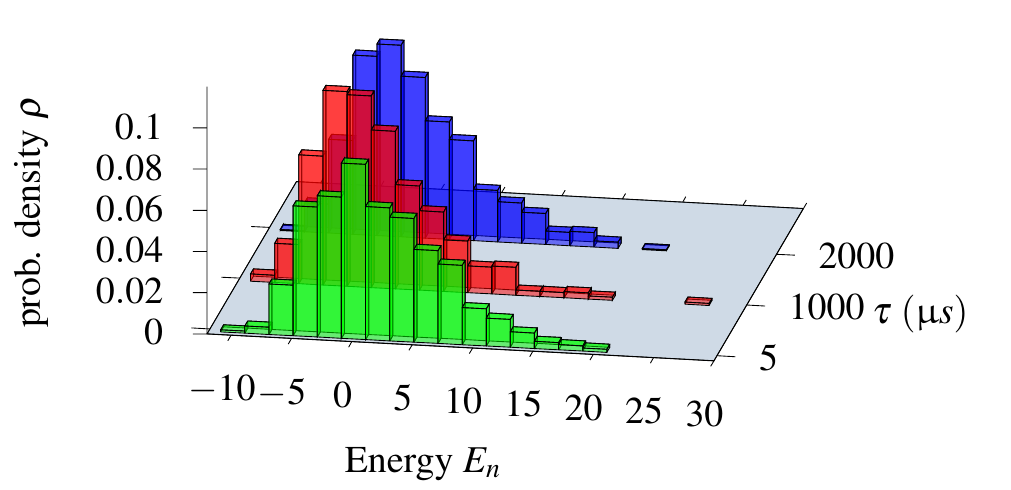}}
	\caption{Distribution of the energies corresponding to the states (solutions), that are sampled by the D-Wave $2000$Q quantum annealer of $48$ logical quantum bits instance of line No. $216$. In particular, $1000$
	samples were taken for each annealing time, and the strength of embedding was set to $css = 2.0$. This device is still very noisy and prone to errors, so the sample contains excited states.
}
\label{fig::DWhists}
\end{figure*}

In the case of line No. $191$, a more detailed analysis of the low-energy spectra of the solutions was possible due to the generality of the
brute-force simulation. The results are presented in
Fig.~\ref{fig::brute_force}. We shall find later on that the
D-Wave solutions managed to get into the ``green'' tail of feasible
solutions, but high degeneracy of higher-energy states may impose some
risk of the quantum annealing ending up in more frequently appearing
excited states (see Fig.~\ref{fig::DWhists}).

\subsubsection{Classical algorithms for the linear (integer programming) IP model and QUBO}\label{sec::clasical_heurisitcs}

We expect classical algorithms for QUBOs to achieve the ground state
of~\eqref{eq::effqubo} or at least low excited states
equivalent to the ground state with respect to the dispatching
problem. It is important to mention that hereafter we transform the
original QUBO coding into the Chimera graph coding (see
Section~\ref{sec::Ising_based_solvers}). This makes the algorithm ready
for processing on a real quantum annealer.

\begin{figure}
	\begin{center}
		\includegraphics[width=0.45\textwidth]{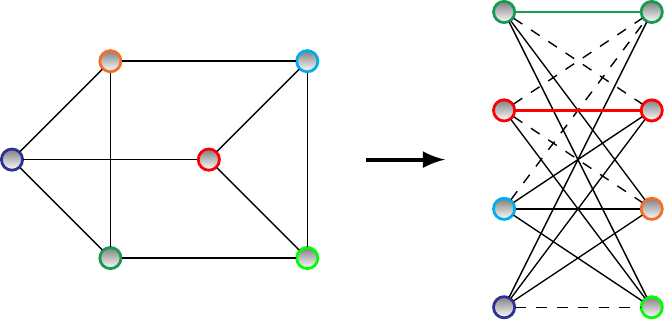}
	\end{center}
	\caption{Embedding of a simple, six-qubit problem. Left: graph of
      the original problem. Right: problem embedded into a unit cell
      of Chimera.  Here, different colors correspond to different
      logical variables. Apparently, the original problem does not map
      directly onto Chimera as it contains cycles of length
      3. Therefore, two chains have to be introduced. Couplings
      corresponding to inner-chain penalties are marked with the same
      color as the variable they correspond to.}
	\label{fig:h_embedding}
\end{figure}

As to a simple example of the embedding, we refer to the problem with $4$ quantum bits that has been discussed in
Section~\ref{sec::simple_ex}. In that case, the mapping was
trivial. In a case of $6$ quantum bits, for instance (by setting
$d_{\text{max}} = 2$), we will have additional terms
in~\eqref{eq::simple_psum},~\eqref{eq::simple_pair},
and~\eqref{eq::simple_pen}.  Hence, the larger problems cannot be
directly mapped onto the Chimera graph, so the embedding procedure is
required, as illustrated in Fig.~\ref{fig:h_embedding}. This
illustrates the basic idea of how the embedding is performed in even
larger models.

As to the model parameters, recall that for the particular QUBO, we
have opted for $p_{\text{pair}} = p_{\text{sum}} = 1.75$ or
$p_{\text{pair}} = 2.2 \ p_{\text{sum}} = 2.7$ for line No. $216$
and $p_{\text{pair}} = p_{\text{sum}} = 1.75$  for line
No. $191$.
Let us present the solutions of the two state-of-the-art
numerical methods, which we shall later compare with the experimental
results obtained by running the D-Wave $2000$Q quantum annealers. The
first solver is developed `in-house' and is based on tensor network
techniques~\cite{rams2018heuristic}. The idea behind this solver is to
represent the probability of finding a given configuration by a
quantum annealing processor as a PEPS tensor
network~\cite{rams2018heuristic}. This allows an
efficient bound-and-branch strategy to be applied in order to find
$M \ll 2^N$ candidates for the low-energy states, where $N$ is the
number of physical quantum bits on the Chimera graph. In principle,
such a heuristic method should work well for rather simple QUBO problems,
i.e., those in which the $Q$ matrix in~\eqref{eq:qubogen} has some
identical or zero terms; this corresponds to the so-called weak
entanglement regime. It can be shown that this is the case in our problem
[see also the simple example of the $Q$ matrix
in~\eqref{eq::simple_Q}].  Furthermore, heuristic parameters such as
the Boltzmann temperature $(\beta)$ can be provided, allowing one to zoom
in on the low-energy spectrum depending on the problem in question.
We set $\beta = 4$, which is quite a typical setting, as discussed
in~\cite{rams2018heuristic}. Although even better solutions may
potentially be achieved by further tuning this parameter, we demonstrated
that this default setting is satisfactory from the dispatching point
of view. The second classical solver is CPLEX~\cite{cplex} (version
12.9.0.0). In our work, we have used the
DOcplex Mathematical Programming package (DOcplex.MP) for Python. In what follows "CPLEX" refers to the QUBO solver of this package.

In order to have a fair comparison with a traditional approach, we have also formulated our model as a linear integer program, this is decribed in Section~\ref{sec::linear_integer} of the Supplemental material in detail. We have implemented the linear integer model with the PuLP package~\cite{PulpDoc} and solved with its
default solver (CBC MILP Solver Version: 2.9.0). All instances were solved
to the optimal solution in 0.03 seconds on an average computer. This was in line with
our expectations as our problems are small.  Our goal is, however, not
to outperform either CPLEX or the standard linear solver but to demonstrate the applicability of quantum hardware; at the
present state of the art, we need the well-established solvers to
produce results for comparisons and reference.

Concerning the results of another railway
line (No. $191$), the values of the objective function
~\eqref{eq::penalty_prticular} are given
Table~\ref{tab::tarfet_f_q}. We also include the values of our
objective function for the FLFS, FCFS, and AMCC optimal solutions.

The agreement with the linear integer programming approach provides the
argument that the
CPLEX results refer to the ground state of the QUBO. We are interested
in the results being equivalent to those of CPLEX and the linear
solver from the dispatching point of view. These results are marked
in blue in Table~\ref{tab::tarfet_f_q}.  The tensor network approach
yields equivalent solutions to those of CPLEX.  However, the tensor
network sometimes returns the excited states of the QUBO, as can be
observed in case $3$. This is caused by the fact that the tensor
network method is based on approximations. This demonstrates that even
some low-energy excited states encode a satisfactory solution.
Interestingly, the results of the AMCC are also equivalent to those
CPLEX in cases $2$,$3$, and $4$ but different in case $1$. This is due
to the fact that the AMCC needs to have a specific objective
function, whereas in our approach we can choose this function more flexibly. Specifically, in case $1$, the meet-and-pass situation of trains IC$1$
and Ks$2$ at station $10$ yields the lowest maximum secondary
delay, so it is optimal from the AMCC point of view. (Note that two
trains have secondary delays: Ks$2$ and Ks$3$ in this case). As
discussed earlier, in this approach Ks$2$ is prioritized, as it is the
train leaving the modeled network segment and one of the goals is
to limit delays propagating further from this
segment.  The train diagrams based on the CPLEX solutions are depicted
in Fig.~\ref{fig::large-sim-cplx}.

In case $3$, observe that the objective function in
Table~\ref{tab::tarfet_f_q} from the tensor network solution differs
from the minimum (yet the solution is still equivalent to the optimal
one). To explain this, observe that there are numerous possibilities of
additional train delays that do not affect the dispatching situation.
An example of such a situation is a train having its stopover
extended at the station with no meet and pass or meet and
overtake. Such a situation increases the value of the objective but
does not affect the optimal dispatching solution. The number of
combinations here is relatively high, and this is why such
extended stopovers may be returned by the approximate algorithm. This
is in contrast to the exact FCFS, FLFS, and AMCC heuristics, which do
not allow for such unnecessary delays; the exact heuristics always return
$f(\mathbf{x})$ that is the minimum for the particular
dispatching solution. In case $3$, the FCFS with $f(\mathbf{x}) = 0.95$
does not give the optimal solution from the dispatching point of view, as
opposed to the tensor network with $f(\mathbf{x}) = 1.65$.

\begin{table}
	\centering
	\begin{tabular}{cccccc}
		\multicolumn{2}{ c }{Method} & case $1$ & case $2$ & case $3$ & case
		$4$  \\
		\hline
		{\multirow{2}{*}{QUBO approach} } &
		CPLEX & $\color{blue}0.54$ & $\color{blue}1.40$ & $\color{blue}0.73$ &
		$\color{blue}0.20$ \\
		\cline{2-6}
		& tensor network & $\color{blue}0.54$ & $\color{blue}1.40$ &
		$\color{blue}1.65$ & $\color{blue}0.29$ \\

		\hline
		\multicolumn{2}{ c }{linear integer programming} & $\color{blue}0.54$ &
		$\color{blue}1.40$
		&
		$\color{blue}0.73$ &
		$\color{blue}0.20$ \\
		\hline
		 {\multirow{3}{*}{Simple heuristics} }& AMCC & $0.77$ &
		 $\color{blue}1.30$ & $\color{blue}0.73$ &
		$\color{blue}0.20$ \\
		\cline{2-6}
		& FLFS & $\color{blue}0.54$ & $1.71$ & $\color{blue}0.73$ &
		$\color{blue}0.20$ \\
		\cline{2-6}
		& FCFS & $0.77$ & $\color{blue}1.30$ & $0.95$ & $\color{blue}0.20$  \\
		\hline
	\end{tabular}
	\caption{The values of the objective function $f(\mathbf{x})$ for the
		solutions obtained by the classical calculation of the QUBO, linear
		integer programming approach, and all the heuristics. The
		blue color denotes
		equivalence from the dispatching point of view with the the ground
		state of the QUBO or the output of the linear integer programming. The
		equivalence concerns the same order of trains at each station.}
	\label{tab::tarfet_f_q}
\end{table}

\subsubsection{Quantum annealing on the D-Wave machine}

As described in Section~\ref{subsec:qannealing}, the solver we discuss
(i.e., D-Wave $2000$Q quantum annealer) is probabilistic.
In particular, as the required time to drive the system into
its ground state is unknown, the output is a sample of the
low-energy spectrum from repeated annealing processes, hence it can be
regarded as a heuristic. The solution is thus assumed to be the
element of this sample with the lowest energy (in practice, these elements are
not from the ground states but from low excited states). The likelihood of obtaining
solutions with a lower energy (or the actual ground state) increases with the number of repetitions.

As already mentioned, qubits on the D-Wave's chip are arranged into a
Chimera graph topology. Furthermore, some nodes and edges may be
missing on the physical device, making the topology different even
from an ideal Chimera graph. This requires \emph{minor
  embeddeding} of the problem, mapping logical qubits onto physical
ones. To this end, multiple physical qubits are chained together to
represent a single logical variable, which increases their
connectivity at the cost of the number of available qubits. Such embedding
is performed by introducing an additional \emph{penalty term} that
favors states in which the quantum bits in each chain are aligned in the
same direction. The multiplicative factor governing this process is called
the chain strength, and it should dominate all the coefficients present in the
original problem. (Note that we encounter yet another penalty term at this
point.)  In this work, we set this factor to the maximum
absolute value of the coefficients of the original problem multipled
by a parameter that we call the \emph{chain strength scale} (css). In our
experiment, \emph{css} ranged from $2.0$ to $9.0$. Another
parameter is the annealing time (ranging from $5 \upmu$s to
$2000 \upmu$s). This is the actual duration of the physical annealing
process.

\begin{figure*}[t]
		\subfigure[The optimal solution from Fig.~\ref{fig::dw_spec_27_22}.\label{fig::DWsol_small2000_22_27}]{\includegraphics[scale
		=
		0.6,width=\columnwidth]{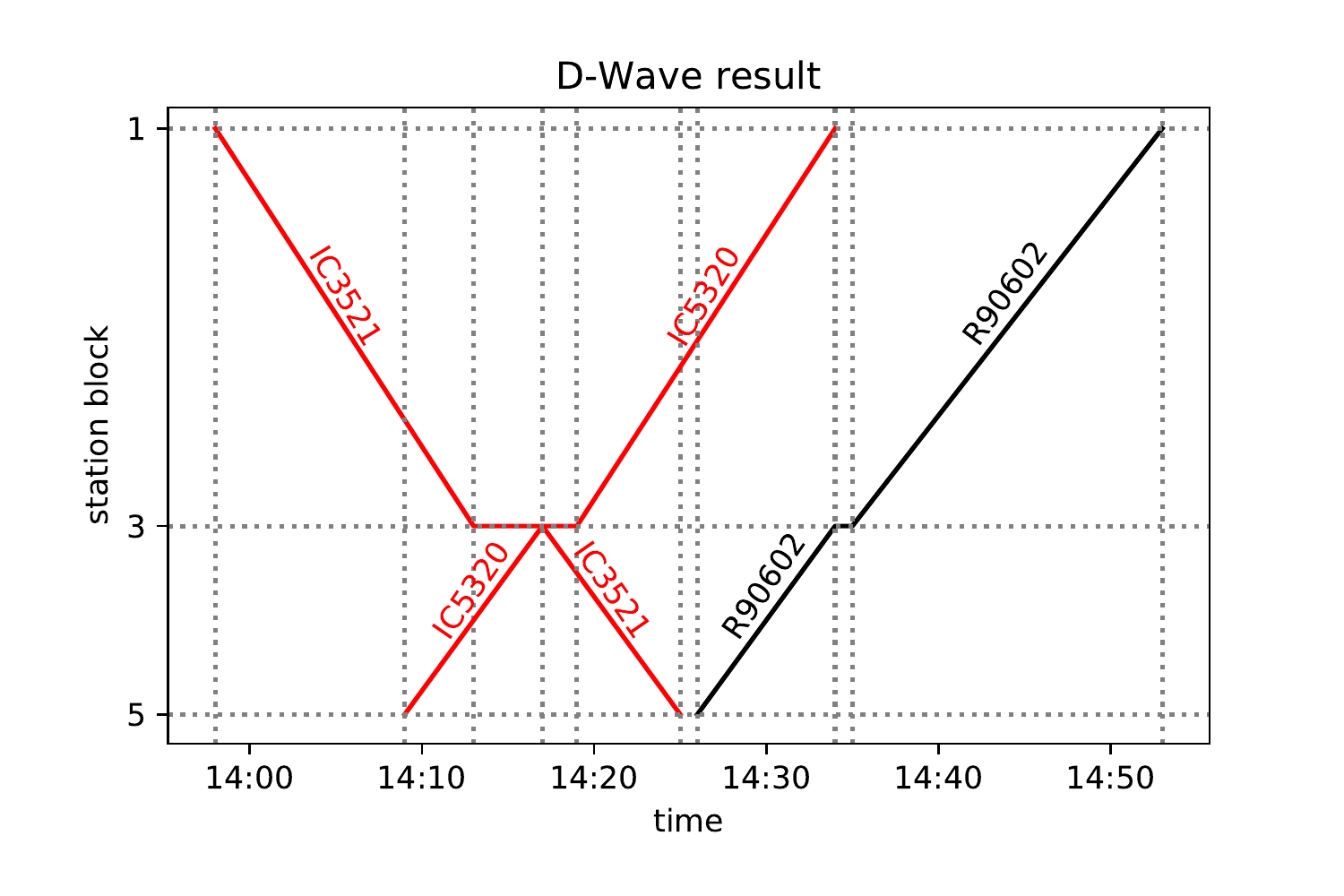}}
	\subfigure[The optimal solution from Fig.~\ref{fig::dw_spec_175_175}. \label{fig::DWsol_small2000_175_175}]{
		\includegraphics[scale =
		0.6,width=\columnwidth]{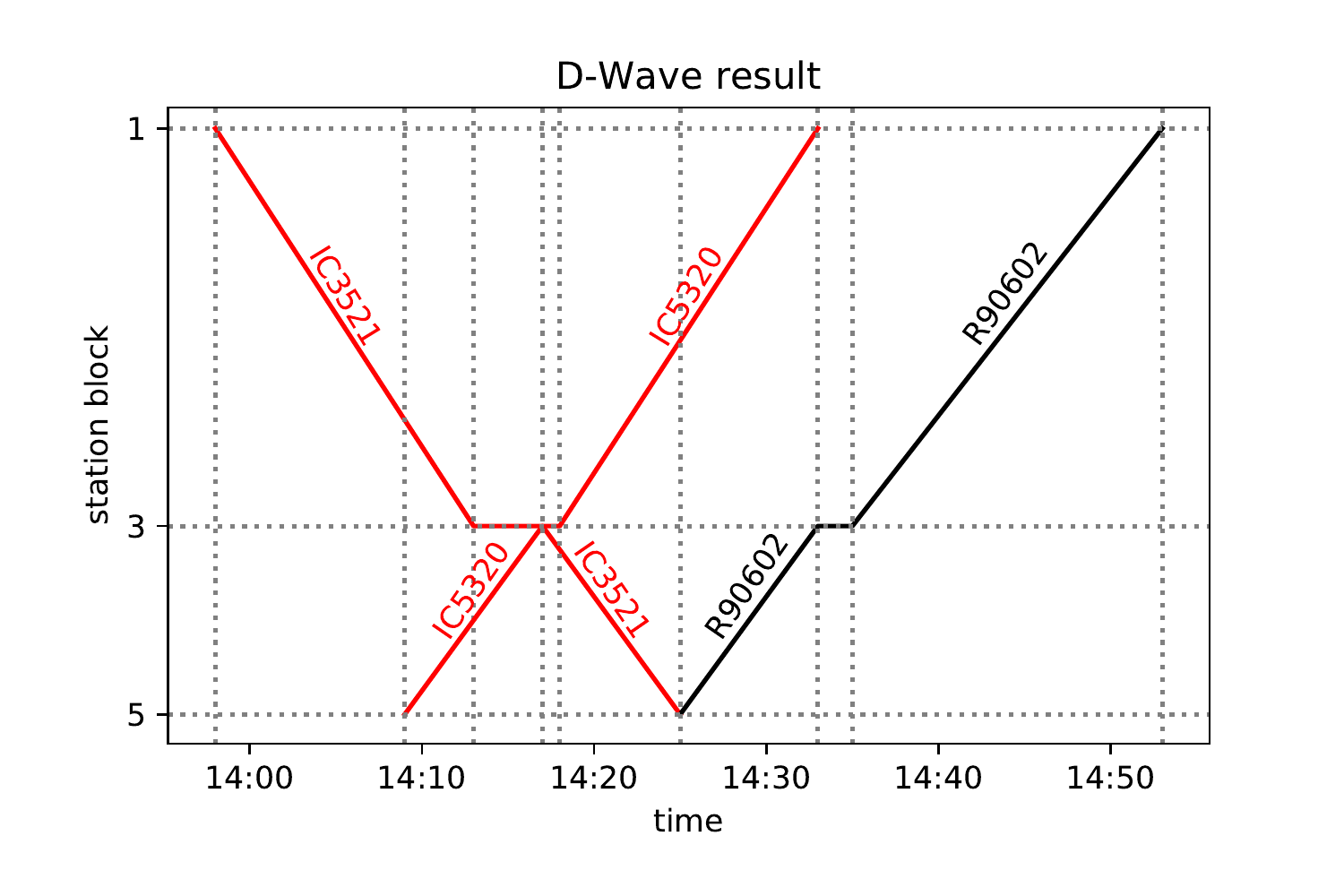}}
	\subfigure[
	$p_{\text{pair}} = 2.7,
	p_{\text{sum}} = 2.2$.\label{fig::dw_spec_27_22}]{\includegraphics[scale
	=
	0.6,width=\columnwidth]{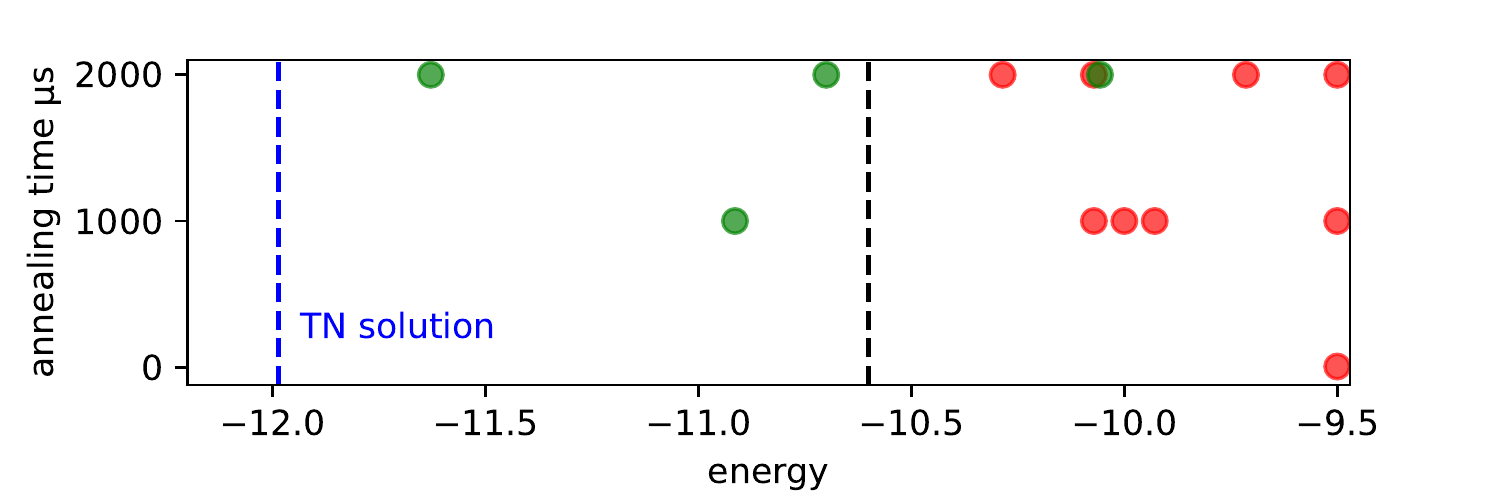}}
	\subfigure[
	$p_{\text{pair}} = p_{\text{sum}} =
	1.75$.\label{fig::dw_spec_175_175}]{\includegraphics[scale
	 =
	0.6,width=\columnwidth]{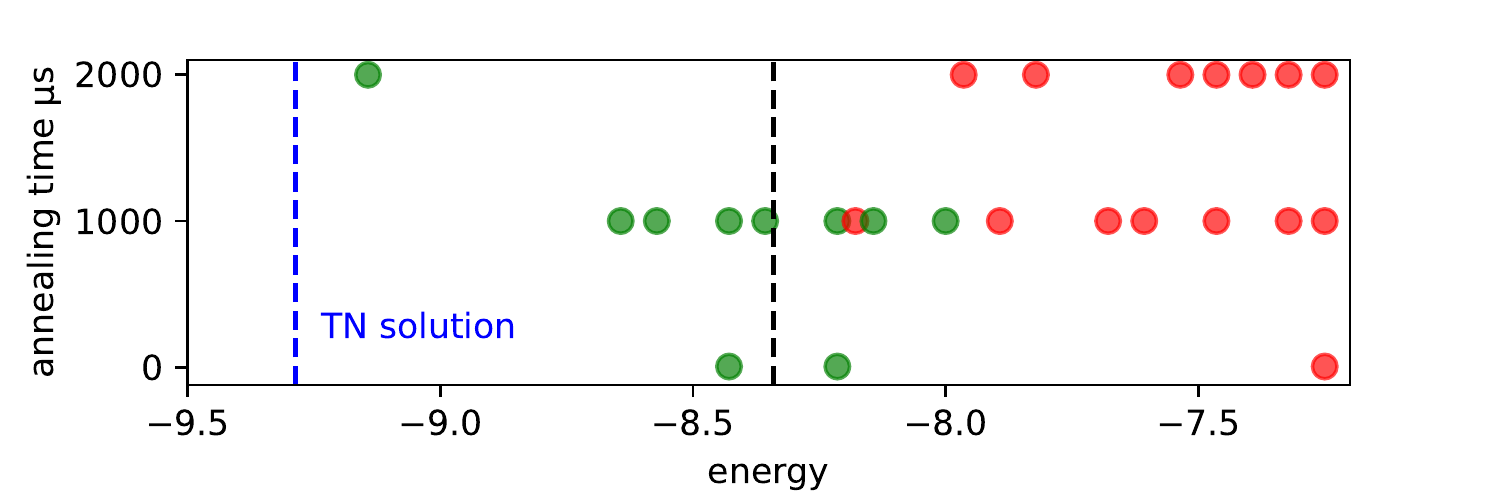}}
  \caption{Train diagrams of the best D-Wave solutions, the lowest energies
    of the quantum annealing on the D-Wave machine (green: feasible, red: not
    feasible), and the
    optimal tensor network solution.\label{fig::bf_spec}}
\end{figure*}

In Figs.~\ref{fig::dw_spec_27_22} and \ref{fig::dw_spec_175_175}, we
present the energies
of the best outcomes of the D-Wave machine for line No. $216$ and various annealing
times. The green dots denote the feasible solutions (and equivalent to the
optimal solution), while the red dots denote solutions that are not feasible. In
general, the
quality of a solution slightly rises with the annealing time;
however, in large examples the best results are for a time
somewhere between $1000 \upmu$s and $2000 \upmu$s. This coincides with the observation
in~\cite{hamerly2019experimental}, in which quantum annealing on the D-Wave machine
was performed on various problems too, and it was demonstrated that for
moderate problem size the performance (in
terms of the probability of success) improves with an annealing
time of up to $1000 \upmu$s. Hence, we have limited ourselves to the annealing times
of the order of magnitude of $1000 \upmu$s in analyzing larger examples.

\begin{figure}
	\subfigure[The minimal energies vs. css for $p_{\text{pair}}
	= 1.75, p_{\text{sum}} = 1.75$.
	\label{fig::DW_vs_css_175_175}]{\includegraphics[scale =
	0.6]{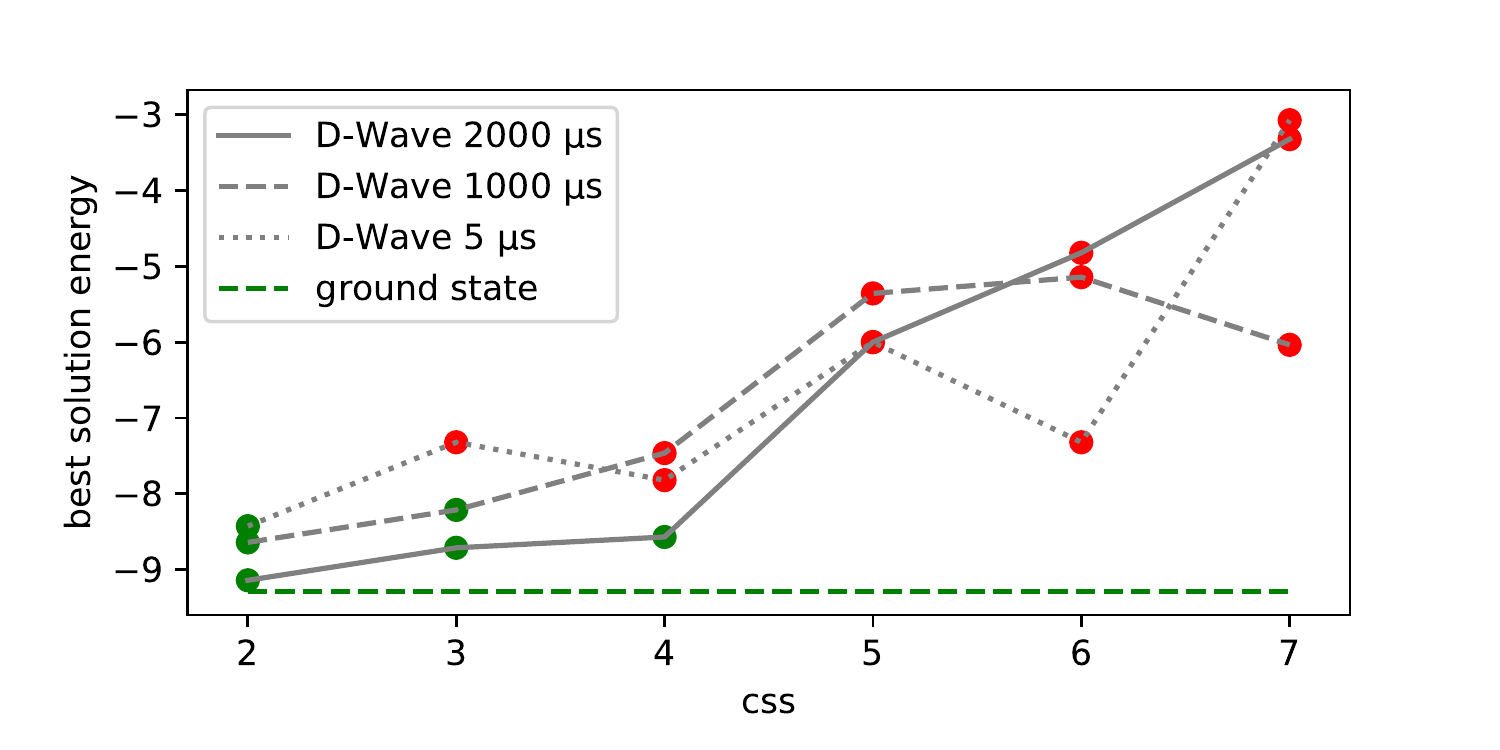}}
	\subfigure[The minimal energies vs. css for $p_{\text{pair}}
	= 2.2, p_{\text{sum}} =
	2.7$.\label{fig::DW_vs_css_22_27}]{\includegraphics[scale =
	0.6]{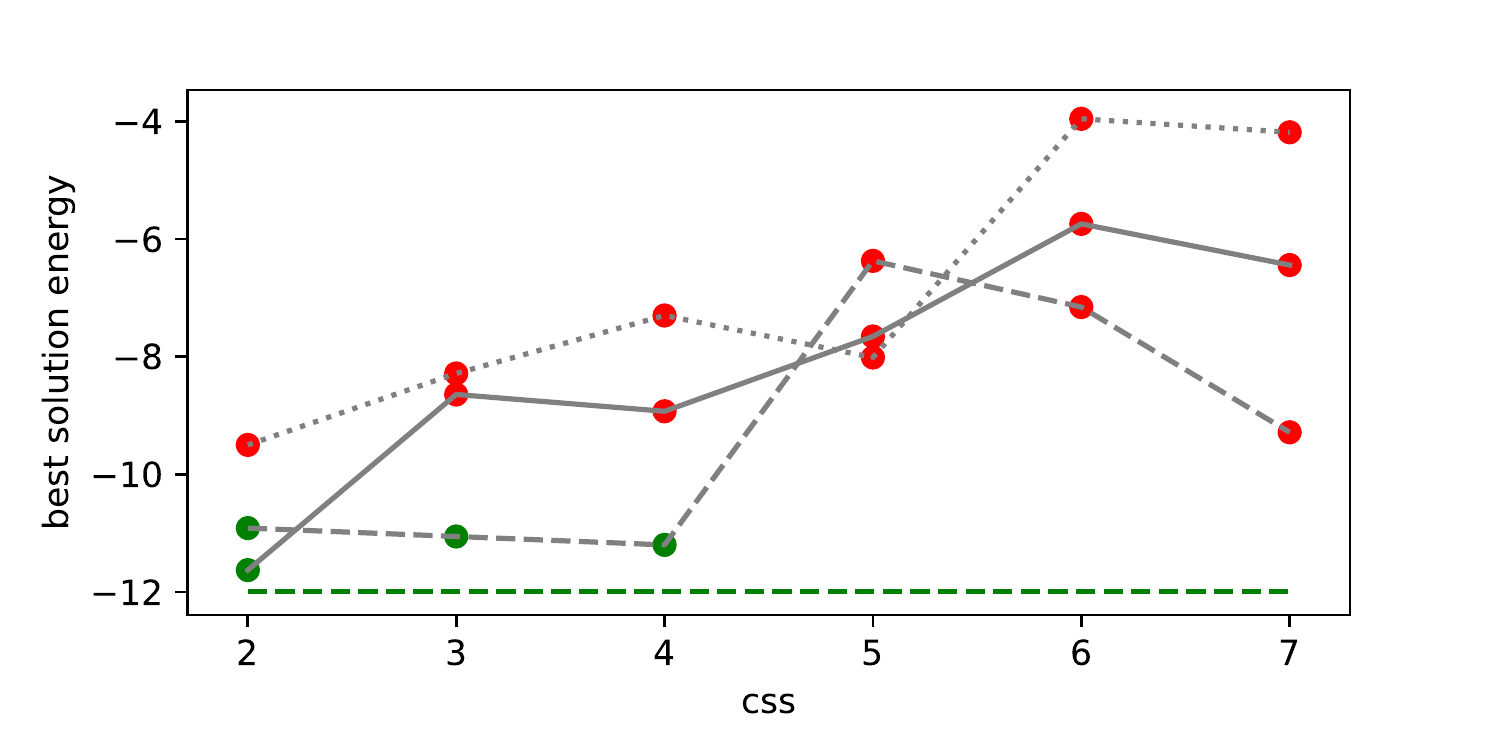}}
	\caption{Line No. 216, with the minimal energy from the D-Wave quantum annealer,
	using $1000$ runs.
	Green
	dots indicates the feasible solutions, while the red dots denote the unfeasible ones. In
	general,
	the energy rises as
	the css strength rises.
	We do not observe that the different settings of $p_{\text{pair}}$ and
	$p_{\text{sum}}$ improve the
	feasibility; see Fig.~\ref{fig::DW_vs_css_22_27}.}\label{fig::DW_vs_css}
\end{figure}

Rather counterintuitively, setting lower penalty coefficients of
$p_{\text{sum}} = p_{\text{pair}} = 1.75$ for the hard constraints,
resulted in samples containing more feasible solutions. For this reason,
we had kept this penalty setting for the analysis of the
larger case.  The embedding strength was set to $css = 2$ in this
case, i.e., the lowest possible value. This has proven to be a good choice,
as demonstrated in Fig.~\ref{fig::DW_vs_css}. The best D-Wave
solutions are presented in the form of train diagrams in
Figs.~\ref{fig::DWsol_small2000_22_27} and \ref{fig::DWsol_small2000_175_175}.

The quality of the solutions in relation to the css strength in the various
parameter settings is presented in Fig.~\ref{fig::DW_vs_css}. We
observed that in our cases, the quality of the solution degraded with
an increase in css. This is unusual, as increasing the css strength typically yields
more solutions without broken chains that do not need to be
post-processed to obtain a feasible solution of the original
problem. This may be caused by the fact that the large coupling of the
embedding may cause the constraints to appear as a small perturbation
in the physical QUBO. These perturbations, as discussed earlier, may be hidden in the
noise of the D-Wave $2000$Q annealer.

Hence, we set $css = 2.0$ (the lowest possible value) for the further
investigations. Some examples of the penalty and objective function
values are presented in Table~\ref{tab::tarfet_penaltes_small2000}.
Again, it appears that the higher the values of $p_{\text{sum}}$ and
$p_{\text{pair}}$, the higher the values of $f(\mathbf{x})$. This may be caused by the objective function being lost in
the noise of the D-Wave $2000$Q annealer.

\begin{figure}[t!]
	\subfigure[Best D-Wave solutions (these are the lowest excited states we have
	recorded). Red dots indicate that the solutions are not
	feasible.]{\includegraphics[scale =
	0.6]{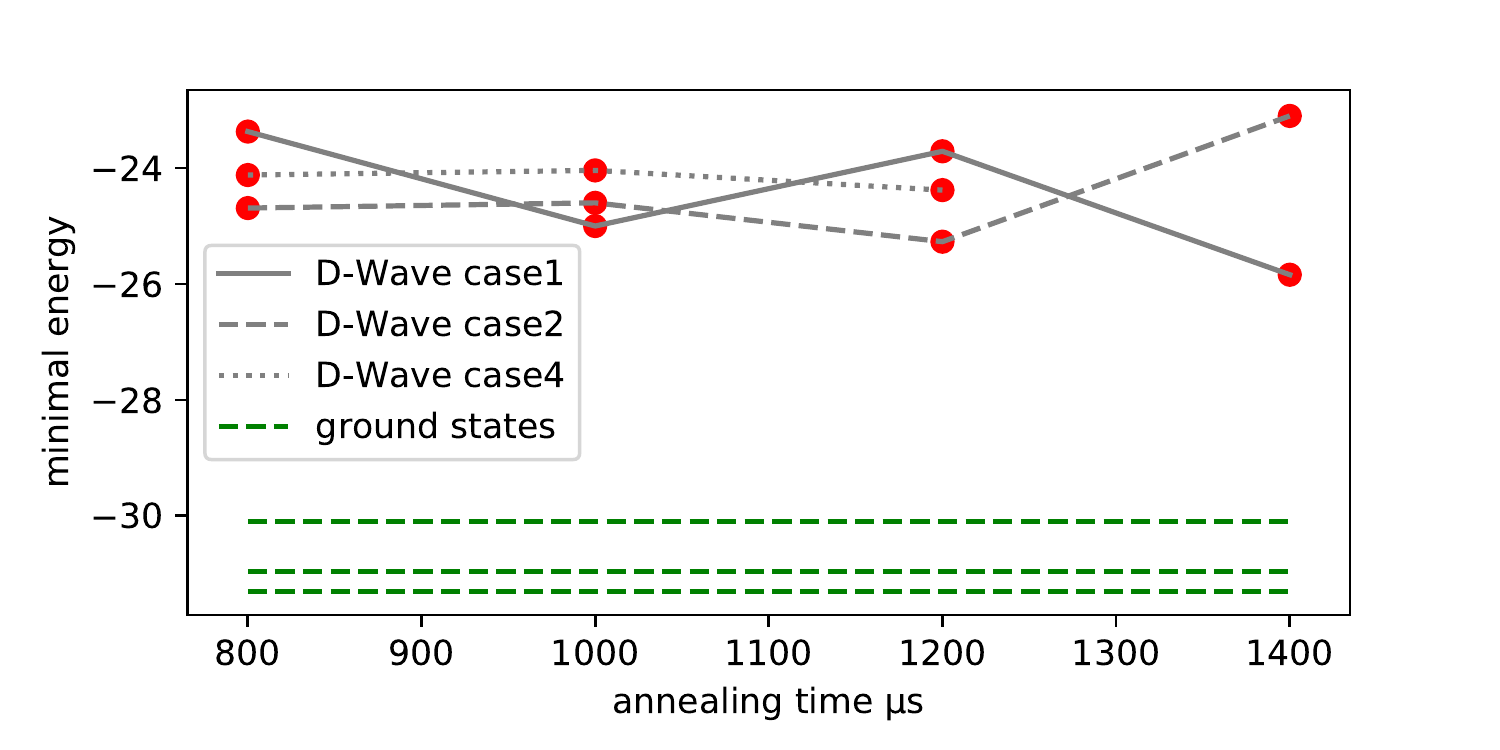}}
	\subfigure[Comparison of the objective and hard penalty between the D-Wave
	outcome and the expected bahaviour (supplied by the
	CPLEX).\label{fig::DW_penalties_w}]{\includegraphics[scale =
	0.6]{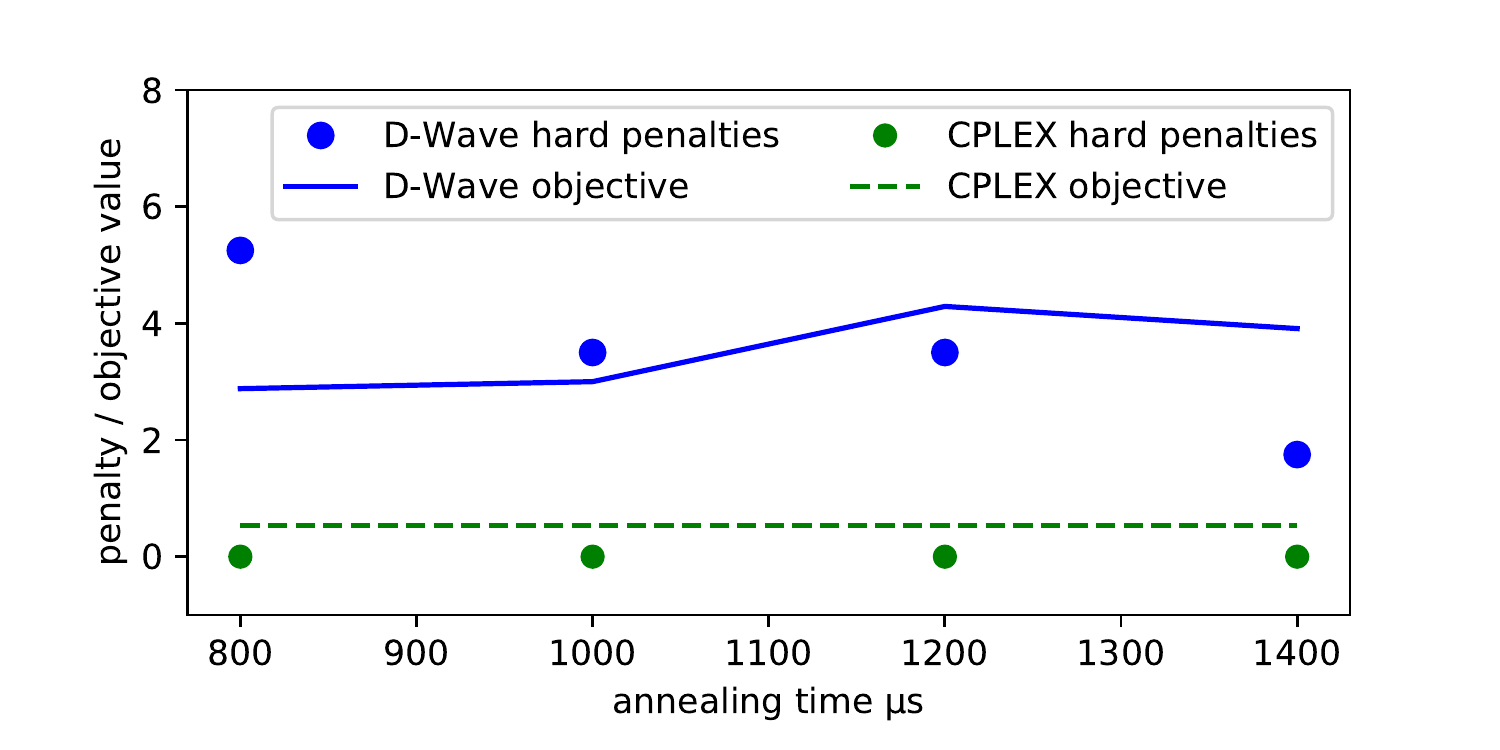}}
	\caption{Line No. 191, with the minimal energy
	from the D-Wave annealer at $250$k runs, css = $2.0$, and
	$p_{\text{pair}} = 1.75, p_{\text{sum}}
	= 1.75$. The output does not dependent on the annealing time (in the
	investigated range)
	and is still far from the ground state.}\label{fig::w_DW_solutions}
\end{figure}

For railway line No. 191, finding a feasible solution is more
difficult. Hence, we took advantage of the maximal number of runs on
the D-Wave machine, which equals $250\ 000$ runs.  The results of the
lowest energies and penalties are presented in
Fig.~\ref{fig::w_DW_solutions}. We had to skip case $3$ because the higher
number of feasibility constraints prevented finding any
embedding on a real Chimera. Interestingly, recall that we
found the embedding for the ideal Chimera while simulating the
solution (see Section~\ref{sec::clasical_heurisitcs}). Hence, the
failure in the case of the real graph is possibly caused by the fact that
some of the required connections or nodes are missing from the real
Chimera.  Finding the feasible solution in such a case (while having
non-zero hard constraints penalties) is a problem for further
research. One would expect that increasing the $p_{\text{pair}}$ and $p_{\text{sum}}$
parameters could be helpful. However, it may aggravate the objective function to
an ever greater extent. In
Fig.~\ref{fig::DW_penalties_w},
the values of the objective function $f(\mathbf{x})$ are much
higher than the optimal ones presented in Table~\ref{tab::tarfet_f_q}.

\begin{table}
	\centering
	\begin{tabular}{cccc}
		css & $p_{\text{sum}}, p_{\text{pair}}$ & hard constraints' penalty
		$f'(\mathbf{x})$ &
		$f(\mathbf{x})$ \\
		\hline
		$2$ & $1.75, 1.75$ &
		$0.0$ & $1.36$ \\
		\hline
		$2$ & $2.2,  2.7$ &
		$0.0$ & $1.57$ \\
		\hline
		$4$ & $1.75, 1.75$ &
		$0.0$ & $1.93$ \\
		\hline
		$4$ & $2.2,  2.7$ &
		$2.2$ & $2.07$ \\
		\hline
		$6$ & $1.75, 1.75$ &
		$5.25$ & $0.43$ \\
		\hline
		$6$ & $2.2,  2.7$ &
		$6.6$ & $0.86$ \\
		\hline
	\end{tabular}
	\caption{Line No. 216, with the objective functions and penalties for violating
	the hard
	constraints: see~\eqref{eq::hard_constrain}. Output from the D-Wave quantum
	annealer, for the annealing time of
	$2000 \upmu$s.
		If $f'(\mathbf{x}) > 0$, the solution is not
		feasible. The $p_{\text{sum}} = p_{\text{pair}} = 1.75$ policy gives
		lower objectives.}
	\label{tab::tarfet_penaltes_small2000}
\end{table}
\begin{figure*}[t]
  \subfigure[Case $1$.\label{fig:DW_case1_solution}]{\includegraphics[scale =
  0.6, width=\columnwidth]{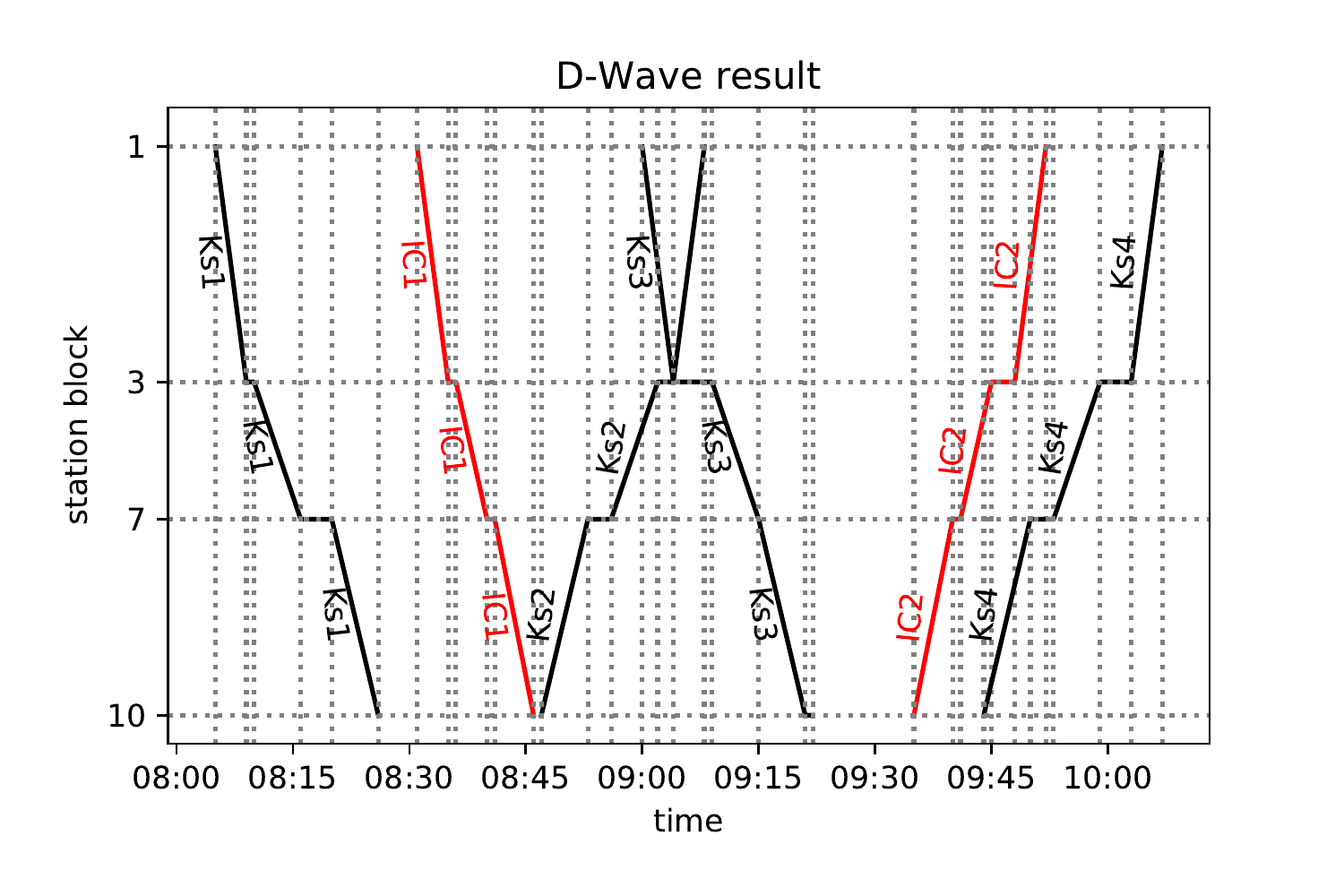}}
  \subfigure[Case $2$.\label{fig:DW_case2_solution}]{\includegraphics[scale =
  0.6, width=\columnwidth]{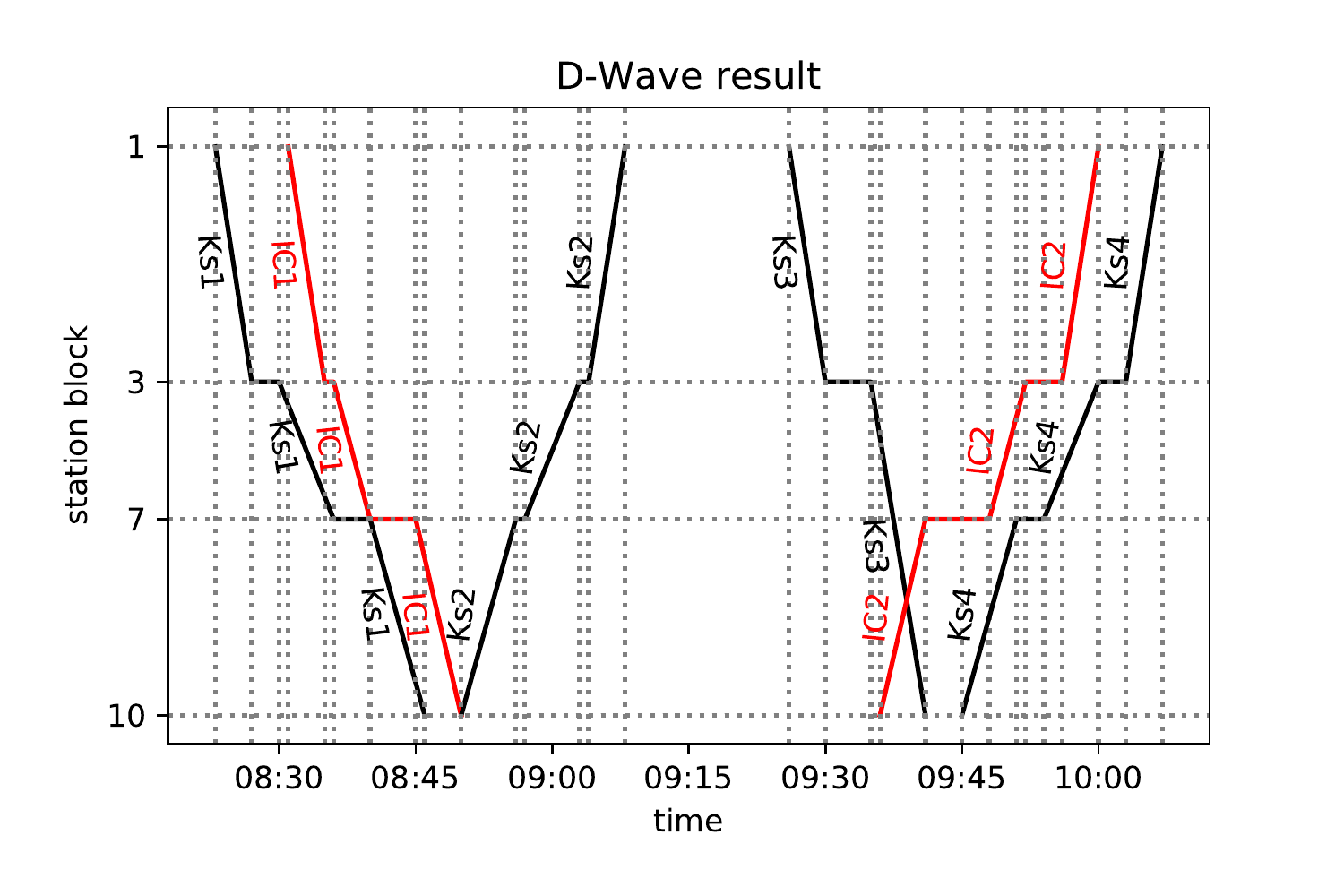}}
  \caption{The best solutions obtained from the D-Wave quantum annealer for line No.
  191.  For
    case 1 (Fig.~\ref{fig:DW_case1_solution}), the annealing time
    is $t = 1400$. The solution is unfeasible since the stay of Ks$3$
    at station $7$ is below $1$ minute. If the solution is corrected
    (i.e., the stay is introduced), it looses its optimailty and reflects a
    dispatching situation different from those obtained from
    FCFS, FLFS, AMCC, CPLEX, or the tensor network. For case $2$
    (Fig.~\ref{fig:DW_case2_solution}), $t = 1200$ is used. The
    solution is unfeasible as Ks$3$ does not stop at station $7$;
    hence, Ks$3$ and IC$2$ are supposed to meet and pass between stations
    $7$ and $10$. It can, however, be amended to an optimal solution:
    shortening the stay of Ks$3$ at station $3$ and shortening the
    stay of IC$2$ at station $7$ (and $3$ if necessary) result in a
    meet-and-pass situation at station $10$, and this is optimal.}
  \label{fig:DW_w_solutions}
\end{figure*}

Although the solutions are not feasible, we select the two in which only
one hard constrain is violated ($f'(\mathbf{x}) = 1.75$); these are
case $1$ with an annealing time of $1400 \upmu$s, and case $2$ with an annealing time of
$1200 \upmu$s. The train diagrams of these solutions
are presented in Fig.~\ref{fig:DW_w_solutions}. Note that both these
diagrams can easily be modified by the dispatcher to obtain a feasible
solution. The case in Fig.~\ref{fig:DW_case1_solution} can be amended
by adding the lacking $1$-minute stay of Ks$3$ in station
$7$. Here the solution would not be optimal, and it would be different
from the optimal one from CPLEX, the tensor network, and FLFS, as well as from the non-optimal yet feasible ones returned by FCFS and AMCC.  The
case in Fig.~\ref{fig:DW_case2_solution} can be upgraded by shortening
the stays of Ks$3$ and IC$2$ and letting them meet and pass at station
$10$. Here the solution would be optimal and equivalent to those
achieved with the tensor networks and CPLEX.

The real D-Wave quantum annealing is tied to some parameters of both
the particular QUBO and the machine itself. We achieved the best
results for a coupling constant css = $2.0$ for the small
example in Fig.~\ref{fig::DW_vs_css_175_175}; the same observation was made
for the large example. This was not expected as the coupling between
quantum bits representing a single classical bit was rather weak. Here we probably took advantage of the possible variations within
the realization of a logical bit. This observation demonstrates that
the embedding selection may be meaningful in searching for the
convergence toward proper solutions lying in the low-energy part of
the spectrum. For the small cases, we observed a feasible
solution for a relatively small number of samples (equal to $1$k). For the
larger case, the number of samples had to be increased to the maximal
possible (equal to $250$k) and still we did not reach any feasible
solution. The conclusion is that the impact of the noise
amplifies strongly with the size of the problem. The convergence of the
best obtained solution toward the optimal one with the given sample size is
complex, and an in-deep statistical analysis of sampling the annealer's
real distribution is required.

As demonstrated in Fig~\ref{fig::DW_penalties_w}, in some cases only a
single hard constraint was violated. This may suggest that we are near
the region of feasible solutions. However, the objective function
values are still far from the optimal ones achieved by means of
simulations (see Table~\ref{tab::tarfet_f_q}). To elucidate the
interplay between penalties, we refer to Fig.~\ref{fig:DW_w_solutions},
in which the solutions are not feasible but can be easily corrected by
the dispatcher to obtain feasible ones. In Fig.~\ref{fig:DW_case2_solution},
the corrected solution would be optimal, while in
Fig.~\ref{fig:DW_case1_solution} it would not be different from all the other
achieved solutions. Hence, the current quantum annealer
would rather sample the excited part of the QUBO spectrum, which can lead
to unusual solutions. Such solutions, however, can be still be used by
the dispatcher for some particular reason not encoded directly in the
model. Such reasons include unexpected dispatching problems,
rolling stock emergency, and non-standard requirements.

Let us also mention the characteristics of our QUBO problems as they
are important features from the point of view of quantum
methods. Table~\ref{tab::graph} summarizes the problem sizes and the
densities of edges in the case of each problem instance.
\begin{widetext}

  \begin{table}
	\centering
	\begin{tabular}{ccccccc}
		{\multirow{2}{*}{Features} } & line 216 & \multicolumn{5}{ c }{line
		191}
		\\
		\cline{2-7}
           & & case $1$ & case $2$ & case $3$ & case $4$ & enlarged \\
		\hline
		 problem size ($\#$ logical bits) & $48$ & $198$ & $198$ & $198$ &
		 $198$  &
		$594$ \\
		\hline
		$\#$ edges & $395$ & $1851$ & $2038$ & $2180$ & $1831
		$   & $5552$ \\
		\hline
		density (vs. full graph) & $0.35$ & $0.095$ & $0.104$ & $0.111$ &
		$0.094$ & $0.032$ \\
		\hlineB{3}
		embedding into & Chimera & Chimera & Chimera & Ideal Chimera &
		Chimera  & Pegasus \\
		\hline
		approximate $\#$ physical bits & $373$ & $< 2048$ & $< 2048$ & $
		\approx
		2048$ &
		$< 2048$  & $< 5760$ \\
		\hline
	\end{tabular}
	\caption{Graph densities for various problems. As case $3$ is supposed to
	be the most complicated one of cases $1-4$, it has the largest graph
	density.
  }
	\label{tab::graph}
      \end{table}

\end{widetext}

\subsection{Initial studies on the D-Wave Advantage machine}\label{sec::pegasus}

During the preparation of the present paper, a new quantum device,
the D-Wave's Advantage\_system1.1 system (with an underlying topology code-named
Pegasus~\cite{dattani.szalay.19}) became commercially available. Hence,
we performed preliminary experiments with this new architecture to
address a slightly larger example. To that end, we expanded our initial
Goleszów -- Wisła Uzdrowisko (line No. $191$) problem instance to be $3$
times bigger in size.  Furthermore, we investigated nine trains in each direction.

The conflicts were introduced by assuming delays of $20$, $25$, or $30$ minutes
of certain trains entering block section $1$. The control parameters' values
$p_{\text{sum}} = p_{\text{pair}} = 1.75$ and $css = 2$ were not changed.
As a result, the problem was mapped onto a QUBO with $594$ variables
and $5552$ connections. This new setup
changed the underlying embedding only slightly, and we omit a discussion
of the details here. Employing a strategy similar to the one used for our other calculations, we used the solution found by CPLEX as a reference for comparisons.

\begin{figure*}
	\begin{center}
		\includegraphics[scale =
		0.75]{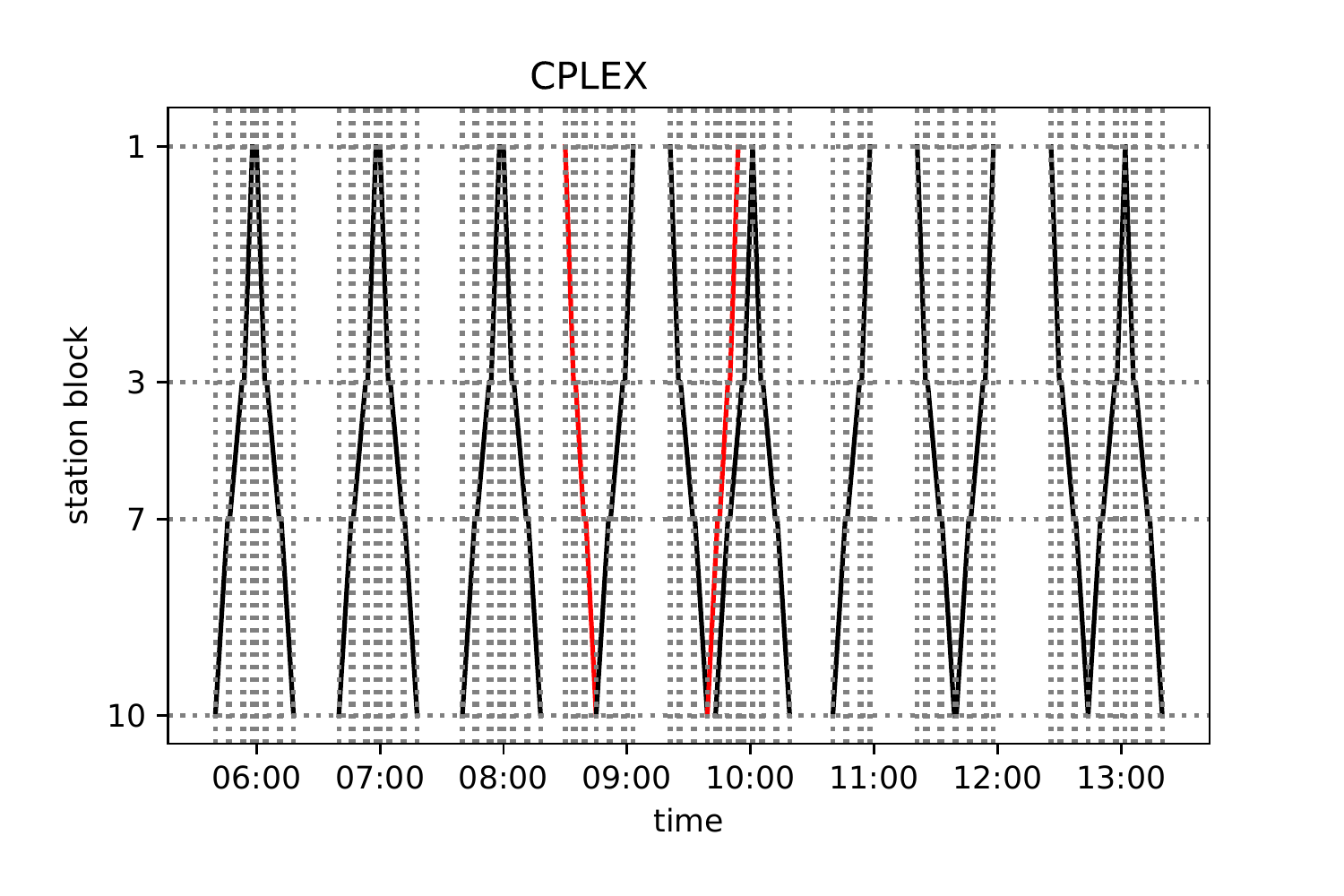}
	\end{center}
	\caption{The CPLEX QUBO solution, coinciding with the linear solver's solution of the $18$-train
		problem.}\label{fig::case7_cplex}
\end{figure*}

After performing $25$k runs, we reached a minimal energy of
$+75.28$ with an annealing time of $1400 \upmu$s. Unfortunately, this is not a feasible solution. The CPLEX calculations, on the other
hand, resulted in an energy of $-92.43$ with an objective function value
$f(\mathbf{x}) = 2.07$ (see Fig.~\ref{fig::case7_cplex}). This is the same solution as
the solution of the linear solver obtained using COIN-OR in 0.02
seconds. This solution is substantially better, and as it coincides with the
linear solver's output, it corresponds to the ground state.
Our preliminary experiments indicate the need for a more detailed investigation
of the new device's behavior (and that of the current model) to determine
whether obtaining solutions with the desired (better) quality is possible.
A part of this problem will likely be eliminated simply by the technological development
of the new annealer. For an intuitive justification, we refer
to~\cite{jalowiecki2020parallel} and
\href{https://www.nature.com/articles/s41598-020-70017-x}{Fig.~$1$}
therein, in which an improvement in the performance between subsequent iterations within one generation
of Chimera-based quantum annealers was observed.

\section{Discussion and conclusions}
\label{sec::conclusions}

\s{The NISQ era \cite{Preskill2018quantum} is here. Early generations of quantum computing hardware have become available that may serve as a stepping stone for the development of practically useful technologies that exhibit genuine quantum advantage. However, until such ``real'' quantum computers it is imperative to demonstrate how and which real-world applications are amendable to be solved on quantum computing hardware.
To this end, we} have introduced a new approach to the single-track line dispatching
problem that can be implemented on a real quantum annealing device
(D-Wave $2000$Q).  \s{Namely, we} have addressed two particular real-life railway
dispatching problems in Poland; many similar
examples exist in other networks, too. \s{Specifically, we} have introduced a QUBO model of the problem that
can be solved with quantum annealing, \s{and we have benchmarked it against} classical algorithms.

The first dispatching problem we considered (the Nidzica -- Olsztynek
section of line No. $216$) was particularly small, \s{and} it was
defined using \s{only} $48$ logical quantum bits (which we were able to embed
into $373$ physical quantum bits of a real quantum processor). The
final state reached by the quantum annealer for this problem was
optimal for many parameter settings. This highlights that small-sized
dispatching problems are already within reach of near-term quantum
annealers. In addition, the limited size of the problem made it
possible to analyze the QUBO with a greedy brute-force search
algorithm, which revealed details of the behavior of the spectrum that
cannot be exactly calculated for bigger instances.

Our second set of dispatching problems (the Goleszów -- Wisła Uzdrowisko
section of line No. $191$) was larger and needed $198$ logical
quantum bits. Here, the number of physical quantum bits depends on the
number of constraints in each of the analyzed cases. We were able to
embed all four dispatching cases of the No. $191$ railway line into an ideal
Chimera graph ($2048$ physical quantum bits) using a state-of-the-art
embedding algorithm. We succeeded in solving these instances with
classical solvers for QUBOs.  Meanwhile, on the physical device (whose
graph is not perfect and lacks several quantum bits and couplings), we
were able to embed only three out of the four cases (case $3$, with the highest number \s{of} conflicts, could not be embedded).  We expect
that such obstacles will become less restrictive as new embedding algorithms
are being developed for both the current Chimera topology and the newest
D-Wave Pegasus; see~\cite{zbinden2020embedding,pelofske2020decomposition}.
Therefore, it is not unreasonable to expect that the range of problems
that can be embedded so that they can be solved on physical hardware
will substantially increase in the near future. Unfortunately, the D-Wave
$2000$Q solutions of our second problem appeared to be far from
optimal. This is attributable to the noise that is still present in the
current quantum machine.

We have successfully solved our model using \s{several} algorithms for
QUBOs running on classical computers, notably the novel tensor network
method. This introduces additional possibilities, namely, that of QUBO
modeling and the use of quantum-motivated classical algorithms. Although
these possibilities obviously do not promise a breakthrough in scalability, they
are essential for the validation and assessment of the results of real
quantum annealing. In addition, they can yield practically useful
results.\s{In fact, hybrid quantum-classical computing is a promising avenue of research, which has recently seen significant development \cite{Endo2021JPSJ,Ding2021SN}.}

We are aware that the examples of the single-track railway dispatching
problem discussed in the paper can be regarded as trivial from the
point of view of professional dispatchers. This is also reflected by
the efficiency of the conventional linear solver they may use. Our intention, however, was to provide a proof-of-concept
demonstration of the applicability of quantum annealing in this
field. This goal has been achieved: we have described a suitable model
and succeeded in solving certain instances.

Due to the small size of the current quantum annealing processors, our
implementation is limited: quantum annealing is an emerging
technology. Owing to the significant efforts put into the development of quantum annealers, the
addressable problem sizes are about to increase, and the quality of the
samples will also improve. With the development of the technology, \s{it is not far-fetched to realize} that at some point \s{soon} quantum annealers will be able to
compete with or even outperform classical solvers.

\begin{acknowledgments}
The research was supported by the Ministry of Innovation and
Technology and the National Research, Development and Innovation
Office within the Quantum Information National Laboratory of
Hungary; and the Foundation for Polish Science (FNP) under grant number
TEAM NET POIR.04.04.00-00-17C1/18-00 (KD and BG); and the National
Science Centre (NCN), Poland, under project number 2016/22/E/ST6/00062
(KJ); and by the Silesian University
of Technology Rector’s Grant no. BKM-745/RT2/2021 12/020/BKM$_{-}$2021/0213 (KK). MK acknowledges the support of the National Research,
Development, and Innovation Office of Hungary under project numbers
K133882 and K124351. We gratefully acknowledge the support of NVIDIA
Corporation, which donated the Titan V GPU used for this research.

\end{acknowledgments}

%

\end{document}


\title{Quantum annealing in the NISQ era: railway conflict management\\SUPPLEMENTAL MATERIAL}

\author{Krzysztof Domino}
\email{kdomino@iitis.pl}
\affiliation{Institute of Theoretical and Applied Informatics, Polish
  Academy of Sciences, Ba{\l}tycka 5, 44-100 Gliwice, Poland}
\author{M\'aty\'as Koniorczyk}
\email{koniorczyk.matyas@wigner.hu}
\affiliation{Wigner Research Centre, H-1525 Budapest, Konkoly-Thege M. \'ut 29-33, Budapest, Hungary}
\author{Krzysztof Krawiec}
\email{krzysztof.krawiec@polsl.pl}
\affiliation{Silesian University of Technology, Faculty of Transport and Aviation Engineering, Gliwice, Poland} 
\author{Konrad Ja\l{}owiecki}
\email{dexter2206@gmail.com}
\affiliation{Institute of Physics, University of Silesia, Katowice, Poland}
\author{Sebastian Deffner}
\email{deffner@umbc.edu}
\affiliation{Department of Physics, University of Maryland Baltimore County, Baltimore, MD 21250, USA}
\author{Bart\l{}omiej Gardas}
\email{bartek.gardas@gmail.com}
\affiliation{Institute of Theoretical and Applied Informatics, Polish
  Academy of Sciences, Ba{\l}tycka 5, 44-100 Gliwice, Poland}
\date{\today}

\maketitle

The present supplemental material contains details which are not necessary for the basic understanding of the results. They are needed, however, to specify the solved model in full detail and illustrate further the soltuions. In particular, in Section~\ref{sec::ourmodel} we describe the model's complete notation and its constraints in a way common in the Operations Research literature. Section~\ref{sec::linear_integer} is devoted to a linear integer programming formulation of the model which is used for a comparison with a classical approach. Section~\ref{sec::qubo} provides all the details of the quadratic unconstrained binary (QUBO) formulaton. I Section~\ref{sec::classalg} we comment further on classical algorithms for solving QUBO (or Ising) problems. Finally, Section~\ref{sec::appendix} contains further computational results concerning particular railway situtations.

\section{Our model: details}
\label{sec::ourmodel} 

In this section, we introduce the model of the railway line and the
dispatching conditions. Table~\ref{tab::symbols} provides a
comprehensive summary of the notation used.
\begin{table}[]
	\centering
	\begin{tabular}{lp{0.5\textwidth}}
		\textbf{symbol} & \textbf{description / explanation}  
		\\ \hline
		$A_{j,s}$ & discretized set of all possible delays of train $j$ at station $s$ \\ \hline
		$\mathcal{H}(t)$, $\mathcal{H}_0$, $\mathcal{H}_p$ & time-dependent Hamiltonian 
		of the annealing process and its time-independent 
		components \\ \hline
		$t \in [0,T]$ & quantum annealing time \\ 
		\hline
		$\hat \sigma^x, \hat \sigma^z$ & Pauli matrices \\ \hline
		$j\in \mathcal{J}$ & trains (jobs) \\ \hline
		$\mathcal{J}^0$ ($\mathcal{J}^1$) & trains heading in a given (opposite) 
		direction 
		\\ \hline
		$m \in \mathcal{M}$ & blocks (machines) \\  \hline
		$s \in \mathcal{S}$ & station blocks \\  \hline
		$l \in \mathcal{L}$ & line blocks \\  \hline
		$M_j, (S_j)$ & the sequence of blocks (station blocks) in the route of 
		$j$\\ \hline
		$s_{j,1}, s_{j,k}, s_{j, \text{end}}$ & the first, $k$-th, and last station 
		block in the route of $j$ \\ \hline
		$m_{j,1}, m_{j,k}, m_{j, \text{end}}$ & the first, $k$-th, and last
		block in the route of $j$ \\ \hline
		$ S_j = (s_{j, 1}, s_{j, 2}, \ldots, 
		s_{j, \text{end}})$ & a sequence of all station blocks in $j$'s route  \\ 
		\hline
		$S_j^*$, ($S_j^{**}$) & a sequence of station blocks in $j$'s route 
		without the 
		last (last two) elements \\ \hline
		$S_{j,j'}$ & a common path of $j$ and $j'$, ordered according to $j$'s path \\ \hline
		$S^*_{j,j'}$ & a common path of $j$ and $j'$ excluding the last block, ordered according to $j$'s path\\ \hline
		$\rho_j(m), \rho_j(s)$ & the subsequent block (station block) in $j$'s 
		route \\ \hline
		$\pi_j(m), \pi_j(s)$ & the preceding block (station block) in $j$ 's
		route \\ \hline
		$t_{\text{out}}(j,s)$, ($t_{\text{in}}(j,s)$) & time of leaving (entering) 
		station block $s$ by train $j$\\ \hline		
		$t^{\text{timetable}}_{\text{out}}(j,s)$ & timetable time of leaving $s$ by 
		$j$ \\ \hline
			$p_{\text{timetable}}(j,m)$, $p_{\text{min}}(j,m)$ & 
		timetable and minimum time of $j$ passing $m$ \\ \hline
		$d(j,s)$ & delay of $j$ leaving $s$ \\ \hline
		$d_U(j,s)$ & primary (unavoidable) delay of $j$ leaving $s$ \\ \hline
		$d_s(j,s)$ & secondary delay of $j$ leaving $s$ \\ 
		\hline
		$d_{\text{max}}(j)$ & maximum possible (acceptable) secondary delay for train 
		$j$\\ \hline
		$\tau_{(1)}(j,...) \tau_{(2)}(j, ...)$ & minimum time for train $j$ to 
		give way to another train going in the same (opposite) direction \\ 
		\hline
		$x, (\mathbf{x}$) &  binary decision variable (vector of binary decision variables), e.g., 
		$x_{j,s,d} = x_i$ is $1$ if $j$ leaves $s$ with a delay $d$ and 
		$0$ otherwise, $i \in \{1,2,\ldots, n\}$\\ \hline
		$Q\in {\mathbb{R}}^{n\times n}$ & symmetric QUBO matrix, where $n$ is the 
		number 
		of logical quantum bits. \\ \hline
		$f(\mathbf{x})$ & objective function; the  weighted sum of secondary delays  \\ \hline
	\end{tabular}
	\caption{Notation summary}\label{tab::symbols}
\end{table}

\subsection{Railway line}\label{sec::model_line}

We assume a railway line $\mathcal{M}$ 
to be a set of block sections: segments that can be occupied by at most one train at a time, as discussed in the paper. These 
are either \emph{line blocks}
or \emph{station blocks}; both are also refereed to as \emph{block sections} or
just \emph{blocks}. The set of line blocks are denoted by $\mathcal{L}$, and the 
set
of station blocks by $\mathcal{S}$. This model also incorporates
sidings or double-track sections by treating them as station
blocks.

Trains can only meet and pass (M-P) or meet and overtake (M-O) at
stations. We follow the buffer approach by treating each station as a
block that can be occupied by up to $b$ trains at a time, where $b$ is the
number of tracks at the station. The other blocks can be occupied
by only one train at a time.

The set of trains is denoted by $\mathcal{J}$ and is split into the
subset of trains traveling in a given direction $\mathcal{J}^0$ and the subset
of trains going in the opposite direction $\mathcal{J}^1$:
\begin{equation}
\label{sup:test}
\mathcal{J}^0 \cup \mathcal{J}^1 = \mathcal{J} \text{ and }  \mathcal{J}^0 \cap 
\mathcal{J}^1 = \emptyset.
\end{equation}

Let $j\in \mathcal{J}$ be a particular train. Its route is a sequence
of blocks $M_j = (m_{j,1}, m_{j,2}, \ldots, m_{j,\text{end}})$, where
$m_{j,1}$ is the starting block and $m_{j, \text{end}}$ is the ending
block. Each block (from $M_j$) is passed by train $j$ once and
only once (i.e., we do not consider recirculation). Given a train $j$
and a block $m_{j,k}$, the preceding block is
$\pi_j(m_{j,k}) = m_{j,k-1}$, while the subsequent block is
$\rho_j(m_{j,k}) = m_{j,k+1}$. We assume that neither
$\rho_j(m_{j, \text{end}})$ nor $\pi_j(m_{j, 1})$ belongs to the
analyzed network segment.

We assume that a route can be defined solely by a sequence of station
blocks $S_j = (s_{j, 1}, s_{j, 2}, \ldots, s_{j, \text{end}})$, where
M-P and M-O may occur (i.e., there are no alternative
routes between stations).  Similarly to blocks in general, for a train
$j$ and a station block $s_{j,k}$, we denote the preceding station
block as $\pi_j(s_{j,k}) = s_{j,k-1}$, and the subsequent station
block as $\rho_j(s_{j,k}) = s_{j,k+1}$. It is convenient to assume
that all train paths start and end at stations; hence
we have $s_{j, 1} = m_{j, 1}$ and
$s_{j, \text{end}} = m_{j, \text{end}}$.

\subsection{Delay representation}\label{sec::model_delays}

Ideally, the time $t_{\text{out}}(j,s)$ when train $j$ leaves station block 
$s$ 
should be the time prescribed by the timetable, $t^{\text{timetable}}_{\text{out}}(j,s)$. If, however, 
$t_{\text{out}}(j,s) > 
t^{\text{timetable}}_{\text{out}}(j,s)$, there is 
a delay in leaving station block $s$:
\begin{equation}\label{eq::delay}
d(j,s) = t_{\text{out}}(j,s) - t^{\text{timetable}}_{\text{out}}(j,s).
\end{equation}

Primary or unavoidable delays (as defined
in Section~\ref{problemdescr}) are denoted by $d_U(j,s)$. If an
already delayed train enters a railway line $\mathcal{M}$, the initial
delay will appear at the first block $d_U(j,s_{j,1})$.
The unavoidable delay propagates along the line, thereby providing a
lower bound of the overall delay. Unavoidable delays are
non-negative, so we have
\begin{equation}\label{eq::duprop}
d_U(j,\rho_j(s)) = \max \{d_U(j, s) - \alpha(j,s, \rho_j(s)), 0 
\},
\end{equation}
where $\alpha(j,s, \rho_j(s))$ accounts for the possible time reserve
in passing the sequence of blocks, starting from the one directly
after $s$ and ending at station block $\rho_j(s)$. In the same way,
the unavoidable delays are propagated due to the minimum times of the rolling stock circulation at terminals. Importantly, all unavoidable delays can be
computed prior to the optimization.

The secondary delay $d_S(j,s)$ is denoted by
\begin{equation}
d_S(j,s) = d(j,s) - d_U(j,s).
\end{equation}
We introduce upper bounds $d_{\text{max}}(j)$ of the secondary delays
as parameters of the model. Their values can either be determined
manually (maximum acceptable secondary delays of the given trains) or be obtained
by using some fast heuristics such as the first come first served
(FCFS) or first leave first served (FLFS) approach (which will be discussed
later). Setting them too low, however, can result in an unfeasible
model.

Having established the upper and lower bounds,
\begin{equation}\label{eq::dlim} d_U(j,s) \leq d(j,s) \leq
  d_U(j,s)+d_{\text{max}}(j),
\end{equation}
we can use the (integer) values of the delays as decision
variables. The bounds ensure that these variables remains in a finite range. In what follows, we shall call this description, in terms of the
discretized delays as decision variables, ``delay representation''; it will be very convenient from the QUBO modeling
point of view.

\subsection{Dispatching conditions}\label{sec::sispcond}

Consider a train $j$ whose path $M_j$ consists of both station and
line blocks. We assume that the leaving time of the given block equals
the entering time of the subsequent block:
\begin{equation}\label{eq::inout}
t_{\text{out}}(j,m) = t_{\text{in}}(j,\rho_j(m)).
\end{equation}
(This is a slight simplification as there is a finite time in which the
train is located in both blocks.) For each train $j\in \mathcal{J}$ and
each block $m \in M_j$, two kinds of passing times are assigned: a nominal
(timetable) $p_{\text{timetable}}(j,m)$ and a minimum
$p_{\text{min}}(j,m)$. Note that the latter can be smaller or equal to
$p_{\text{timetable}}(j,m)$ (as there can be a reserve).

We address common dispatching conditions, including: the minimum
passing time condition, the single block occupation condition, the
deadlock condition, the rolling stock circulation condition 
at the terminal, and the capacity condition.

\begin{condition}\label{cond::minpasstime}
\textbf{The minimum passing time condition}. The leaving time from the block 
section cannot be lower than the sum of the entering time and the minimum 
passing time: 
\begin{equation}\label{eq::pminuneqiality}
t_{\text{out}}(j,m) \geq t_{\text{in}}(j, m) + p_{\text{min}}(j,m).
\end{equation}
For subsequent station blocks $s = m_{j,k}$ and $\rho_j(s) = 
m_{j,l}$, we have
\begin{equation}
t_{\text{out}}(j,\rho_j(s)) \geq t_{\text{out}}(j, s) + \sum_{i = k+1}^l p_{\text{min}}(j,m_{j,i})
=  t_{\text{out}}(j, s) + \sum_{i = k+1}^l 
p_{\text{timetable}}(j,m_i) - \alpha(j, s, \rho_j(s)),
\end{equation}
where $\alpha(j, s, \rho_j(s))$ is the time reserve mentioned
before. In the delay representation, this
condition takes the simple form
\begin{equation}\label{eq::toutinequal}
d(j,\rho_j(s)) \geq d(j, s) - \alpha(j, s, \rho_j(s)).
\end{equation}
(Compare this with~\eqref{eq::duprop}, where we have an equal sign for the lower 
limit.)
\end{condition}

\begin{condition}\label{cond::2trains1way}
  \textbf{The single block occupation condition.}  Let $j$ and $j'$ be
  two trains heading in the same direction and sharing their
  routes between station $s$ and subsequent station
  $\rho_j(s)$. If train $j$ leaves station block $s$ at time
  $t_{\text{out}}(j,s)$, the subsequent
  ($t_{\text{out}}(j',s) \geq t_{\text{out}}(j,s)$) train $j'$ can
  leave this block at a time for which the following equation is
  fulfilled:
\begin{equation}
t_{\text{out}}(j',s) \geq  t_{\text{out}}(j,s) + 
\tau_{(1)}(j,s, \rho_j(s)),
\end{equation}
where $\tau_{(1)}(j,s , \rho_j(s))$ is the time required for train $j$ to 
give way to train $j'$ on the route between station block $s$ and 
subsequent station block $\rho_j(s)$.
In the delay representation we have: 
\begin{equation}
d(j',s)+t^{\text{timetable}}_{\text{out}}(j', s) \geq  
d(j,s) + t^{\text{timetable}}_{\text{out}}(j, s) + 
\tau_{(1)}(j,s, \rho_j(s))
\end{equation}
or
\begin{equation}
d(j',s) \geq  
d(j,s) + t^{\text{timetable}}_{\text{out}}(j, s) - t^{\text{timetable}}_{\text{out}}(j', s) + 
\tau_{(1)}(j,s, \rho_j(s)).
\end{equation}
Hence, taking $\Delta(j, s, j', s) = t^{\text{timetable}}_{\text{out}}(j, s) - 
t^{\text{timetable}}_{\text{out}}(j', s)$, we get
\begin{equation}\label{eq::tau1}
d(j',s) \geq  
d(j,s) + \Delta(j, s, j', s) +
\tau_{(1)}(j,s, \rho_j(s)).
\end{equation}
As mentioned before, the condition in~\eqref{eq::tau1} needs to be tested for 
$t_{\text{out}}(j',s) \geq t_{\text{out}}(j,s)$, i.e., $d(j',s) \geq d(j,s) + \Delta(j, s, j', 
s)$; otherwise trains must be investigated in the reversed order.

The actual form of $\tau_{(1)}(j,s,\rho_j(s))$ depends on the
dispatching details of the particular problem. We assume that all the
time reserves are realized on stations. Consequently,
$\tau_{(1)}(j,s,\rho_j(s))$ is delay independent, which makes the
problem tractable. 
\end{condition}

\begin{condition}\label{cond::2trains2way}
  \textbf{The deadlock condition.} Assume that two trains $j$ and $j'$
  are heading in opposite directions on a route determined by
  subsequent station blocks $s$ and $\rho_j(s)$ in the path of
  train $j$. In the path of $j'$, these are reversed, so $j$ goes
  $s \rightarrow \rho_j(s)$, while $j'$ goes
  $\rho_j(s) \rightarrow s$. Assume for now that the train $j$ will
  enter the common block section before $j'$. (This condition must also be
  checked in the reverse order.) Let
  $\tau_{(2)}(j, s, \rho_j(s))$ be the time required for train $j$
  to get from station block $s$ to $\rho_j(s)$. Given this, the
  deadlock condition can be stated as follows:
	\begin{equation}
	t_{\text{out}}(j',\rho_j(s)) \geq t_{\text{in}}(j, \rho_j(s)),
	\end{equation}
	i.e.,
\begin{equation}
	 t_{\text{out}}(j', \rho_j(s)) \geq t_{\text{out}}(j,s) + \tau_{(2)}(j, s, \rho_j(s)).
\end{equation}
In the delay representation,
\begin{equation}
d(j',\rho_j(s)) + t^{\text{timetable}}_{\text{out}}(j', \rho_j(s)) \geq d(j, s) + 
t^{\text{timetable}}_{\text{out}}(j, s) + \tau_{(2)}(j, s, \rho_j(s))
\end{equation}
and 
\begin{equation}
d(j', \rho_j(s))\geq  d(j,s) + t^{\text{timetable}}_{\text{out}}(j, s) - 
t^{\text{timetable}}_{\text{out}}(j', \rho_j(s)) 
+ \tau_{(2)}(j, s, \rho_j(s)).
\end{equation}
Hence, taking $\Delta(j, s, j', \rho_j(s)) = 
t^{\text{timetable}}_{\text{out}}(j, s) - 
t^{\text{timetable}}_{\text{out}}(j', \rho_j(s))$, we get:
\begin{equation}\label{eq::tau2}
d(j', \rho_j(s)) \geq  d(j,s) + \Delta(j, s, j', \rho_j(s)) 
+ \tau_{(2)}(j, s, \rho_j(s)).
\end{equation}
Again, condition~\eqref{eq::tau2} needs to be tested for
$t_{\text{out}}(j',\rho_j(s)) \geq t_{\text{out}}(j,s)$; otherwise
trains must be investigated in the reversed order.

Further, similarly to Condition~\ref{cond::2trains1way}, the form of
$\tau_{(2)}(j, s, \rho_j(s))$ depends on the dispatching details
resulting from the formulation of the problem. Again, as all time
reserves are assumed to be realized at stations,
$\tau_{(2)}(j, s, \rho_j(s))$ is delay independent, which makes the
problem more tractable.
\end{condition}

As mentioned before, the particular form of the $\tau$s are problem
dependent; we propose the following approach to this.
Suppose that train $j$ departs from station
$s$ to subsequent station $\rho_j(s)$, passing the blocks
$m_{k}, m_{k+1}, \ldots , m_{l-1}, m_{l}$, where $s = m_{k}$ and
$m_{l} = \rho_j(s)$. The subsequent train proceeding in the same direction is 
allowed to leave at 
least after
\begin{equation}\label{eq::tau1_fact}
\tau_{(1)}\left(j, s \right) = \max_{i \in \{k+1, \ldots, l-1\}} \left(
t_{\text{in}}^{\text{timetable}}(j,m_{i+1}) -
t_{\text{in}}^{\text{timetable}}(j, m_i) \right).
\end{equation}
The subsequent train proceeding in the opposite direction is allowed to leave 
at least after
\begin{equation}\label{rem::tin}
\tau_{(2)}\left(j, s \right) = \sum_{i \in \{k+1, \ldots, l-1\}} \left(
t_{\text{in}}^{\text{timetable}}(j,m_{i+1}) -
t_{\text{in}}^{\text{timetable}}(j, m_i) \right) \equiv  
t_{\text{in}}^{\text{timetable}}(j,\rho_j(s)) -
t_{\text{out}}^{\text{timetable}}(j, s).
\end{equation}

Referring to the minimum and maximum delay conditions -- see~\eqref{eq::dlim} --
there are
pairs of trains for which either Condition~\ref{cond::2trains1way}, or
Condition~\ref{cond::2trains2way}, is always fulfilled. This
observation simplifies our QUBO representation of the problem. 

\begin{condition}\label{cond::rollingstock}\textbf{Rolling stock circulation 
condition at the terminal.}
If train $j$ with a given train set assigned terminates at a
station where the next train $j'$ of the same train set starts its
course (after turnover),
    i.e., $s_{j, end} = s_{1, j'}$, the following  
	condition arises:
	\begin{equation}
		t_{\text{out}}(j',s_{j', 1}) > t_{\text{in}}(j,s_{j, end}) + 
		\Delta(j,j'),
	\end{equation}
	where $\Delta(j,j')$ is the minimum turnover time. In 
	the delay representation, we have
	\begin{equation}
		d(j',1) + t^{\text{timetable}}_{\text{out}}(j', 1) > d(j,s_{j, end-1}) 
		+ 
		t^{\text{timetable}}_{\text{out}}(j, s_{j, end-1}) + \tau_{(2)}\left(j, 
		s_{j, 
		end-1}) \right) + \Delta(j,j').
	\end{equation}	
	Hence, taking $R(j,j') = t^{\text{timetable}}_{\text{out}}(j', 1) - 
	t^{\text{timetable}}_{\text{out}}(j, s_{j, end-1}) - 
	\tau_{(2)}\left(j, s_{j, 
		end-1}) \right) - \Delta(j,j')$, we get
	\begin{equation}\label{eq::circdelays}	
		d(j',1)  > d(j,s_{j, end-1}) - R(j,j').
	\end{equation}
  \end{condition}

\begin{condition}\label{cond::capacity} \textbf{The capacity condition.}
  Here we include the buffer approach of handling stations in our
  model. Suppose we have a station block $s$, capable of handling up
  to $b$ trains at a time. Let
  $\{j_1, j_2, \ldots, j_{b+1} \} \subset \mathcal{J}$ be any
  $b+1$-tuple of trains.  No time $t$ may exist for which all the
  conditions below are simultaneously fulfilled:
	\begin{eqnarray}
	t_{\text{in}}(j_1, s) &\leq t \leq t_{\text{out}}(j_1, s) \nonumber \\
	\ldots \nonumber \\
	t_{\text{in}}(j_{b+1}, s) &\leq t \leq t_{\text{out}}(j_{b+1}, s).
	\end{eqnarray}
	In the delay representation,
	\begin{eqnarray}
	d(j_1,\pi_{j_1}(s)) + t^{\text{timetable}}_{\text{out}}(j_1,\pi_{j_1}(s)) + 
	\tau_{(2)}\left(j_1, \pi_j(s)\right) &\leq t 
	\leq d(j_1,s)  + t^{\text{timetable}}_{\text{out}}(j_1,s) \nonumber \\
	\ldots \nonumber \\
	d(j_{b+1},\pi_{j_1}(s)) + 
	t^{\text{timetable}}_{\text{out}}(j_{b+1},\pi_{j_{b+1}}(s)) + 
	\tau_{(2)}\left(j_{b+1}, \pi_j(s)\right) &\leq t 
	\leq d(j_{b+1},s)  + t^{\text{timetable}}_{\text{out}}(j_{b+1},s).
	\label{eq::dlimit}
	\end{eqnarray}
\end{condition}

As a consequence of Condition~\ref{cond::capacity}, many new
constraints may arise. These may make the calculations more complex,
even exceeding the capacity of the current quantum computers. In our
particular problem instances, we will temporarily ignore this condition, but we will verify the solutions against it.

Finally, it is worth observing that
Conditions~\ref{cond::minpasstime} - \ref{cond::capacity} refer to
station blocks only; line blocks do not appear. As we have a
single-track line, there is no need to analyze line blocks in the
optimization algorithm: the decisions are made at the stations. The
leaving time from the ending (station) block does not have to be
analyzed either.

\section{Linear integer programming approach}\label{sec::linear_integer}

Before proceeding to the QUBO approach we describe a linear integer
programming formulation, too. This is in the line with the standard
treatment of railway dispatching problems; meanwhile, it is formulated
so that it is compares easily with the QUBO approach. It will 
therefore be used as a reference for comparisons.

Similarly to the model in~\cite{lange_approaches_2018}, we opt for
using precedence variables as it is very suitable for a single-track
railway model. We introduce the binary decision variables $y_{j,j',k}$
so that they have a value of $1$ if the train $j$ occupies the
particular part of the track (denoted by $k$) before train $j'$, and
are zero otherwise. 

Train delays will be represented with discrete decision variables
$d(j, s)$ that fulfil~\eqref{eq::dlim}. (The discretization is not necessary, but it is practical for the comparison with the QUBO
results, as the discretization is required there and our
particular problem instances were found to be tractable with a
standard solver.)

Note that the ordering of the train departures is uniquely described
by the precedence variables ($y$s), but for each configuration there
is still some freedom in determining the value of the delay variables
($d$s). For the solution to be valid, the values of the $y$s and $d$s
should be consistent; this will be ensured by the constraints.

The constraints are the following. The constraints in~\eqref{eq::toutinequal}, and~\eqref{eq::circdelays} are linear; hence,
they can directly be included in the model. The single block
occupation condition, see~\eqref{eq::tau1}, is expressed in terms
of the precedence and delay variables:
 \begin{equation}\label{eq::sbo_cond}
d(j', s) + M \cdot (1-y_{j,j',s}) \geq d(j,s) + \Delta(j,s,j' s) + 
\tau_{(1)}(j,s,\rho_j(s)),
\end{equation}
where $y_{j,j',s}$ determines the order of trains $j$ and $j'$ leaving
station $s$, and $M$ is an arbitrary large number. For two trains $j$
and $j'$ heading in opposite directions, the deadlock condition is
to be prescribed. For trains with a common path between subsequent
stations $k \rightarrow \{s, \rho_j(s)\}$, the requirement in~\eqref{eq::tau2} takes the following form:
\begin{equation}\label{eq::dl_cond}
d(j', s) + M \cdot (1-y_{j,j',k}) \geq d(j,s) + \Delta(j,s,j' \rho_j(s)) + 
\tau_{(2)}(j,s,\rho_j(s)),
\end{equation}
where $y_{j,j',k}$ determines which train enters the common path
first.

Finally, as to the objective function, the weighted sum of secondary
delays (or the total weighted tardiness in the scheduling terminology)
will be minimized, which is also inherently linear:
\begin{equation}\label{eq::penalty_linear}
\min \sum_j \frac{d(j, s_{j, \text{end}-1}) - d_U(j, s_{j, 
		\text{end}-1})}{d_\text{max}(j)} w_j,
\end{equation}
where $w_j$ is the weight reflecting the train's priority.

Although the defined linear model is suitable for the given railway
environment, our intention is to use a solver that inputs QUBOs. For
this purpose, an integer program is not a good choice; hence, we
construct an alternative quadratic model with binary decision
variables.

\section{QUBO formulation of our model}\label{sec::qubo}
We construct a QUBO model that can be solved either by quantum
annealers or by classical algorithms inspired by them. After
presenting a constrained $0$-$1$ representation, we employ a penalty
method to move the constraints to the effective objective function to
get an unconstrained problem. This is maybe the most challenging step,
not only in our present work, but also in logical programming using
QUBOs.

\subsection{0-1 program representation}\label{sec::01programming}

As a step toward a QUBO model, we formulate our problem entirely in
terms of binary decision variables. We achieve this by the
discretization of time, i.e., the discretization of the delay variables. Hence, we
need to set a delay resolution step. We opt for a resolution of one
minute as this is reasonable from train timetabling point of view (and
the generalization is straightforward).  Given such a
representation,~\eqref{eq::dlim} can be rewritten into the following
form:
\begin{equation}\label{eq::dlims_discrete}
d(j,s) \in A_{j,s} = \{d_U(j, s),d_U(j, s)+1,\ldots, d_U(j, s)+d_{\text{max}}(j)\},
\end{equation}
where $A_{j,s}$ is a discretized set of all possible delays of train $j$ at 
station $s$.

For the QUBO representation, we introduce the binary decision variables
\begin{equation}
x_{s,j,d} \in \{0,1\},\label{eq:qubovarsdef}
\end{equation}
which take the value of $1$ if train $j$ leaves station block
$s$ at delay $d$, and zero otherwise. These variables will also be
referred to as ``QUBO variables.'' Their vector is
$\mathbf{x} \in \{0,1\}^n$. Each variable is assigned a
logical quantum bit. Hence solving the problem requires $n$ of
these bits. The number $n$ depends on the size of the system
and is dependent on the number of trains and stations and the value of the maximum 
secondary delay.

We assume that each train 
leaves each station block once and only once:
\begin{equation}\label{eq::Q1}
\forall_{j} \forall_{s \in S_j} \sum_{d \in A_{j,s}} 
x_{s,j,d} = 1.
\end{equation}
\begin{remark}\label{rem::sstar}
Observe that Conditions~\ref{cond::2trains1way} and
\ref{cond::2trains2way} (the single block occupation condition and the deadlock 
condition) refer to the subsequent stations in train
$j$ path -- $s$ and $\rho_j(s)$. (Recall that $\rho_j(s_{j, 
\text{end}})$ does not exist in our model.) Time of entering of $\rho_j(s)$ is 
computed from $x_{s, j, d}$ and $\tau_{(1)}(j,s, 
\rho_j(s))$, but it does not refer to $x_{\rho_j(s), j, d}$.
Hence we do not need to investigate the leaving time from the 
last block of the train's path. We assume that the arrival time at this 
block can be computed from the leaving time from the penultimate block 
and the passing time. (Of course, our goal is to reduce the number of
QUBO variables in the analysis.) Here, delays at the end of the route are
investigated on leaving the penultimate station of the analyzed
route. 
\end{remark}

Let $S_{j,j'}$ be the sequence of blocks in the common route of trains $j$ and 
$j'$. If both these trains are traveling in the same direction, the order of 
blocks in 
$S_{j,j'}$ is 
straightforward. Alternatively, we need to regard the block sequence of train 
$j$ as the reversed sequence of blocks of train $j'$.
Therefore, we introduce $S_{j,j'}^* = S_{j,j'} \setminus \{s_{j, 
\text{end}}\}$ for
Conditions \ref{cond::2trains1way} and \ref{cond::2trains2way}. 
Condition~\ref{cond::2trains1way} states that two trains traveling in the 
same 
direction are not allowed to appear at the same block section.
In particular, from~\eqref{eq::tau1} it follows that
\begin{equation}\label{eq::Q3}
\forall_{(j,j') \in \mathcal{J}^0 (\mathcal{J}^1)}
\forall_{s \in S^*_{j,j'}}
\sum_{d \in A_{j,s}}\left(\sum_{d' \in B(d) \cap A_{j', s}} 
x_{j, s, d} x_{j', s, d'}\right) = 0,
\end{equation}
where $B(d) = \{d + \Delta(j, s, j', s), d + \Delta(j, s, j', s)+ 1,\ldots, d + 
\Delta(j, 
s, j', s) + \tau_{(1)}(j,s, \rho_j(s))-1 
\}$ is a set of delays that violates Condition~\ref{cond::2trains1way}.

Assume now that two trains $j$ and $j'$ are heading in opposite directions. 
From~\eqref{eq::tau2} it follows that
\begin{equation}\label{eq::Q31}
\forall_{j \in \mathcal{J}^0(\mathcal{J}^1),  j' \in 
\mathcal{J}^1(\mathcal{J}^0) }
\forall_{s \in S^*_{j,j'}}
\sum_{d \in A_{j,s}}\left(\sum_{d' \in C(d) \cap A_{j', \rho_j(s)}}
x_{j, s, d} 
x_{j', \rho_j(s), d'}\right) = 0
\end{equation}
where $C(d) = \{d(j,s) + \Delta(j, s, j', \rho_j(s)), d(j,s) + \Delta(j, s, j', 
\rho_j(s))  + 1, \ldots,  d(j,s) + \Delta(j, s, j', \rho_j(s)) + 
\tau_{(2)}(j,s, \rho_j(s))-1\}$.

We do not need to examine delays
  when leaving the ending station of the train's path; see Remark~\ref{rem::sstar}. For the minimum 
  passing 
  time -- Condition~\ref{cond::minpasstime} -- we introduce $S_j^{**} = S_j 
  \setminus 
  \{s_{j, end}, s_{j, 
  end-1}\}$. From~\eqref{eq::toutinequal} we have:
\begin{equation}\label{eq::Q2}
\forall_{j} \forall_{s \in S_j^{**}} 
\sum_{d \in A_{j,s}} 
\left(
\sum_{ d' \in D(d) \cap A_{j, \rho_j(s)}} x_{j, s, d} 
x_{j, \rho_j(s), d'} \right) = 0,
\end{equation}
where $D(d) = \{0, 1, \ldots,  d  - \alpha(j, s, \rho_j(s)) -1\}$.

Following the the rolling stock circulation (Condition~\ref{cond::rollingstock}) 
we have, from~\eqref{eq::circdelays},
\begin{equation}\label{eq::circ}
\forall_{j,j' \in \text{terminal pairs}}  
\sum_{d \in A_{j,s_{(j, end-1)}}} 
\sum_{ d' \in E(d) \cap A_{j', 1}} x_{j, s_{(j, end-1)}, d} 
\cdot x_{j', s_{(j,' 1)}, d'}  = 0,
\end{equation}
where $E(d) = \{0, 1, \ldots, d - R(j,j')\}$.

The objective of the algorithm is to schedule trains so that
secondary delays are minimized. The general objective function can be
written in the following form:
\begin{equation}
f(d, j, s) = \hat{f}\left(\hat{d}, j, s 
\right), 
\end{equation}
where $\hat{d} = \frac{d(j,s)- 
	d_U(j,s)}{d_{\text{max}}(j)}.$
  As discussed in Section~\ref{problemdescr}, primary delays
  ($d_U$) are unavoidable, so they are not relevant for the objective. Recall that upper bounds of the secondary delays $d_{\text{max}}(j)$ have been introduced as parameters, see~\eqref{eq::dlim}. Thus we require $\hat{f}(\hat{d},j,s)$ to
  obey the following conditions:
\begin{equation}
\hat{f}(\hat{d}, j, s) = \begin{cases} 0 &\text{ if } \hat{d} = 0, \\  
\max_{\hat{d} 
\in [0,1]} 
\hat{f}(\hat{d},j,s) &\text{ if } \hat{d} = 1, \\ \text{is 
non-decreasing in } \hat{d} &\text{ if } \hat{d} \in (0,1). \end{cases}
\end{equation} 
This non-decreasing property reflects that higher delays cannot contribute to a lower extent to the objective.
Finally, our objective function will be linear:
\begin{equation}\label{eq::penalty}
f(\mathbf{x}) = \sum_{j \in J} \sum_{s \in S_j^*} \sum_{d \in A_{j,s}}
f(d, j, s)  \cdot x_{j,s,d},
\end{equation}
where $f(d, j, s)$ are the weights.

Apart from the constraints discussed above, the penalty function can be
chosen deliberately, which adds some relevant flexibility to the model.
By selecting the appropriate $\hat{f}(\hat{d},j,s)$, various
dispatching policies can be represented. This ensures freedom of
choice in striving for the best suited dispatching solution. Let us
mention just a few of them:
\begin{enumerate}
\item For a quasi-minimization of the maximum secondary delays, one may 
opt for a
strongly increasing convex function in $\hat{d}$, such as an exponential or 
geometrical.
\item To minimize the number of delayed trains, one may opt for the step function 
$\hat{d}$.
\item To minimize the sum of delays, one may opt for a linear 
function in $\hat{d}$.
\item Subsequent trains can be assigned various weights for the delays on which their 
priorities depend.
\item A subset of stations can be selected as the only relevant stations from
  the point of view of delays. For practical reasons, we analyze delays
  on penultimate stations -- see Remark~\ref{rem::sstar}.
\end{enumerate}
For our particular dispatching problems, we select the policies set out
in Points 
$3-5$.
 
\subsection{A remark on the penalty coefficients}
\label{sec::quboimplem_remarl}

To get some hint of how to determine the coefficients of
the summands that warrant feasibility, let us first consider a direct
search solution of a QUBO of the form in~\eqref{eq:qubogen}.  This
amounts to evaluating the objective function with all possible values
of the decision variables. In our effective QUBO in~\eqref{eq::effqubo}, 
the total matrix $Q$ is a sum of the terms in~\eqref{eq:pairpen} and~\eqref{eq:sumpen} and the original
objective function of~\eqref{eq:linobj}. So we have a sum of
three QUBOs, and the objective function value is linear in
the matrix of QUBOs. Hence, the objective value will be the sum of the
original objective function value and the values of the
summands representing the constraints.

The feasibility terms have a negative minimum $-L$ because of the
omitted 0th order terms when using~\eqref{eq:sumpen} (instead 
of~\eqref{eq:sumpenq} if the solution is feasible).
For each element of the outer sum in~\eqref{eq:sumpen}, the value $p_{\text{sum}}$ contributes 
to $L$, hence $L = p_{\text{sum}} \cdot (\text{the number of linear constraints})$.
The value
\begin{equation}\label{eq::hard_constrain}
f''(\mathbf{x}) = \mathcal{P}_{\text{pair}} + \mathcal{P}_{\text{sum}} - L
 \end{equation}
 will be zero if the solution is feasible, and non-zero otherwise. We will
 call it the hard constraints' penalty.

 If there is solution in which the ``cost'' of violating
 some hard constraints is lower than the particular objective
 function value, the effective QUBO may yield a minimum that is
 unfeasible. A way to avoid this is to ensure that the lowest
 violation of any hard constraint has a larger contribution to $f'(\mathbf{x})$ than a violation of all soft constraints (encoded in the 
 objective $f(\mathbf{x})$) of a
 given feasible (not necessarily optimal) solution. Such a solution
 can be obtained by some fast heuristics.

 This suggests that one should assign high coefficients to the hard
 constraints. If one employs a direct search algorithm calculating the
 values of the objective very accurately, this approach can work out
 easily.  However, the numerical accuracy is always limited, and other
 inaccuracies of the minimum search can also appear. In the case of a
 quantum annealer, this is due to the noise of the system. What we get
 in reality is not the guaranteed to be absolute minimum but a set of
 samples: vectors for which the effective objective function is close
 to the minimum. If the coefficients are too high, the original
 objective function is just a small perturbation over the feasibility
 violations. Hence, while obtaining strictly feasible solutions, the
 actual minimum can be lost in the noise. Therefore finding the
 appropriate values of $p_{\text{sum}}$ and $p_{\text{pair}}$ amounts
 to finding the values that address both the criteria of both feasibility and
 optimality to a suitable extent.

\subsection{A simple example}\label{sec::simple_ex}

Let us demonstrate our approach in a simple example. Consider two
trains $j \in \{1,2\}$, two stations $s \in \{1,2\}$, and a single
track between them.  The passing time value (scheduled and minimum)
between the stations is $1$ (minute) for both trains.  Train $j=1$
is ready to depart from station $s=1$ (heading to $s=2$) at the same
time as train $j=2$ is ready to depart from station $s=2$ (heading
to $s=1$). Under these circumstances, a conflict appears on a single
track between the stations.

Let the initial delay of both trains be 
$d=d_U=1$.
As one 
of the trains needs 
to wait a minute to 
meet and pass the other one, the maximum acceptable secondary delay is  
$d_{\text{max}} 
= 1$; see~\eqref{eq::dlims_discrete}. 
Taking the QUBO representation as in~\eqref{eq:qubovarsdef} (i.e., 
$x_{s,j,d}$),
we have the 
following $4$ quantum bits: $x_{1,1,1}, x_{1,1,2}$ (train $1$ can leave station 
$1$ at delay $1$ or $2$), $x_{2,2,1}$, and $x_{2,2,2}$ (train $2$ can leave 
station $2$ at delay $1$ or $2$). The linear constraints express that each train 
departs from 
each 
station once and only once, so ~\eqref{eq::Q1} takes the form
\begin{equation}
x_{1,1,1} + x_{1,1,2} = 1 \text{ and } x_{2,2,1} + x_{2,2,2} = 1.
\end{equation}
Referring to ~\eqref{eq:sumpen}, the optimization subproblem is as follows:
\begin{equation}\label{eq::simple_psum}
\mathcal{P}_{\text{sum}}= - p_{\text{sum}} \left(x_{1,1,1}^2 + x_{1,1,2}^2 - 
 x_{1,1,1} x_{1,1,2} -  x_{1,1,2} x_{1,1,1} + x_{2,2,1}^2 + x_{2,2,2}^2 - 
 x_{2,2,1} x_{2,2,2} -  x_{2,2,2} x_{2,2,1}\right),
\end{equation}
with the optimal value equal to $ -L = - 2 p_{\text{sum}}$. 

The quadratic constraint 
is
that the
two trains are not allowed to depart from the stations at the same time, i.e.,
$x_{1,1,1} x_{2,2,1} 
= 0$ and $x_{1,1,2} x_{2,2,2} = 0$. Using 
~\eqref{eq:pairpen}, the 
optimization subproblem takes the following form:
\begin{equation}\label{eq::simple_pair}
\mathcal{P}_{\text{pair}}=p_{\text{pair}} \left(x_{1,1,1} x_{2,2,1} +  
x_{2,2,1}x_{1,1,1} + x_{1,1,2} x_{2,2,2} + x_{2,2,2} x_{1,1,2} \right),
\end{equation}
with the optimal value equal to $0$. Note that since we have only two 
stations in 
this simple example, the minimum passing time condition does not appear 
($S^{**} = \emptyset$). 

Finally, a possible objective function is
\begin{equation}\label{eq::simple_pen}
f(\mathbf{x}) = x_{1,1,2} w_1 + x_{2,2,2} w_2 = x_{1,1,2}^2 w_1 + x_{2,2,2}^2 
w_2,
\end{equation}
where the secondary delay of train $1$ is penalized by $w_1$ and the secondary 
delay of train $2$ is penalized by $w_2$. 

Let the vector of decision variables be denoted by $\mathbf{x} = [x_{1,1,1}, 
x_{1,1,2}, 
x_{2,2,1}, 
x_{2,2,2}]^T$. The QUBO problem can thus be written in the form of 
~\eqref{eq:qubogen}, so
\begin{equation}\label{eq::simple_Q}
Q = \begin{bmatrix}
-p_{\text{sum}} & p_{\text{sum}} & p_{\text{pair}} & 0  \\
p_{\text{sum}} & -p_{\text{sum}}+w_1 & 0 & p_{\text{pair}}  \\
p_{\text{pair}} & 0 & -p_{\text{sum}} & p_{\text{sum}} \\
0 & p_{\text{pair}} & p_{\text{sum}} & -p_{\text{sum}}+w_2
\end{bmatrix}.
\end{equation}

As the solution is parameter dependent, we can use various trains 
prioritization policies. For the sake of demonstration, assume that train 
$j=2$ is assigned 
a higher 
priority than train $j=1$. This implies the assignment of different penalty weights. We set $w_1 = 0.5$ and $w_2 = 1$. 

As discussed in Section~\ref{sec::quboimplem}, to ensure that the calculated 
solution 
is feasible, we 
require that the following conditions are met:
$p_{\text{sum}} > \max\{w_1, w_2\}$ and $p_{\text{pair}} > \max\{w_1, w_2\}$. 
We propose $p_{\text{pair}} = 
p_{\text{sum}}  = 1.75$, so matrix $Q$ to takes following form:
\begin{equation}
Q = \begin{bmatrix}
-1.75 & 1.75 & 1.75 & 0  \\
1.75 & -1.25 & 0 & 1.75  \\
1.75 & 0 & -1.75 & 1.75 \\
0 & 1.75 & 1.75 & -0.75
\end{bmatrix}.
\end{equation}
The optimal solution is $\mathbf{x} = [0,1,1,0]^T$ (train $2$ goes first) with 
$f'(\mathbf{x}) = - 3$. Another feasible solution (not optimal) is 
$\mathbf{x} = [1,0,0,1]^T$ (train $1$ goes first) with $f'(\mathbf{x}) = - 2.5$. 
The other solutions are not feasible: for example, $\mathbf{x} = [1,0,1,0]^T$ is not 
feasible as the two trains are expected to
depart from the stations at the same time, with $f'(\mathbf{x}) = 0$. Observe 
that  
the classical heuristics (such as FCFS and FLFS) do not make a difference 
between the two feasible solutions, as both trains enter the conflict 
segment at the same time and need the same time to pass it. Also, both
solutions have the same value of the secondary delay.

Having formulated our model as a QUBO problem, it is ready to be
solved on a physical quantum annealer or by a suitable algorithm.

\section{Classical algorithms for solving Ising problems}
\label{sec::classalg}

An additional benefit of formulating problems in terms of Ising-type
models is that the existing methods developed in statistical and
solid-state physics for finding ground states of physical systems can
also be used to solve a QUBO on classical hardware. Notably,
variational methods based on finitely correlated states (such as
matrix product states for 1D systems or projected entangled pair
states suitable for 2D graphs) have had a very extensive development in the
past few decades. A quantum information theoretic insight into
density matrix renormalization group methods
(DMRG~\cite{RevModPhys.77.259}) -- being the most powerful numerical
techniques in solid-state physics at that time -- helped in proving
the correctness of DMRG. These methods also led to a more general view of the
problem~\cite{PhysRevB.73.094423}, resulting in many algorithms that
have potential applications in various problems, in particular solving
QUBOs by finding the ground state of a quantum spin glass. We have
used the algorithms presented in~\cite{rams2018heuristic} to solve the models derived in the present
manuscript.

Both quantum computers and the mentioned classical algorithms may not
provide the energy minimum and the corresponding ground state (as it
is not trivial to reach it~\cite{PhysRevA.100.042326}) but another
eigenstate of the problem with an eigenvalue (i.e., a value of the
objective function) close to the minimum. The corresponding states are
referred to as ``excited states.'' Another important point in
interpreting the results of such a solver is the degeneracy of the
solution: the possibility of having multiple equivalent optima.

In analyzing these optima, it is helpful that for up to 50 variables, one
can calculate the exact ground states and the excited states closest
to them using a brute-force search on the spin configurations with
GPU-based high-performance computers. In the present work, we also use
such algorithms, in particular those introduced in
\cite{jaowiecki2019bruteforcing} for benchmarking and evaluating our
results for smaller examples. This way we can compare the exact
spectrum with the results obtained from the D-Wave quantum hardware
and the variational algorithms.

\section{Further examples of solutions}\label{sec::appendix}

In this Section we present all the solutions of the dispatching
problems on line No. $191$ obtained by our algorithms. Both the
CPLEX and tensor network approaches (which are based on QUBO) allow for
rather arbitrary decisions on train prioritization. These approaches
focus on the train delay propagation on subsequent trains, as
illustrated by the comparison of all the solutions of case $1$.
Provided Ks$2$ is delayed, an additional delay of Ks$3$ would happen
(which we call a ``cascade effect''). Furthermore, the tensor network
output in Fig.~\ref{fig::large-sim-tn} demonstrates the degeneracy of
the ground state and the solutions in the low excited state, which,
however, do not have a relevant impact on the dispatching situation. (In our cases all CPLEX solutions ar the same as these of the linear solver.)

\begin{figure}
	\subfigure[Case $1$ -- ``cascade effect''; the delay of Ks$2$ causes 
	a further 
	delay of Ks$3$.\label{fig::c1_FCFS}]{\includegraphics[scale = 
		0.6, width=0.45\columnwidth]{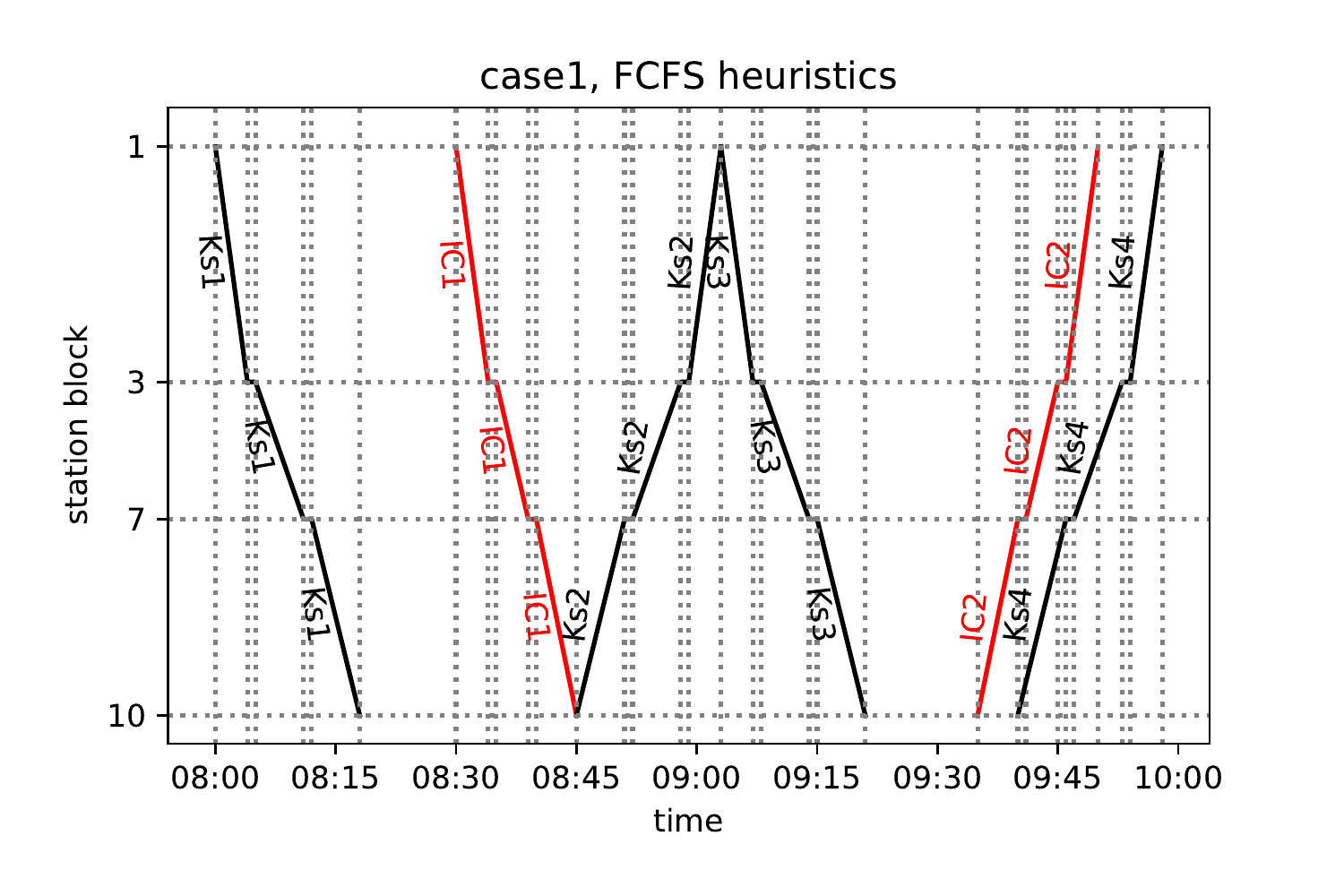}}
	\subfigure[Case $2$ -- optimal solution reached rather ``at 
	random'': probably it is reached because the peoblem is relativelly simple.\label{fig::c2_FCFS}]{\includegraphics[scale = 
		0.6, width=0.45\columnwidth]{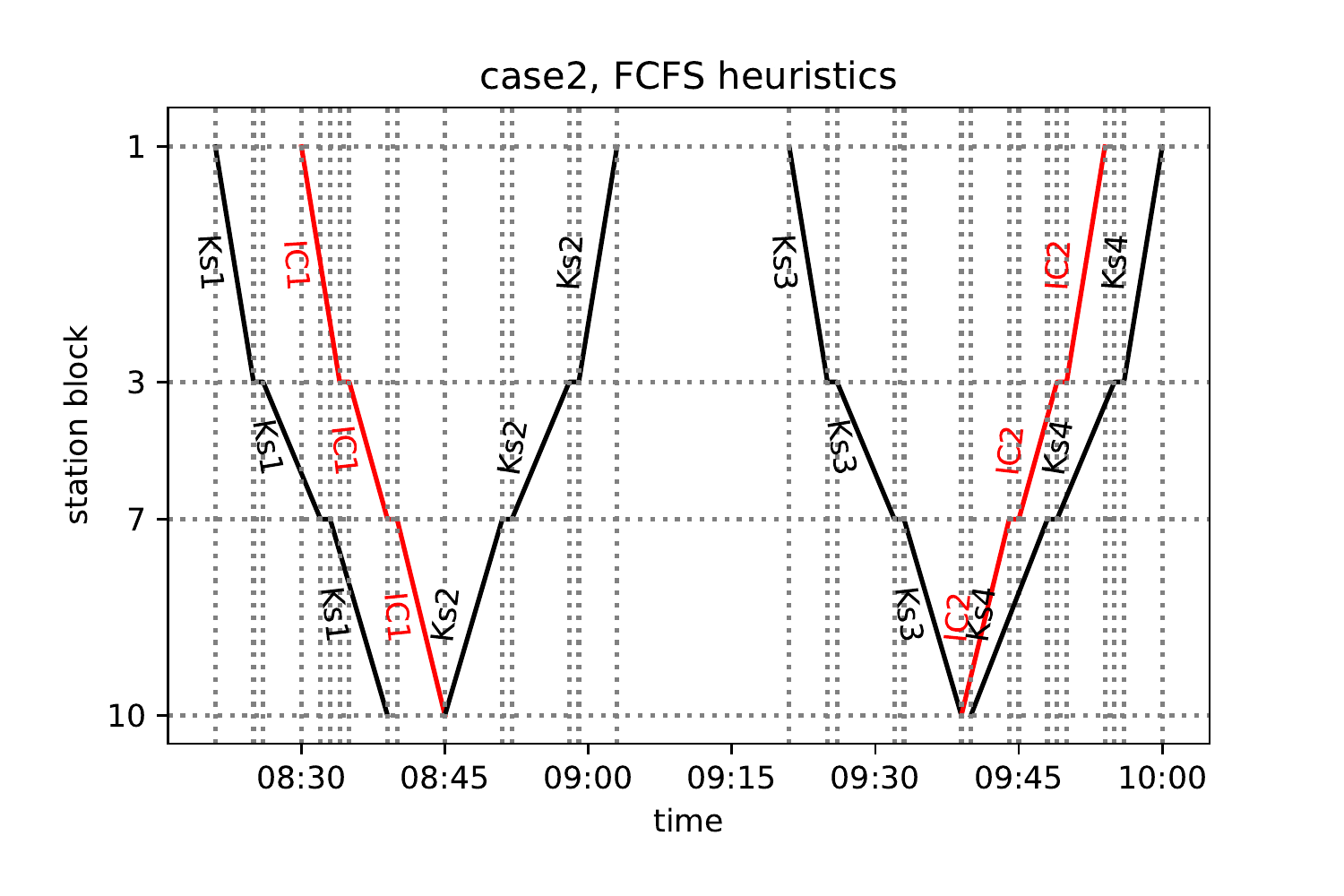}} \\
	\subfigure[Case $3$ -- a problematic solution with undesirably long waiting times of certain trains; observe the stopover of 
	Ks$2$.\label{fig::c3_FCFS}]{\includegraphics[scale = 
		0.6, width=0.45\columnwidth]{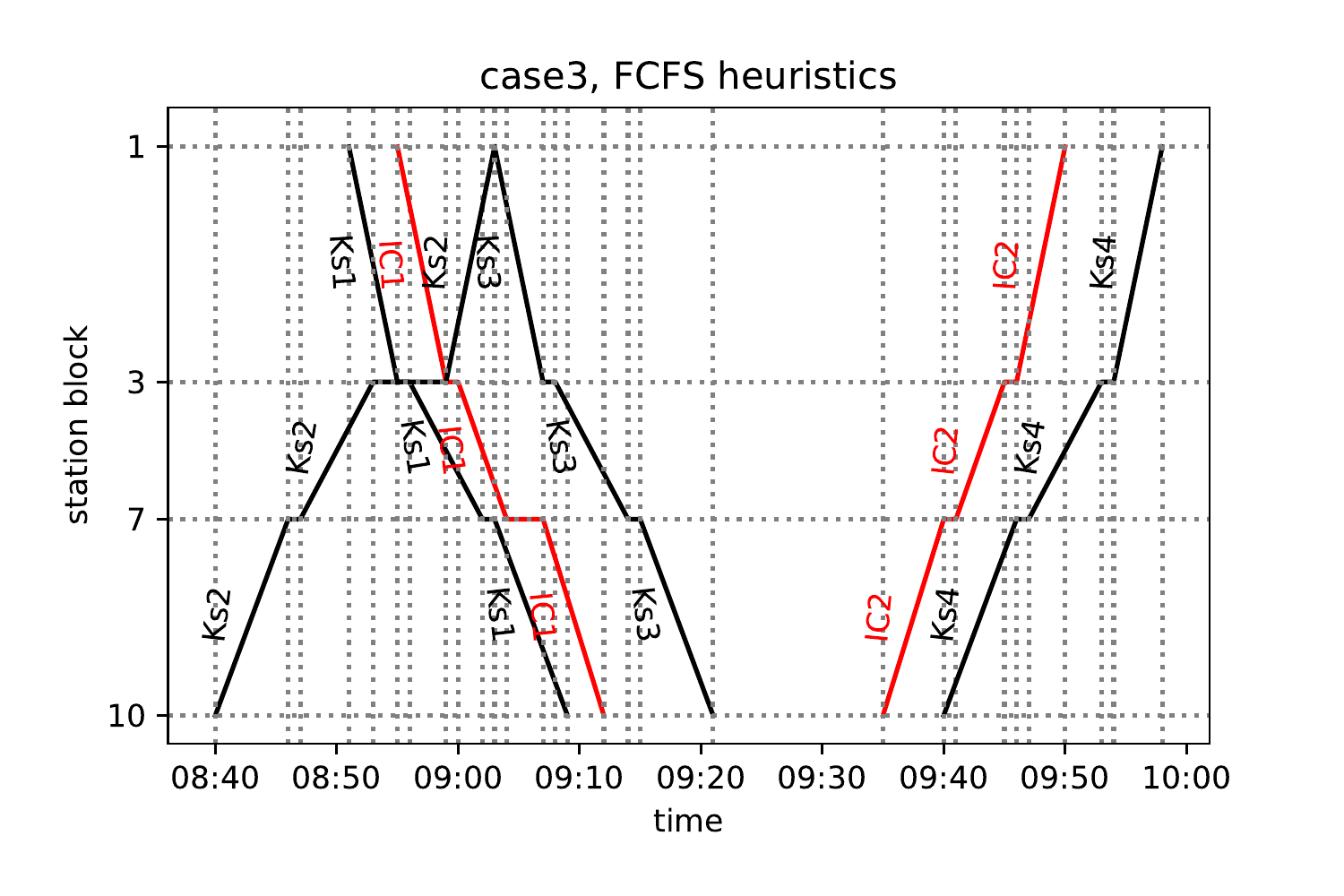}}
	\subfigure[Case $4$ -- optimal solution according to all 
	methods.\label{fig::c4_FCFS}]{\includegraphics[scale = 
		0.6]{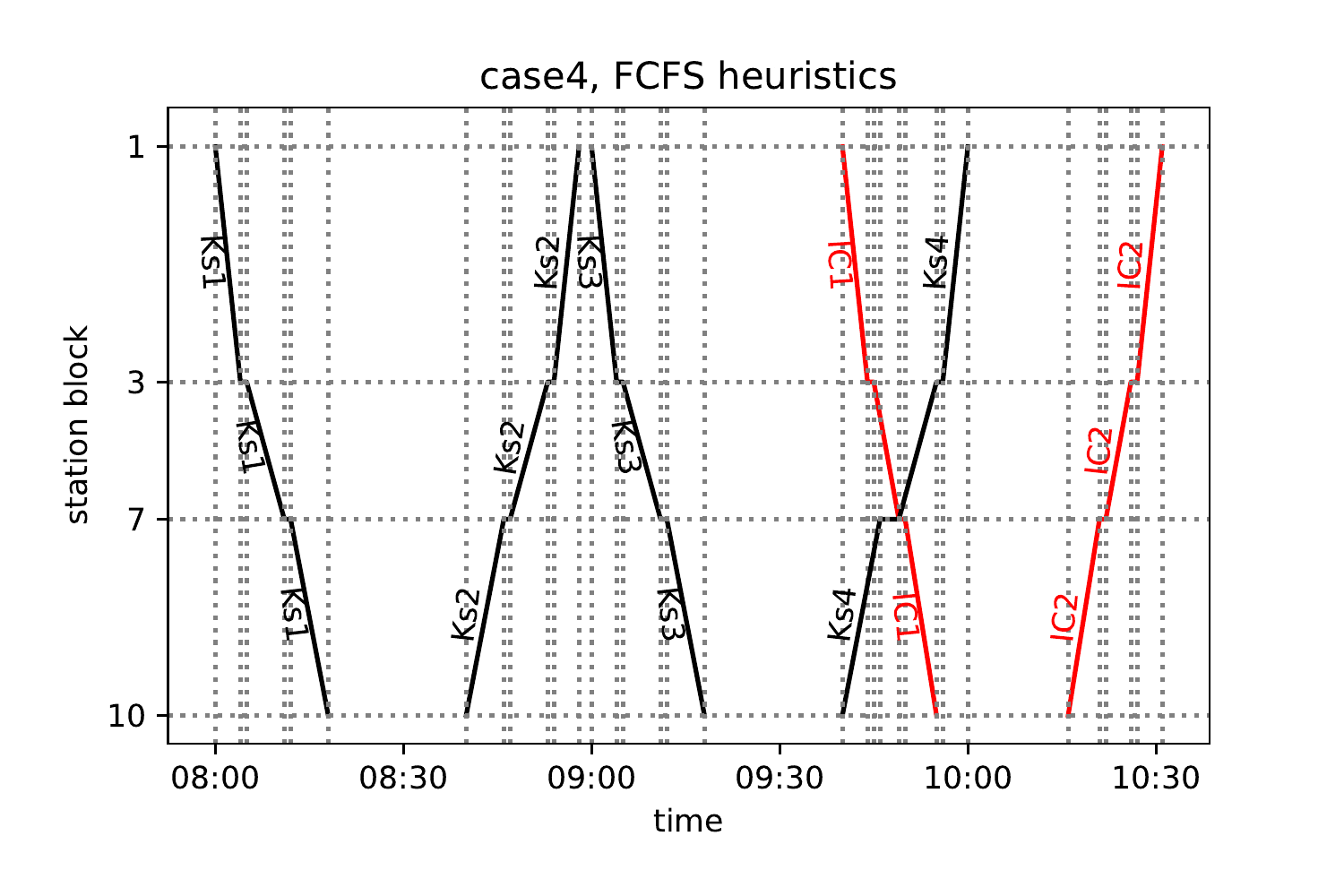}}
	\caption{The FCFS solutions, some with a trouble-causing stopover of 
		a particular 
		train.}\label{fig::large-FCFS}
\end{figure}

\begin{figure} 
	\subfigure[Case $1$ -- optimal solution is reached ``at random,'' as is its 
	duplicate in Fig.~\ref{fig::c2_FLFS}, which is an undesired solution.
	\label{fig::c1_FLFS}]{\includegraphics[scale = 
		0.6, width=0.45\columnwidth]{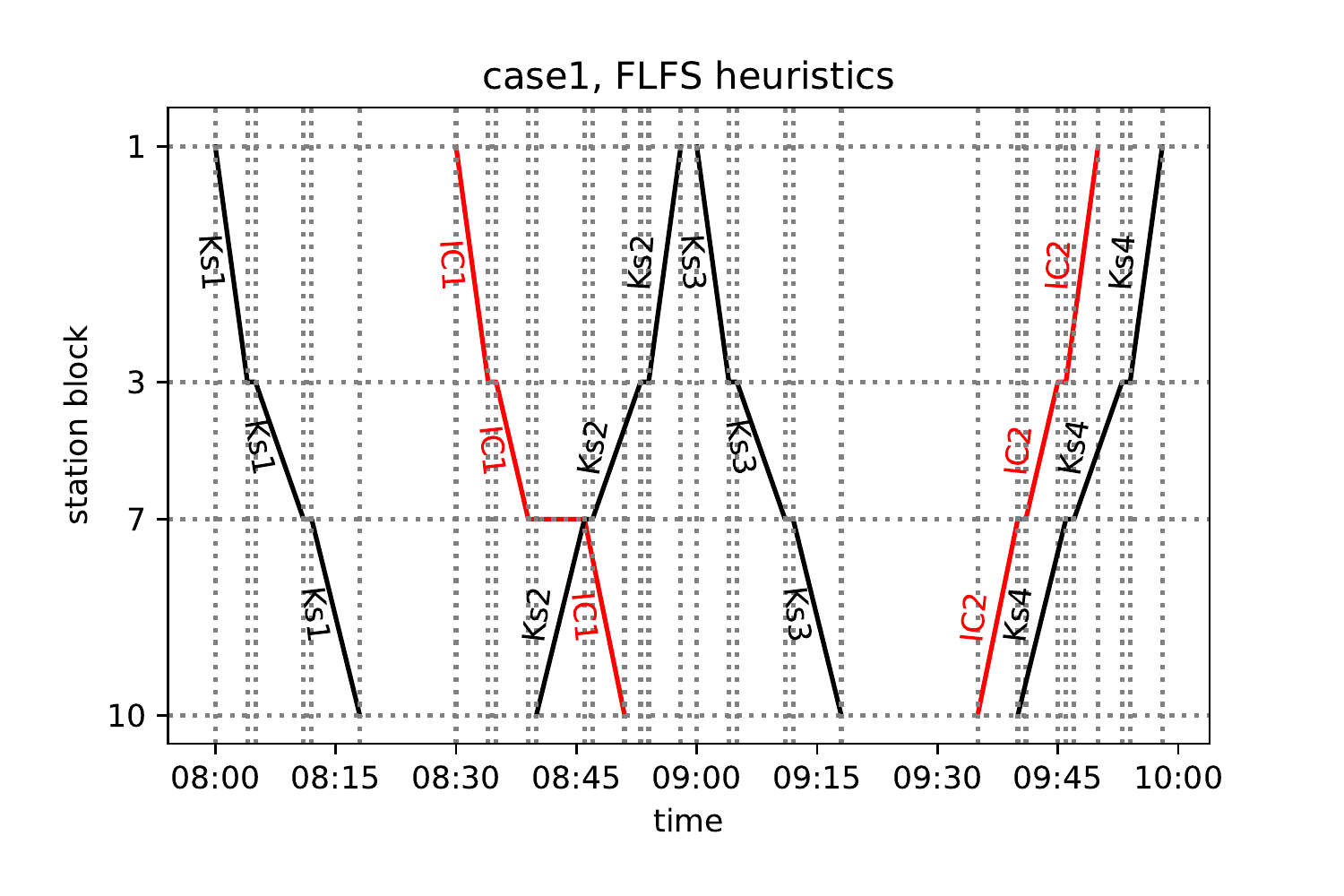}}
	\subfigure[Case $2$ -- duplicate of the solution in 
	Fig.~\ref{fig::c1_FLFS} 
	causing an stopover of Ks$3$. \label{fig::c2_FLFS}]{\includegraphics[scale 
	= 
		0.6, width=0.45\columnwidth]{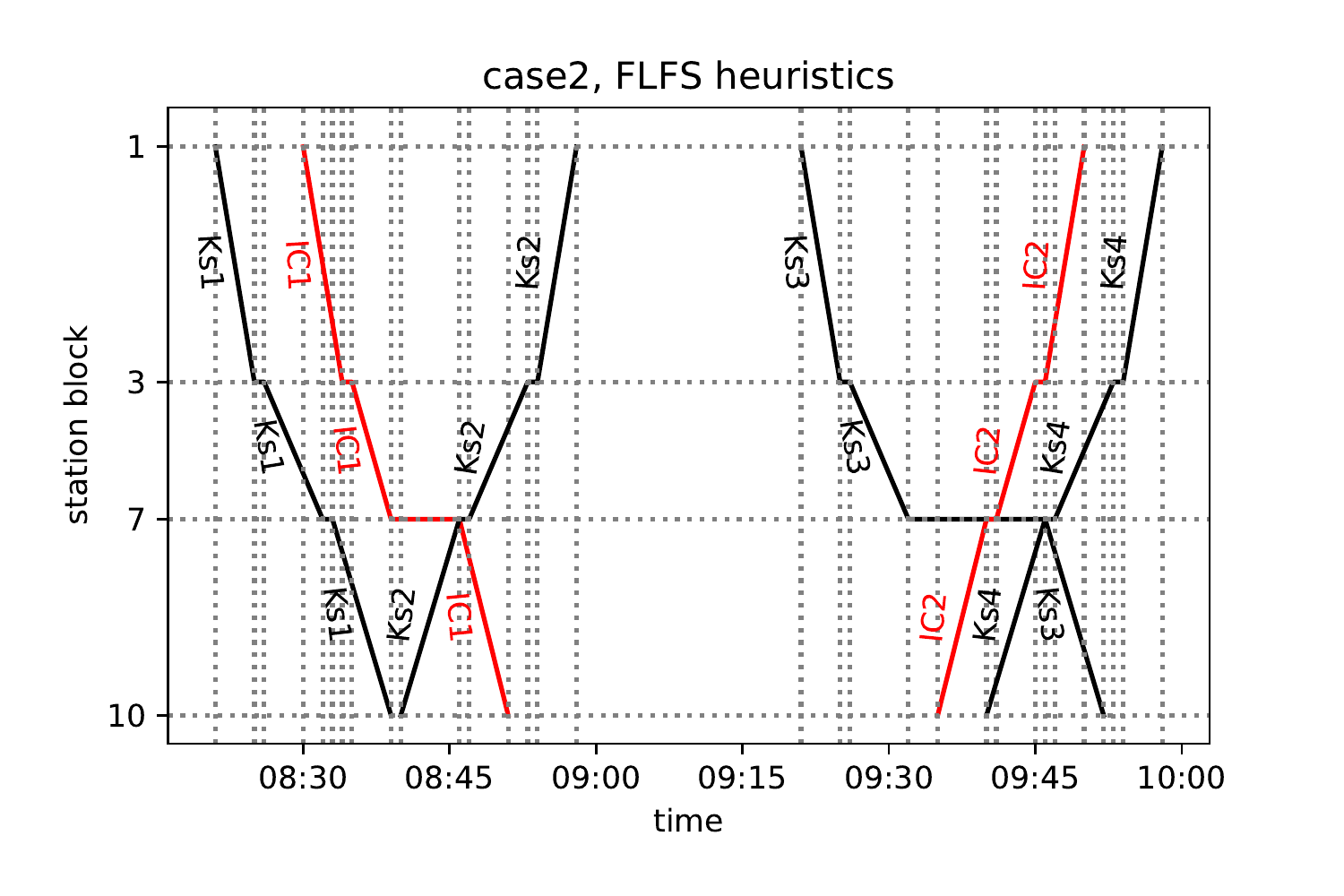}} \\
	\subfigure[Case $3$ -- no unacceptable stopovers.\label{fig::c3_FLFS}]{\includegraphics[scale = 
		0.6, width=0.45\columnwidth]{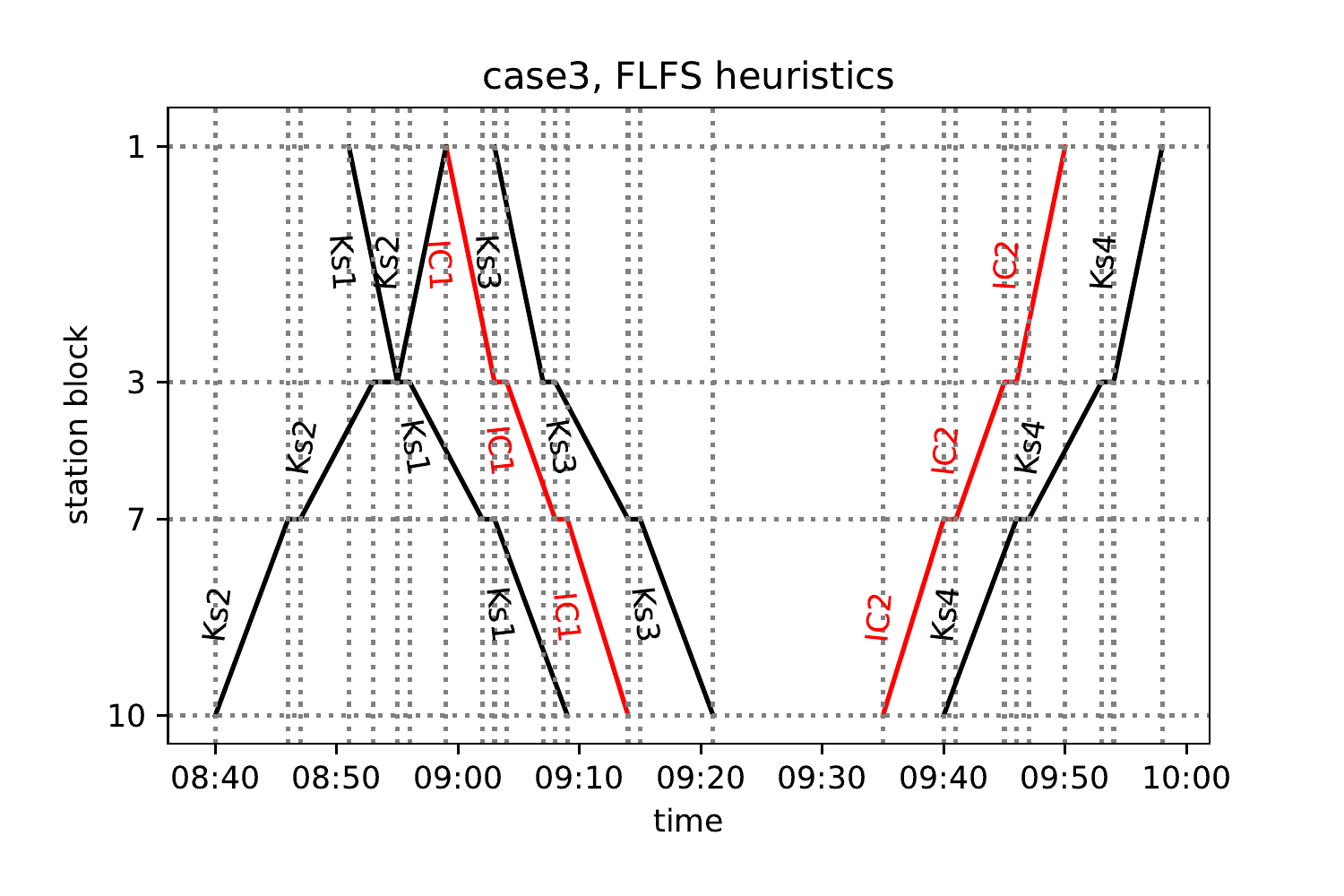}}
	\subfigure[case $4$ -- optimal solution according to all 
	methods.\label{fig::c4_FLFS}]{\includegraphics[scale = 
		0.6, width=0.45\columnwidth]{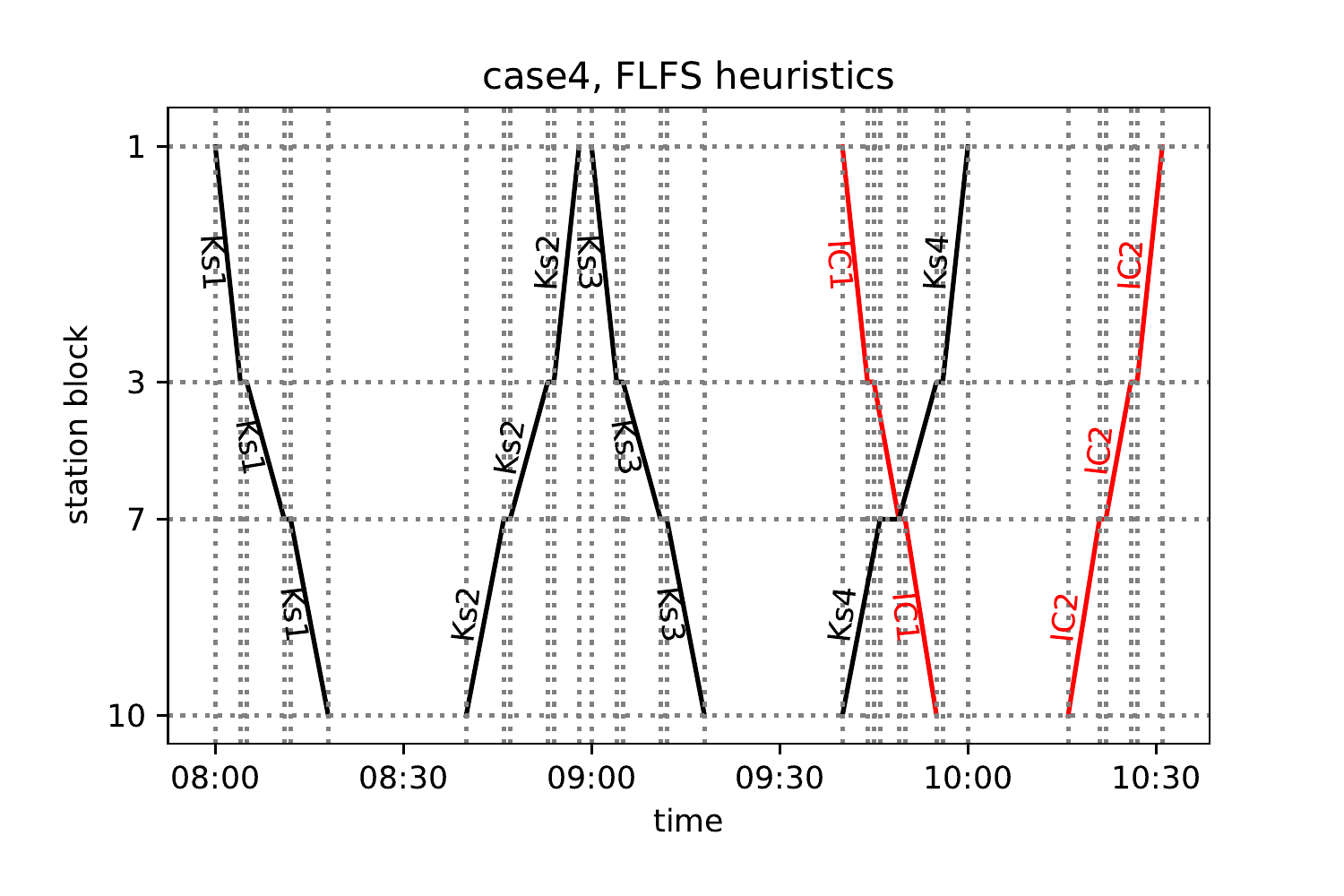}}
	\caption{The FLFS solutions, some with a trouble-causing stopover of 
	a particular 
		train.}\label{fig::large-FLFS}
\end{figure}

Note that the simple heuristics (FCFS, FLFS) sometimes return
trouble-causing solutions. This situation suggests a solution in which one train
needs to have a time-consuming stopover on a particular station; see
Figs.~\ref{fig::c3_FCFS},\ref{fig::c2_FLFS}. (Such problems sometimes
appear in real-life train dispatching too.)  Finally, if the
problem is easily solvable, as in case $4$, all the methods analyzed in
the paper give the same solution. This serves as a quality test of our
method.

\begin{figure}
	\subfigure[Case $1$ -- single conflict, observe that the additinal delay 
	of Ks$2$ will propagate to the delay of Ks$3$.
	\label{fig::c1_conflict}]{\includegraphics[scale = 
		0.6, width=0.45\columnwidth]{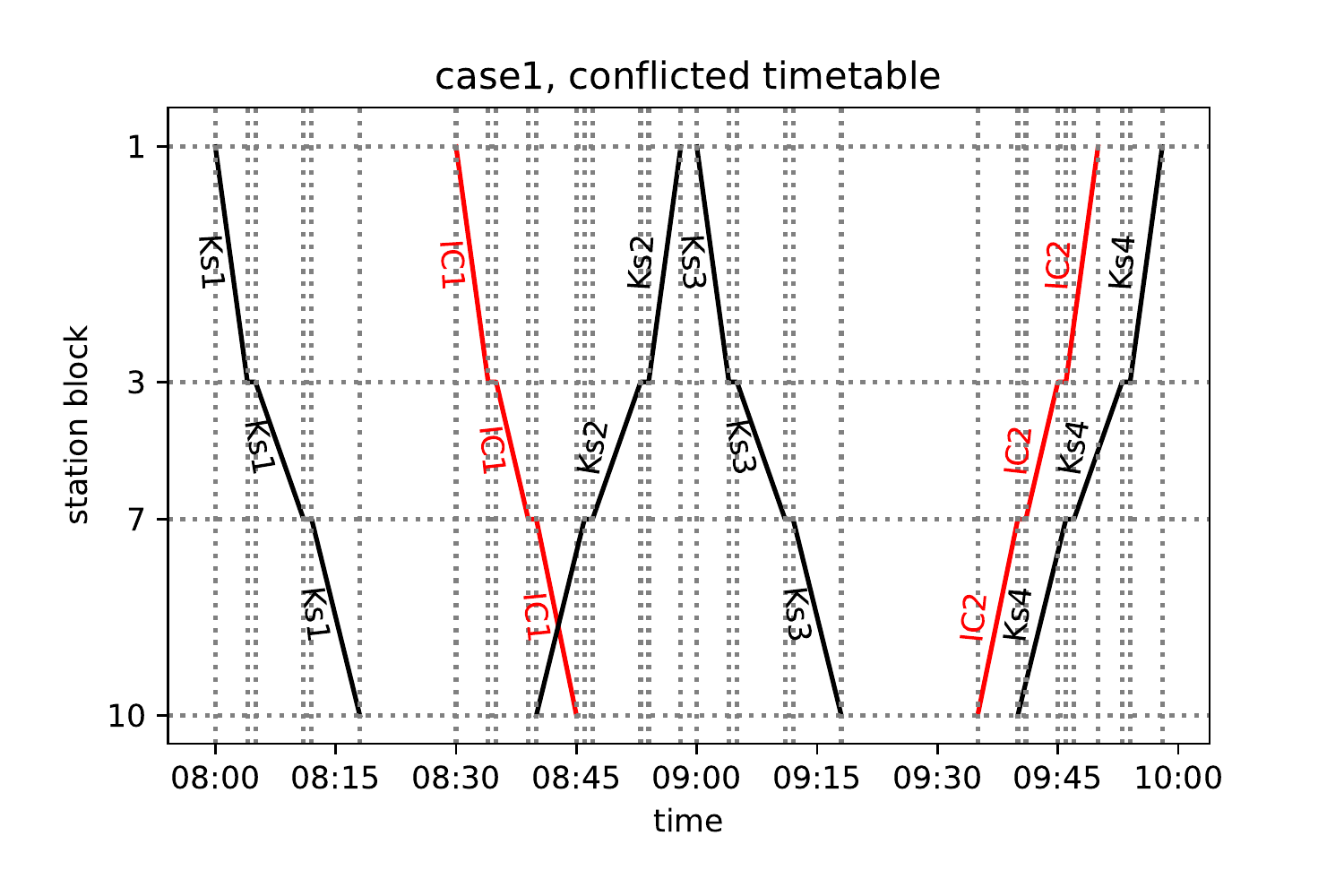}}
	\subfigure[Case $2$ -- two conflicts, simillar to 
	Fig.~\ref{fig::c1_conflict}, but with no impact of Ks$2$ on 
	Ks$3$.\label{fig::c2_conflict}]{\includegraphics[scale = 
		0.6, width=0.45\columnwidth]{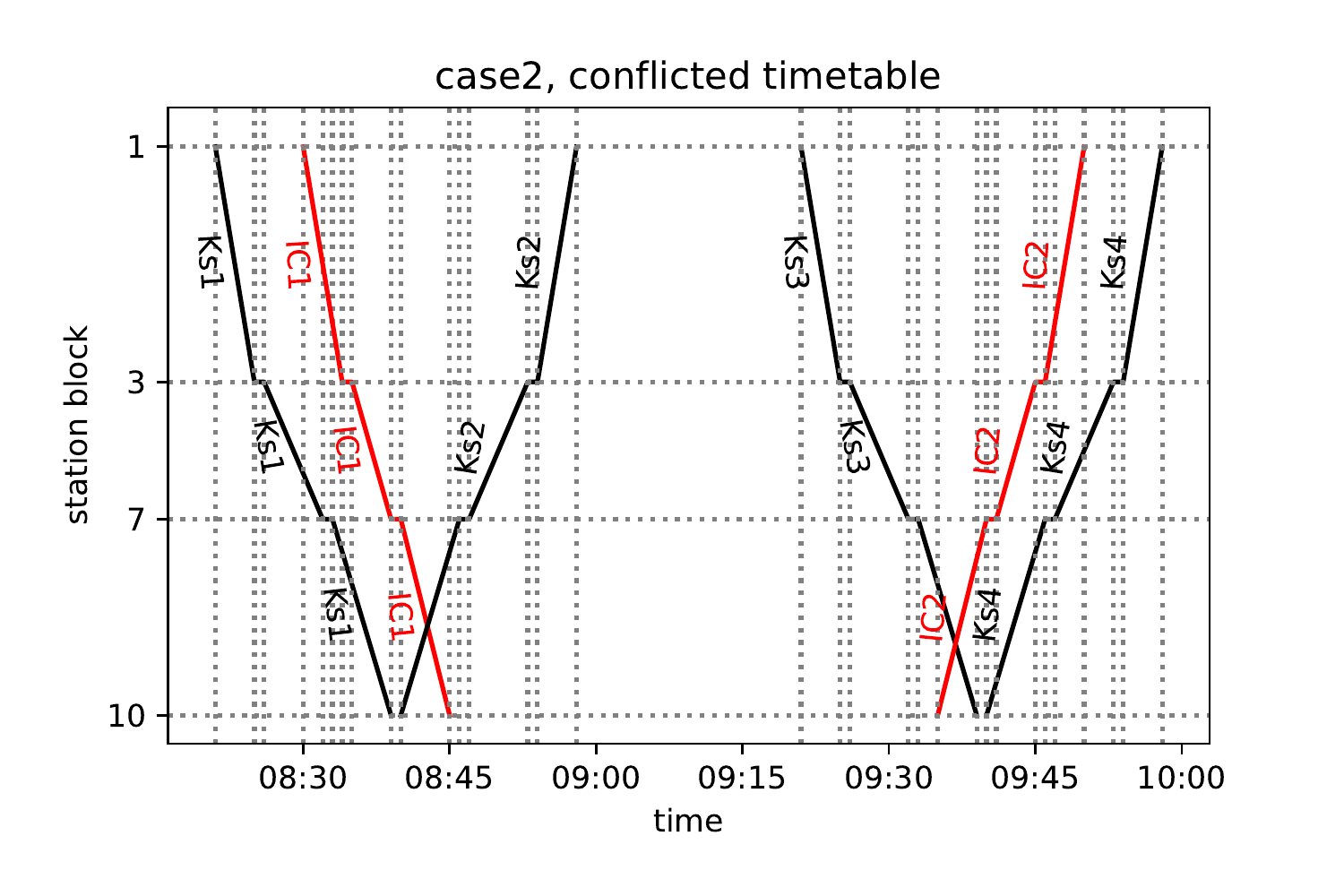}} \\
	\subfigure[Case $3$ -- multiple 
	conflicts.\label{fig::c3_conflict}]{\includegraphics[scale = 
		0.6, width=0.45\columnwidth]{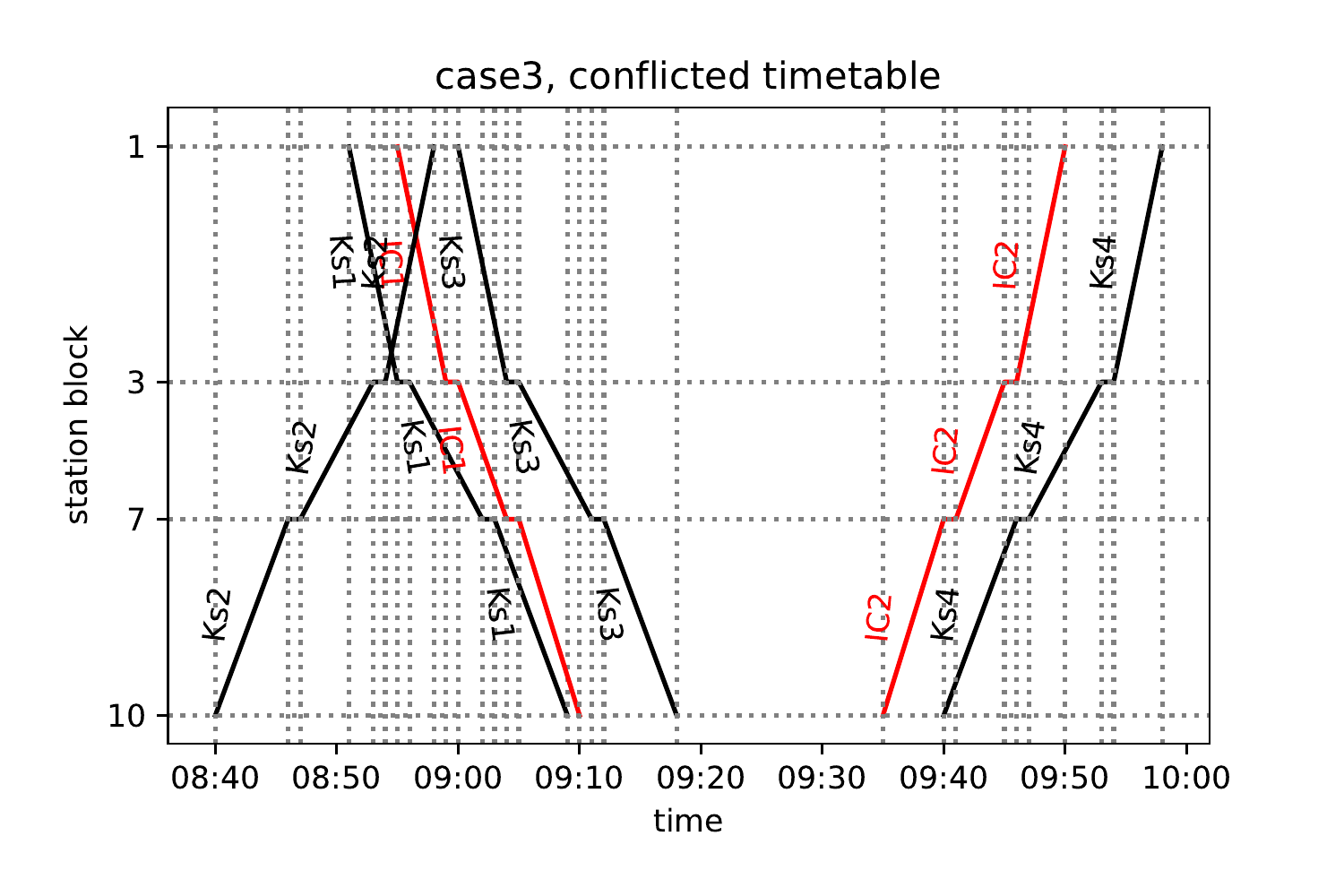}}
	\subfigure[Case $4$ -- conflict that is straightforward to  
	resolve.\label{fig::c4_conflict}]{\includegraphics[scale = 
		0.6, width=0.45\columnwidth]{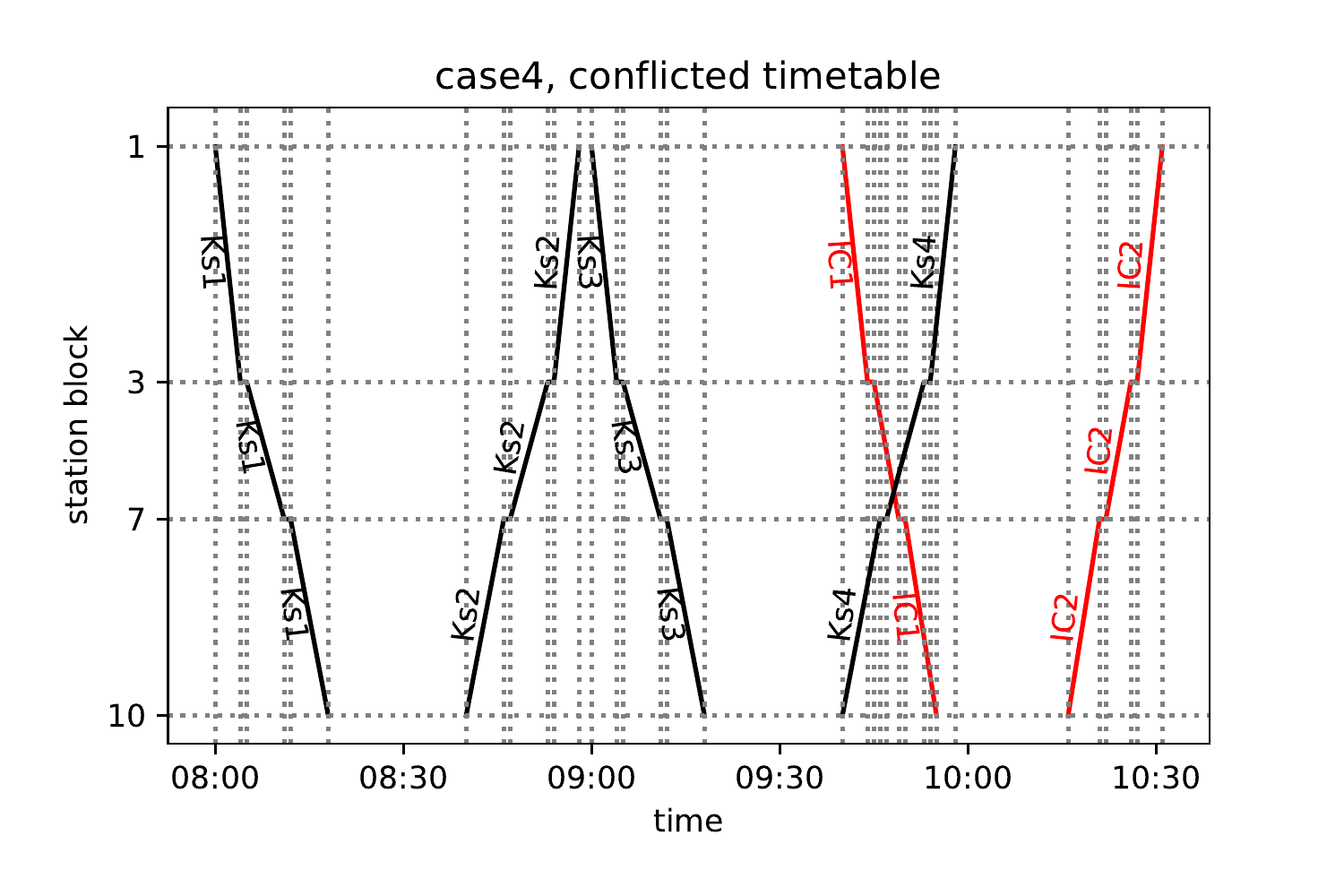}}
	\caption{The conflicted timetables, various types of 
	conflicts.}\label{fig::large-conf}
\end{figure}

\begin{figure}
	\subfigure[Case $1$  -- ``cascade effect,'' the delay of Ks$2$ causes 
	further a
	delay of Ks$3$.\label{fig::c1_AMCC}]{\includegraphics[scale = 
		0.6, width=0.45\columnwidth]{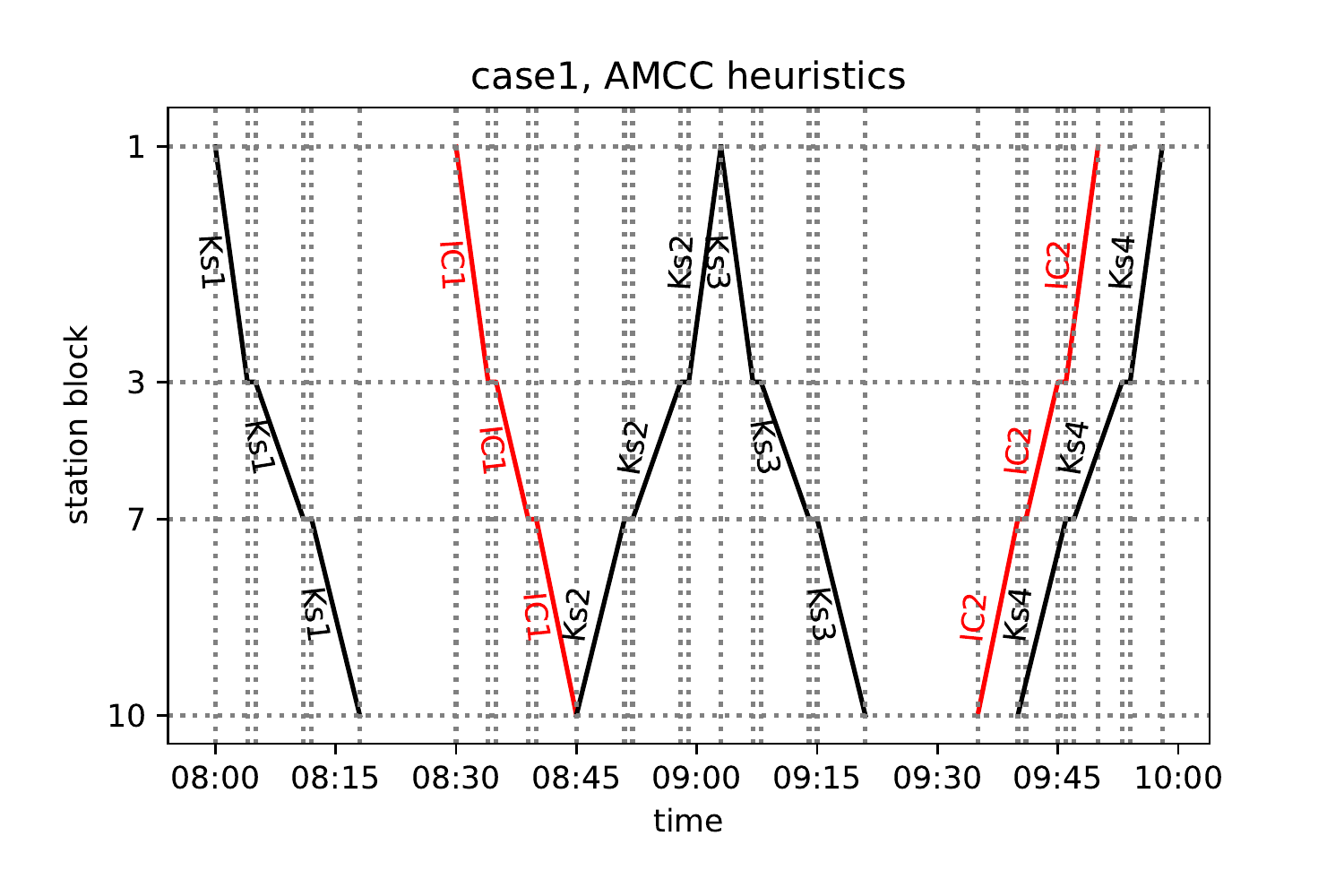}}
	\subfigure[Case $2$ -- no unacceptable
	stopovers.\label{fig::c2_AMCC}]{\includegraphics[scale = 
		0.6, width=0.45\columnwidth]{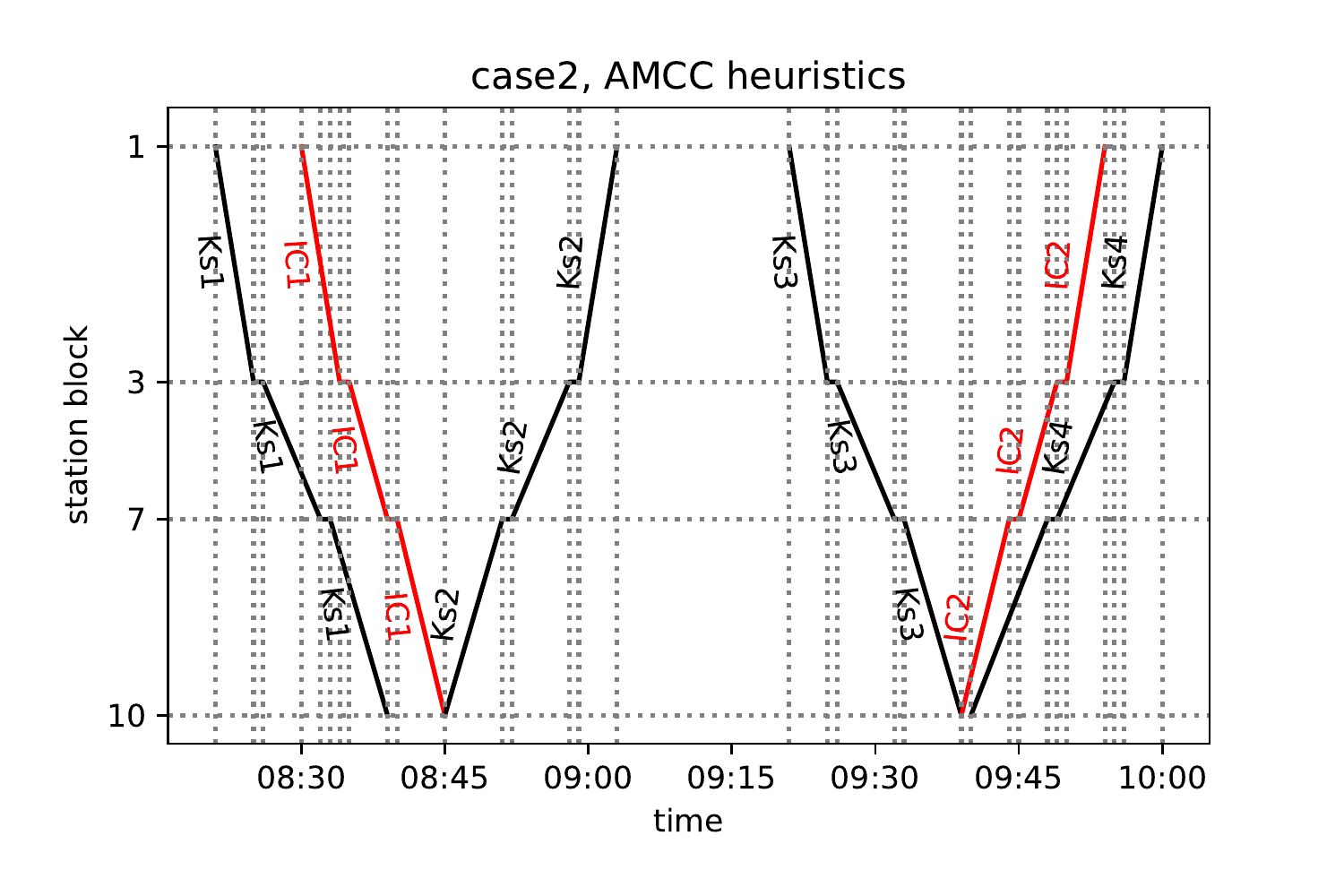}} \\
	\subfigure[Case $3$  -- no unacceptable
	stopovers\label{fig::c3_AMCC}]{\includegraphics[scale = 
		0.6, width=0.45\columnwidth]{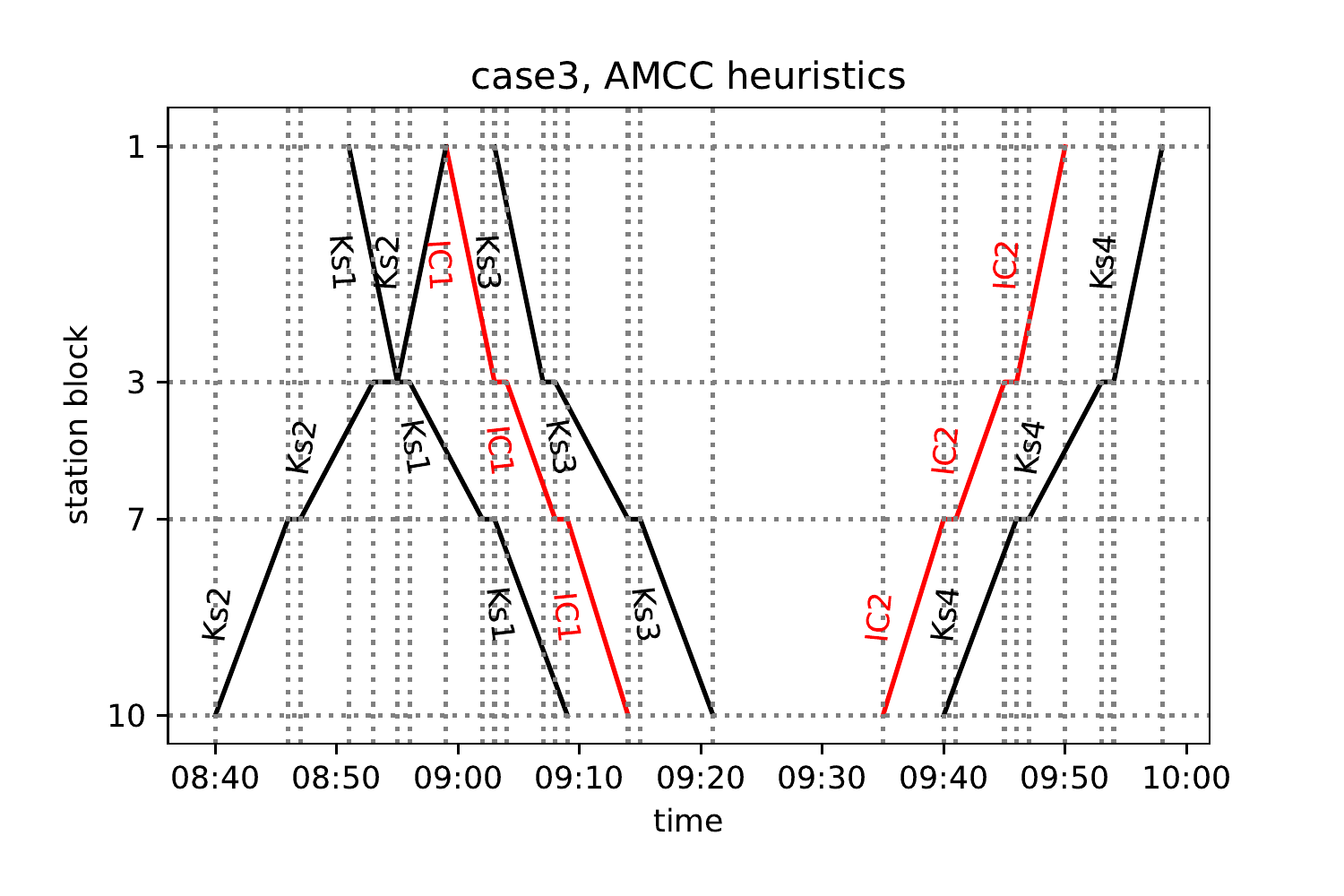}}
	\subfigure[Case $4$ -- optimal solution according to all 
	methods.\label{fig::c4_AMCC}]{\includegraphics[scale = 
		0.6, width=0.45\columnwidth]{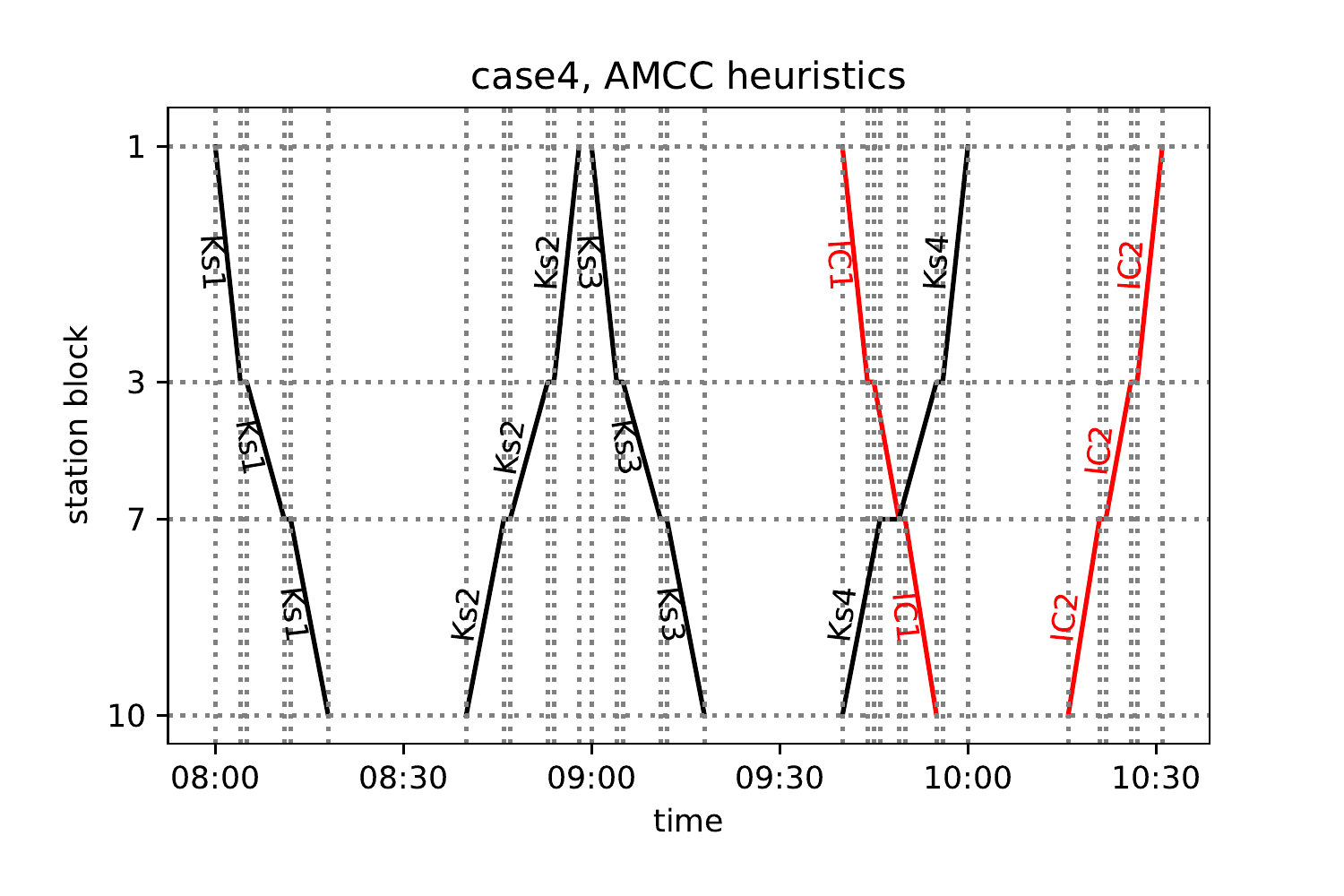}}
	\caption{The AMCC solutions. The minimization of the maximal 
	secondary delays from AMCC excludes unacceptably long stopovers 
	such as those in Figs.~\ref{fig::c3_FCFS} and \ref{fig::c2_FLFS}. However, these solutions 
	do not exclude the 
	propagation of smaller delays among several trains (``cascade effect''); see 
	Fig.~\ref{fig::c1_AMCC}.
	}\label{fig::large-AMCC}
\end{figure}

\begin{figure}
	\subfigure[Case $1$ -- no ``cascade effect'' (Ks$2$ does not delay Ks$3$):
	a consequence of the pioritization of 
	Ks$2$.
	\label{fig::c1_cplex}]{\includegraphics[scale = 
		0.55]{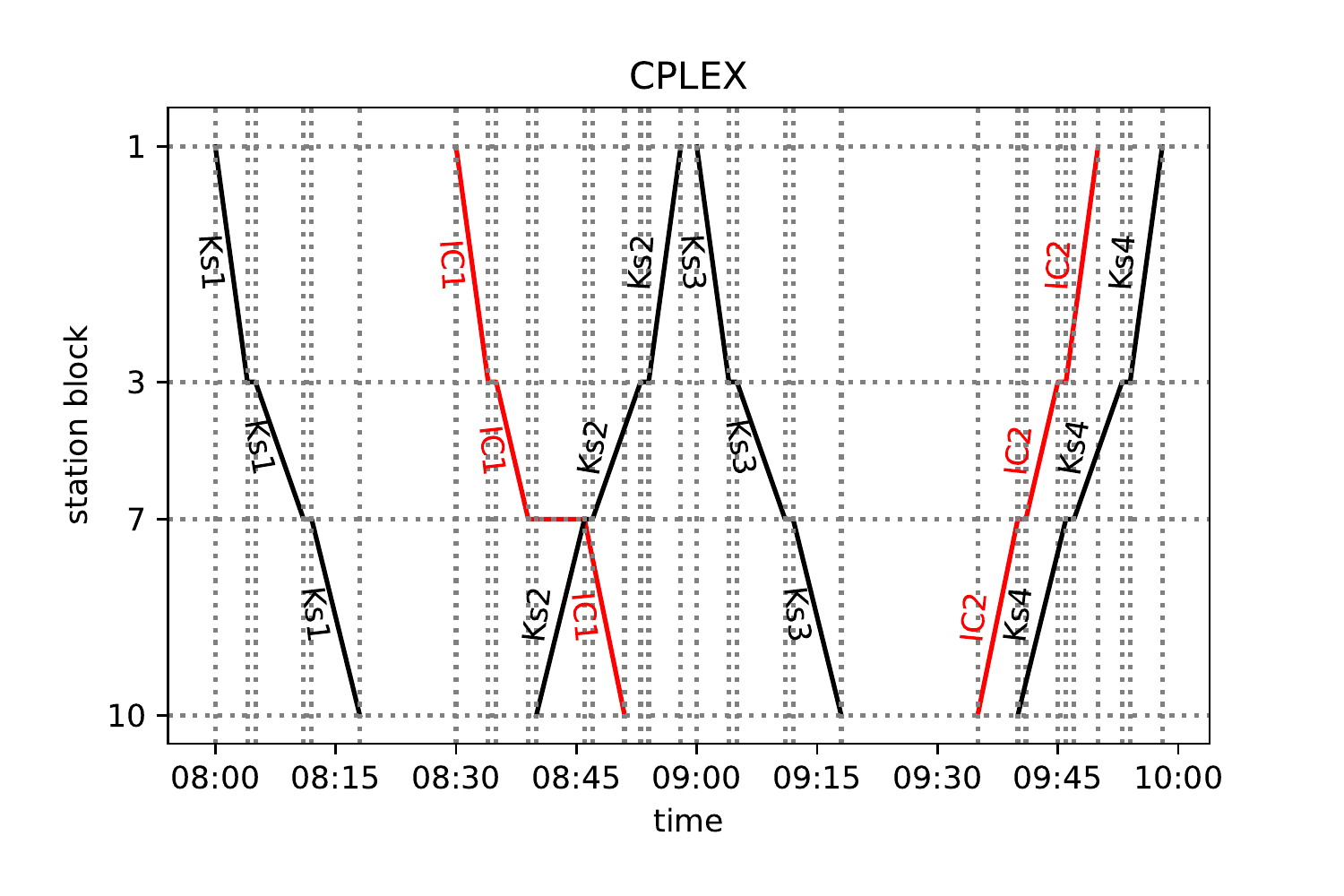}}
	\subfigure[Case $2$ -- no uncacceptable
	stopovers.\label{fig::c2_cplex}]{\includegraphics[scale = 
		0.55]{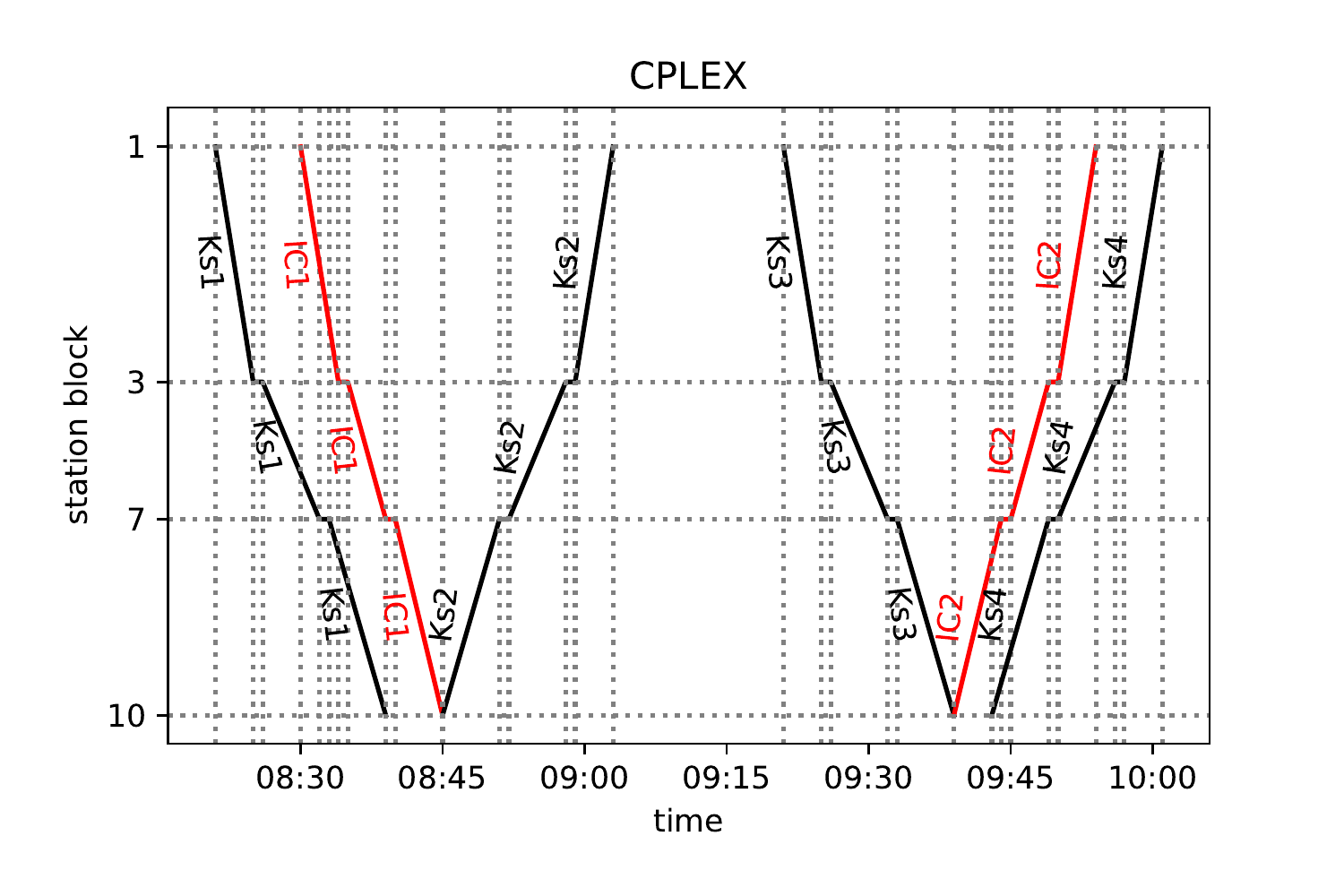}} \\
	\subfigure[Case $3$ -- no unacceptable
	stopovers.\label{fig::c3_cplex}]{\includegraphics[scale = 
		0.55]{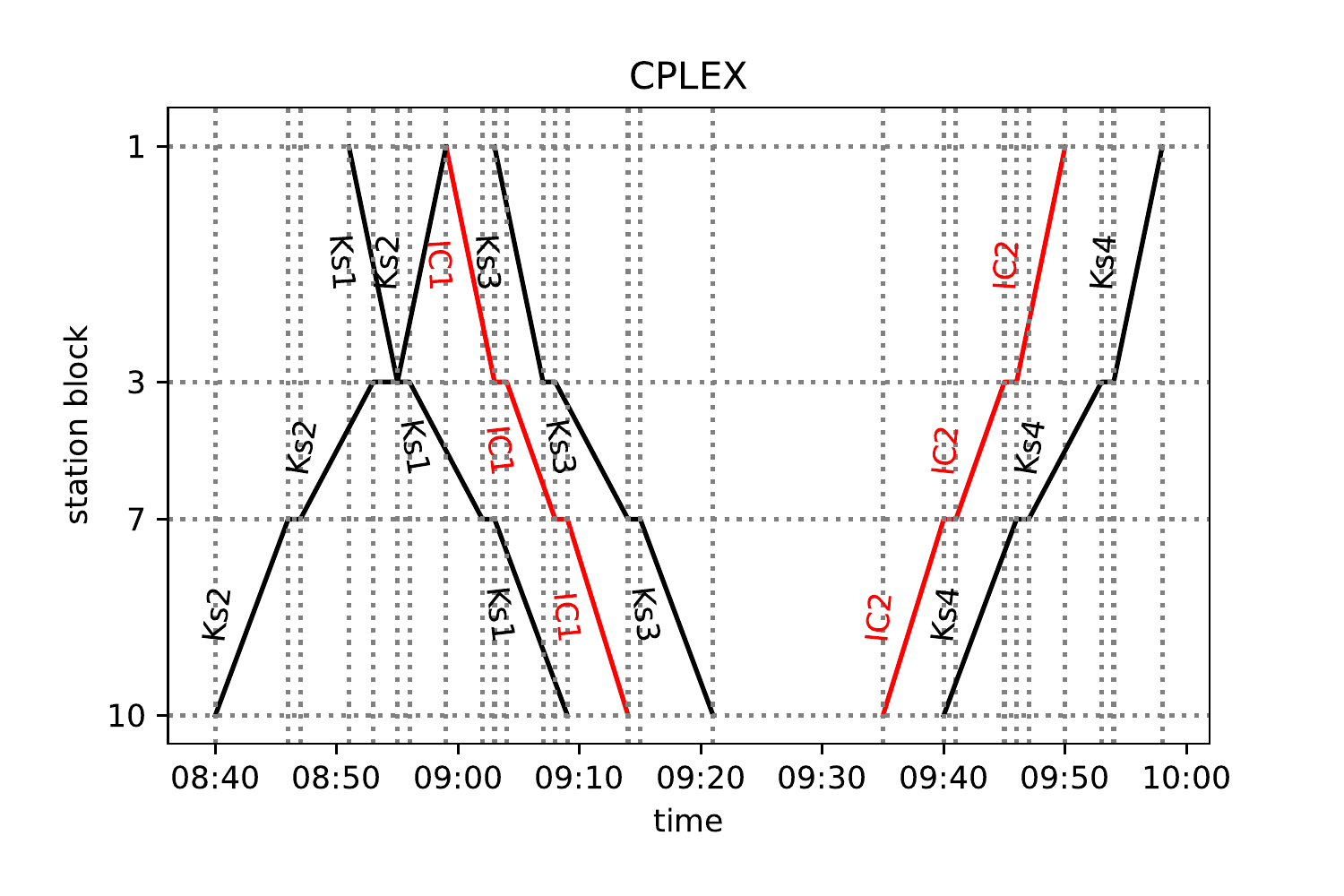}}
	\subfigure[Case $4$ -- optimal solution according to all 
	methods.\label{fig::c4_cplex}]{\includegraphics[scale = 
		0.55]{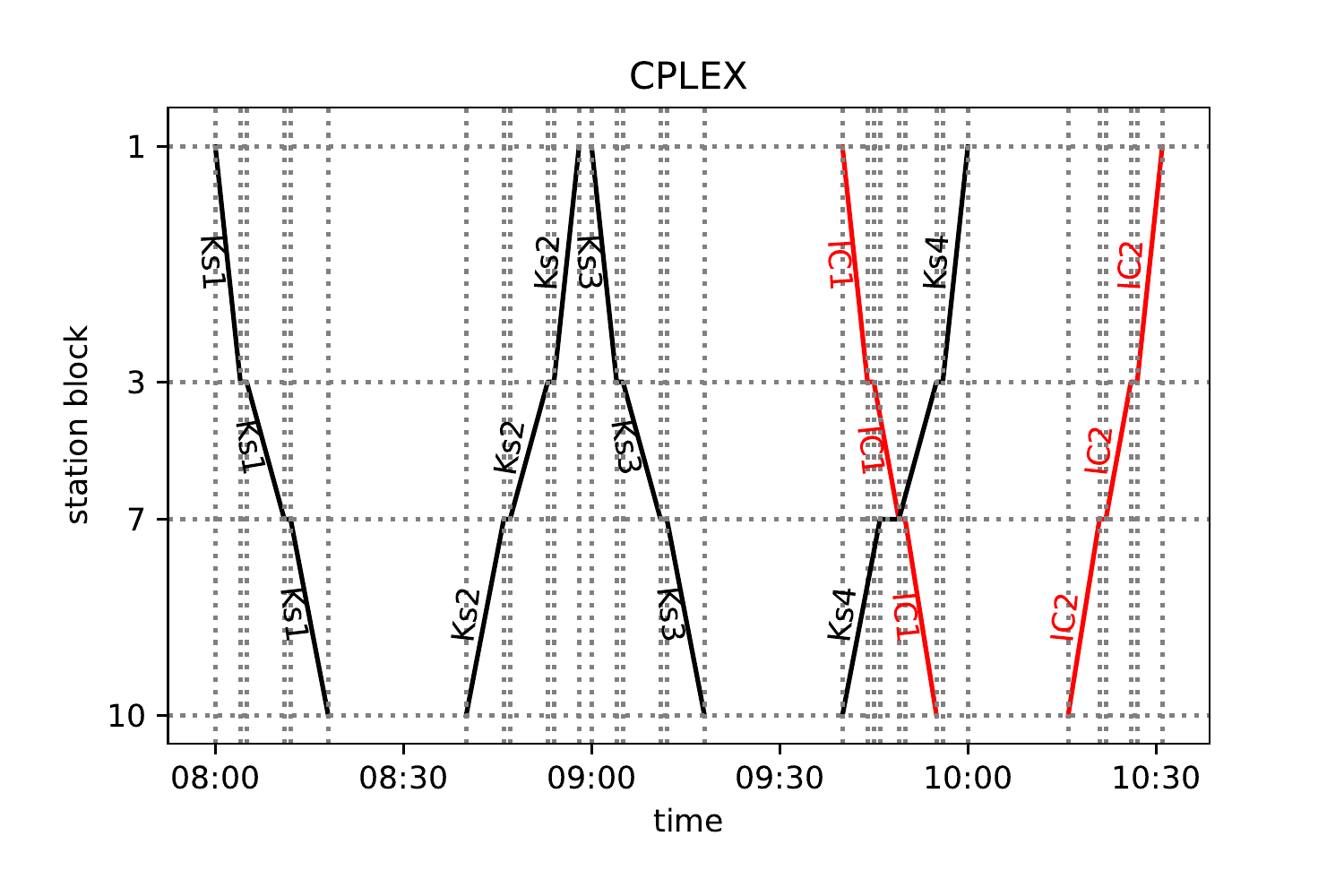}}
	\caption{The CPLEX solutions: exact ground 
	states of the QUBOs. There are no unacceptably long stopovers. Further, the 
	trains' prioritization and the delay propagation to subsequent trains are 
	taken into 
	account. The solutions are the same as these of the linear solver.}\label{fig::large-sim-cplx}
\end{figure}

\begin{figure}
	\subfigure[Case $1$ -- ground 
	state of the QUBO; the degeneracy of the ground state is reflected by a 
	stay of IC$1$ both at block $3$ and at block
	$7$.\label{fig::c1_tn}]{\includegraphics[scale = 
		0.55]{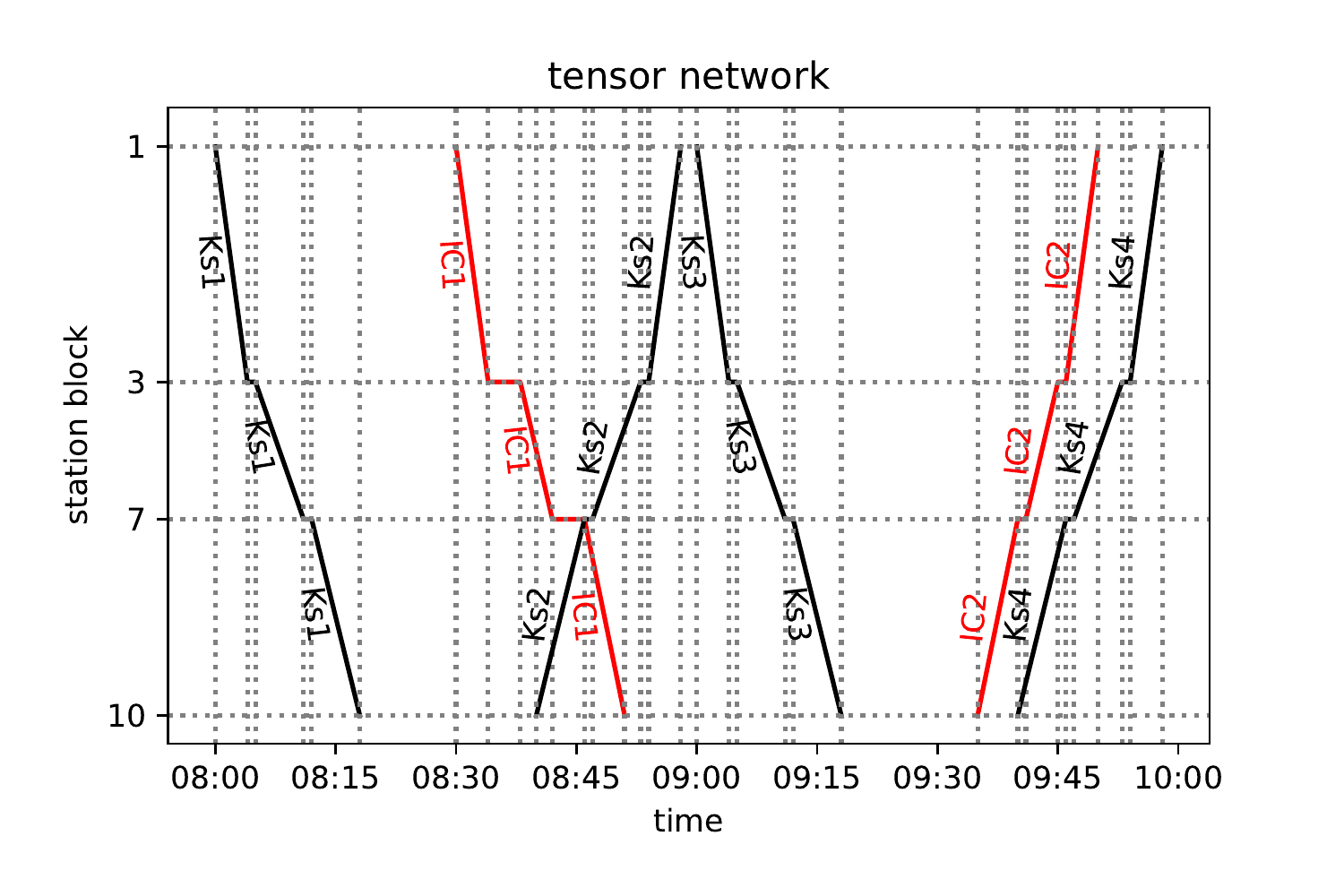}}
	\subfigure[Case $2$ -- ground 
	state of the QUBO.\label{fig::c2_tn}]{\includegraphics[scale = 
		0.55]{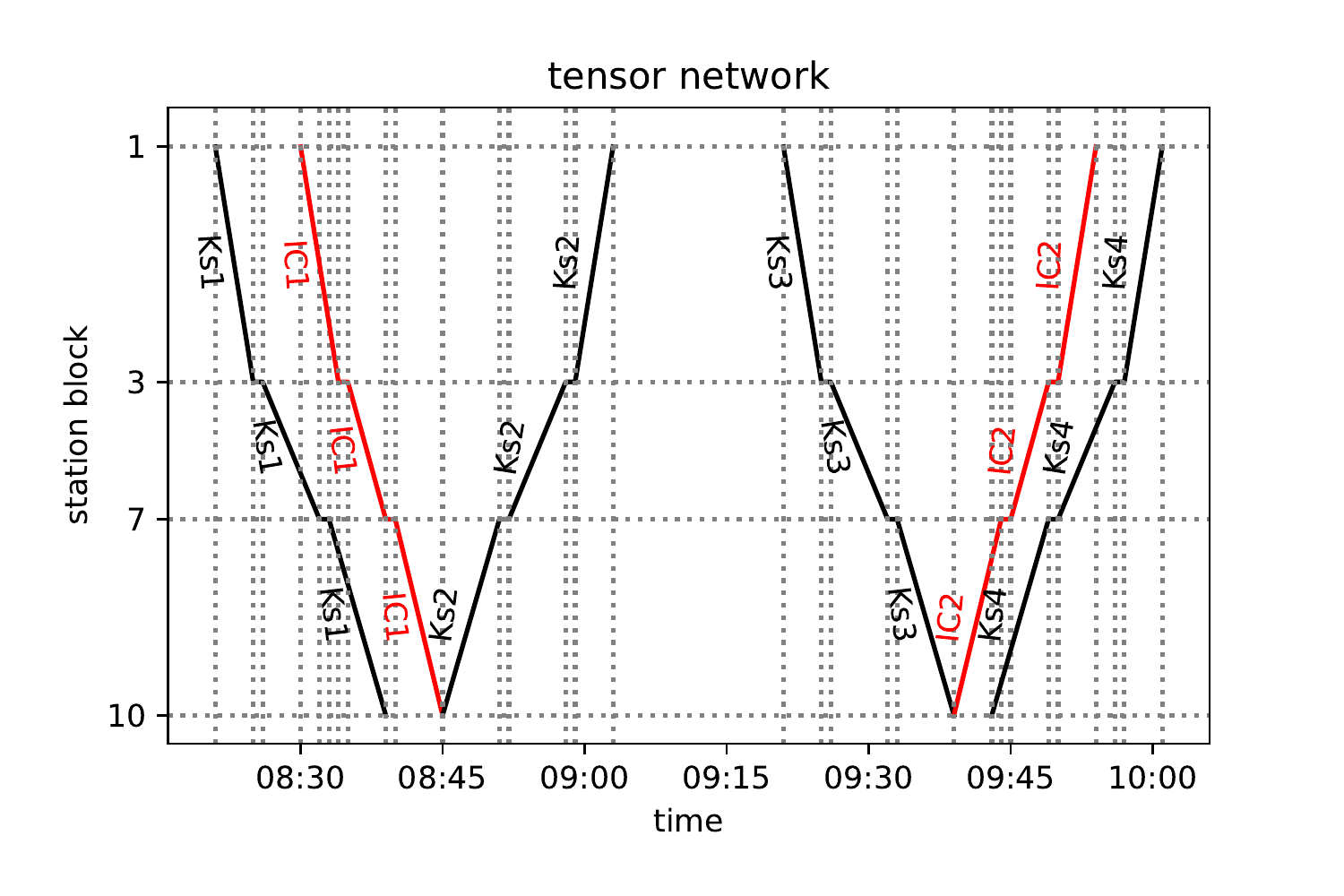}} \\
	\subfigure[Case $3$ -- excited state of the QUBO; 
	notice the slightly longer stay of 
	IC$1$ at block $7$.\label{fig::c3_tn}]{\includegraphics[scale = 
		0.55]{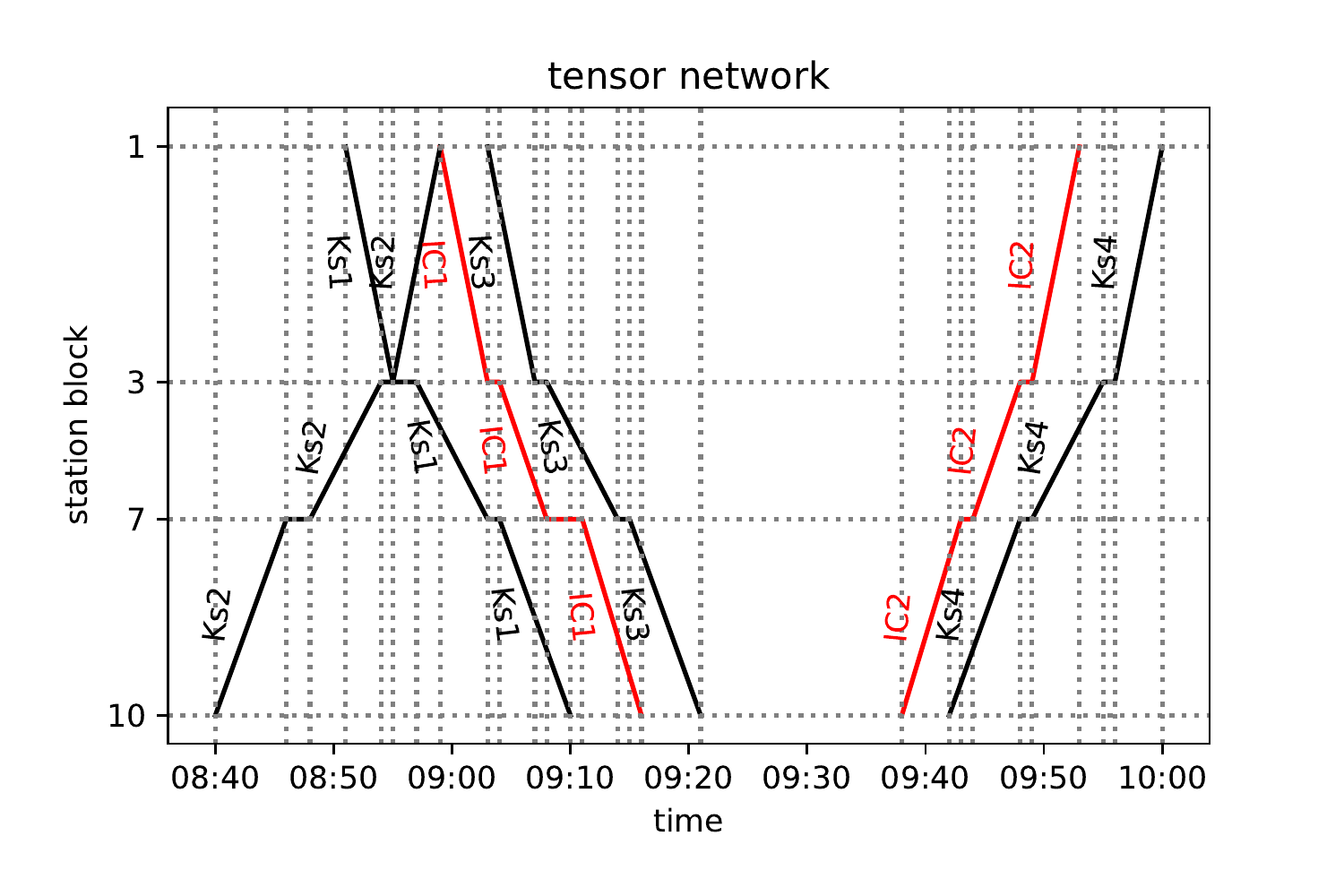}}
	\subfigure[Case $4$ -- excited state of the QUBO; notice the slightly 
	longer stay of 
	Ks$3$. 
	\label{fig::c4_tn}]{\includegraphics[scale = 
		0.55]{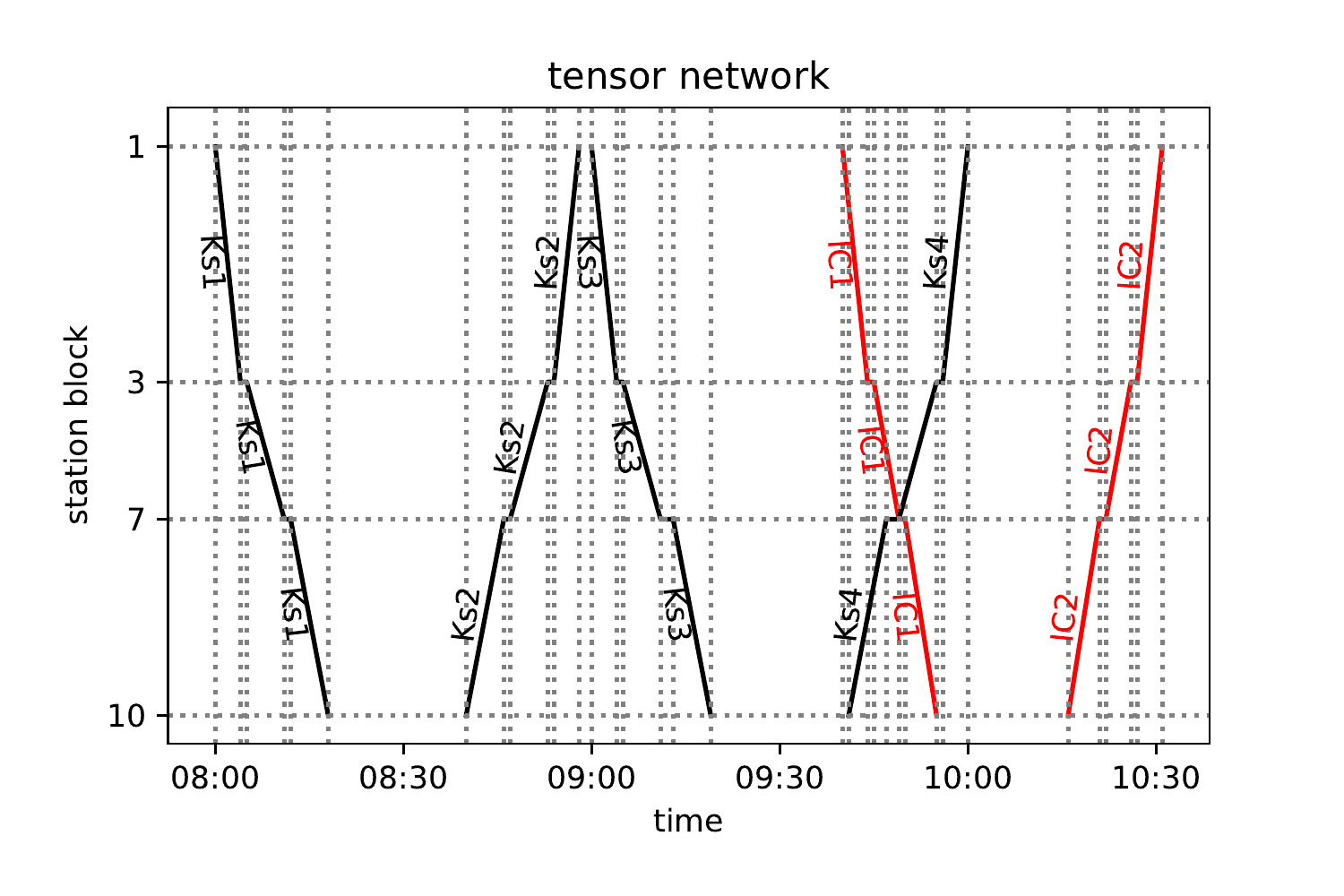}}
      \caption{The tensor network solutions; although the
        exact ground states were not always achieved, the solutions are
        equivalent from the dispatching point of view with to in
        Fig.~\ref{fig::large-sim-cplx}.}\label{fig::large-sim-tn}
\end{figure}

\bibliographystyle{apsrev4-2}
%